\documentclass{article}
\usepackage{a4wide}
\usepackage{amssymb}       
\usepackage{amsthm}        
\usepackage{amsmath}
\usepackage{natbib}

\def\marginpar#1{}
\def\endnote#1{}{}

\def\boldrho{{\boldsymbol\rho}}
\def\boldsigma{{\boldsymbol\sigma}}

\def\boldone{{\mathbf 1}}
\def\boldzero{{\mathbf 0}}

\def\ket#1{{ | #1 \rangle }}
\def\bra#1{{ \langle #1 | }}
\def\braket#1#2{{ \langle #1  |#2 \rangle }}
\def\ketbra#1#2{{\ket{#1}\bra{#2}}}

\theoremstyle{plain}

\newtheorem{thm}{Theorem}
\newtheorem{prop}{Proposition}
\newtheorem{lemma}{Lemma}
\newtheorem{cor}{Corollary}

{
\theoremstyle{definition}       

\newtheorem{defn}{Definition}
\newtheorem{example}{Example}
\newtheorem{rmk}{Remark}
\newtheorem*{prf}{Proof}        
}

\title{On Quantum Statistical Inference}
\author{Ole E. Barndorff-Nielsen\\
MaPhySto
\thanks{MaPhySto is the Centre for Mathematical Physics and Stochastics,
funded by the Danish National Research Foundation, University of Aarhus, Denmark.}\\
\\ Richard D. Gill\\
Mathematical Institute, University of Utrecht and
\\
EURANDOM, Eindhoven, Netherlands\\
\\ Peter E. Jupp\\
School of Mathematics and Statistics, University of St Andrews, U.K.}
\date{October 1, 2001}

\begin{document}
    

\maketitle

\begin{abstract}
Recent developments in the mathematical foundations of quantum mechanics have
brought the theory closer to that of classical probability and statistics. 
On the other hand, the unique character of quantum physics sets
many of the questions addressed apart from those met classically in
stochastics. Furthermore, concurrent advances in experimental techniques 
and in the theory of quantum computation have
led to a strong interest in questions of quantum information, in 
particular in the sense of the amount of information about unknown 
parameters in given observational data or accessible through various possible 
types of measurements. This scenery is outlined. (A shorter version of the paper,
omitting some topics but otherwise much improved, 
is available as \texttt{quant-ph/0307191}).
\end{abstract}

\newpage              
\tableofcontents      
\newpage              

\section{Introduction}\label{s:intro}

In the last two decades, developments of an axiomatic type in the mathematical
foundations of quantum mechanics have brought the theory closer to that of
classical probability and statistics. On the other hand, the
unique character of quantum physics 
(we use the terms `quantum mechanics' and 
`quantum physics' synonymously) 
sets many of the questions addressed apart
from those met classically in stochastics. The key mathematical notion is that
of a \emph{quantum instrument}, which we shall describe in Section
\ref{s:states} and which, for arbitrary quantum experiments, 
specifies the joint probability distribution of the observ\-ational outcome 
of the experiment together with the state of the physical system after the 
experiment. 
Concurrently with these theoretical developments, major advances in
experimental techniques have opened many possibilities for studying small
quantum systems and this has led to considerable current interest in a range
of questions that in essence belong to statist\-ical inference and are concerned
with the amount of information about unknown para\-meters in given observational 
data or accessible through various possible types of measurements.
In quantum physics, the realm of possible experiments is specified
mathematically, and noncommutativity between experiments 
plays a key role. Separate measurements on independent and separate systems 
result in independent observ\-ations, as in classical stochastics. 
However, joint measure\-ments allow for major
increases in statistical information.

The present paper outlines some of these developments and contains suggestions
for additional readings and further work.
We make some new contributions to the theory of quantum statistical inference,
in particular, developing new notions of quantum sufficiency and exhaustivity. 
We give complete but short proofs of the quantum information 
(Cram\'er--Rao) bound and some of its consequences, filling some gaps 
in the proofs in the physics literature. 
The paper does \emph{not} contain practical examples in the 
sense of real data analyses, for several reasons. For one thing, 
the realistic modelling of present-day laboratory experiments in this field 
involves several more layers of complexity (technical, not conceptual)
on top of the picture presented here. The closest we come to real 
data is in our discussion of quantum tomography in Section \ref{ss:tomo}. 
For another thing, the theory in this 
paper is largely concerned with the design rather than the analysis 
of experiments in quantum physics, and there is still a gap between 
what is theoretically possible under the laws of quantum mechanics, 
and what is practically possible in the laboratory, though this gap is 
closing fast.
`Information' is understood throughout in the sense it has in mathematical statistics.
We do not discuss quantum information theory in the sense of
optimal coding and transmission of messages through quantum communication 
channels, nor in the more general sense of quantum information 
processing \citep{green00}.
Within quantum statistics, we concentrate on the topics of estimation and 
of inference. The classic books of \citet{helstrom76} and \citet{holevo82} 
are on the other hand largely devoted to a decision theoretic approach to hypothesis 
testing problems. See \citet{parthasarathy99} and \citet{ogawanagaoka00}
for recent 
contributions to this field. Confusingly, the phrase `maximum 
likelihood estimator' has an unorthodox meaning in the older literature.
In many papers of which we just mention a few recent ones,
\citet{belavkin94,belavkin00,belavkin01} develops a 
continuous time Bayesian filtering approach to estimation and control.

It should be emphasised from the start, that we see quantum mechanics 
as describing \emph{classical probability models} for the outcomes of 
laboratory experiments, or indeed, for the real world outcomes of any 
interactions between `the quantum world' of microscopic particles
and `the real world' in which statisticians analyse data. 
Those probability models may depend on
unknown parameters, and quantum statistics is concerned with 
statistical design and inference concerning those parameters. This 
point of view is commonplace in experimental quantum physics but seems 
to be less common in theoretical physics and in some parts of pure 
mathematics, in particular in the field called `quantum probability', 
where a special nature is claimed for the randomness of quantum 
mechanics, placing it outside the ambit of classical 
probability and statistics. We disagree firmly with this conclusion 
though we do agree that there are fascinating foundational issues in 
quantum mechanics. We develop our stance on these issues further in
Section \ref{quprob}. Quantum mechanics is concerned 
with randomness of the most fundamental nature known to science, and 
probabilists and statisticians definitely should be involved in the game, 
rather than excluded from 
it.
    
In quantum mechanics the state of a physical system is 
described by a non-negative self-adjoint operator $\rho$ (referred to as the 
\textit{state}) with trace $1$, on a separable complex Hilbert space 
$\mathcal{H}$. In accordance with the previous paragraphs, our interest in this 
paper concerns cases where the state is specified only up to some unknown 
parameter $\theta$ and the question is what can be learned about the parameter from 
observation of the system.

Many of the central ideas can be illustrated by finite-dimensional 
quantum systems, the simplest being based on those 
in which $\mathcal{H}$ has (complex) dimension 2.  
We shall often use the phrase `spin-half particle' to refer to such a 
quantum system, as one of the best known examples concerns the 
magnetic moment or spin of the electron, which in appropriate units 
can only take on the values $\pm\frac12$. But a two-dimensional 
state space is also appropriate for modelling the polarisation of one 
photon, and yet another example is provided by an atom at very low temperature
when only its ground state and first excited state are relevant.
The theory of quantum computation is concerned with how a finite collection 
of two-dimensional quantum systems, which are then called \emph{qubits}, 
can be used to carry and manipulate information.
We shall mainly concentrate on such examples.
However, many physical problems 
concern infinite-dimension\-al systems, one area of great current interest 
being quantum tomo\-graphy and quantum holography,
which we shall discuss briefly. 
While the theory for finite-dimensional systems can be outlined in relatively 
simple math\-ematical terms, in general it is necessary to draw on 
advanced aspects of the theory of operators on infinite-dimensional Hilbert 
spaces and we will only outline this, with quantum tomography
in mind, in Section \ref{s:further}.

The paper is organised as follows. Section \ref{s:states} describes
the mathematical structure linking states of a quantum system,
possible measurements on that system, and the resulting state of
the system after measurement. Section \ref{s:lik} introduces quantum
statistical models and notions of quantum score and quantum
information, parallel to the score function and Fisher information
of classical parametric statistical models. In Section \ref{s:qeqtm} we 
introduce quantum exponential models and quantum transformation 
models, again forming a parallel with fundamental classes of
models in classical statistics. In Section \ref{s:exhaust} we describe
the notions of quantum exhaustivity and quantum sufficiency of 
a measurement, relating them to the classical notion of sufficiency. We next,
in Section \ref{s:info}, turn to a study of the relation between 
quantum information and classical
Fisher information, in particular through Cram\'er--Rao type
information bounds. In Section \ref{s:further} we discuss the
infinite-dimensional model of quantum tomography, which poses the challenge
of developing non-parametric quantum information bounds. In
Section \ref{quprob} we discuss the difference between classical and 
quantum probability and statistics, relating them both to 
foundational issues in quantum physics and to emerging quantum 
technologies. Finally in Section \ref{s:conclude} we conclude with
remarks on further topics, in particular, quantum stochastic processes.
The appendix contains some mathematical details.

This paper greatly extends our more mathematical survey 
\citep*{barndorffnielsenetal01} on quantum statistical information.
\citet{gill01,gill01b} contains further introductory material.
Many proofs and further details will be found in \citet*{barndorffnielsenetal02}.
Some general references which we have found extremely useful are the 
books of \citet{isham95}, \citet{peres95}, \citet{gilmore94},
\citet{holevo82,holevo01book}. 
Finally, the Los Alamos National 
Laboratory preprint service for quantum physics, \texttt{quant-ph} at 
\texttt{http://xxx.lanl.gov} is an invaluable resource. 

\section{States, Measurements and Instruments}\label{s:states}

In quantum mechanics the \emph{state} of any physical system to be
investigated is described by an operator $\rho$ on a complex separable Hilbert
space $\mathcal{H}$ such that $\rho$ is non-negative and (hence)%
\marginpar{*}\endnote{R: book: ref nonneg implies SA} 
self-adjoint and has trace $1$.
In this paper (except for Section \ref{s:further}) we
shall restrict attention to the case where $\mathcal{H}$ is
finite-dimensional, and our examples will mainly concern 
the spin of spin-half particles, where the dimension of $\mathcal{H}$ is 
$2$. 
The classic example in this context is the 1922 experiment of Stern and 
Gerlach, see \citet[{\ }Section 1.4]{brantdahmen95}, to determine the
size of the magnetic moment of the electron. The electron was conceived
of as spinning around an axis and therefore behaving as a magnet
pointing in some direction. Mathematically, each electron carries a
vector `magnetic moment'. One might expect the size of the magnetic
moment of all electrons to be the same, but the directions to be
uniformly distributed in space. Stern and Gerlach made a beam of silver
atoms move transversely through a steeply increasing 
vertical magnetic field. A silver atom has 47 electrons 
but it appears that the magnetic
moments of the 46 inner electrons cancel and essentially only one
electron determines the spin of the whole system. Classical physical
reasoning predicts that the beam would emerge spread out vertically
according to the component of the spin of each atom (or
electron) in the direction of the gradient of the magnetic field. 
The spin itself would not be altered by passage through the magnet.
However, amazingly, the emerging beam consisted of just two well
separated components, as if the component of the spin vector in the
vertical direction of each electron could take on only two different
values. 

In this case, $\mathcal{H}$ can be thought of as 
$\mathbb C ^{2}$, i.e.\ as pairs of
complex numbers, and, correspondingly, $\rho$ is a $2\times2$ matrix
$$
\left(
\begin{tabular}
[c]{cc}
$\rho_{11}$ & $\rho_{12}$\\
$\rho_{21}$ & $\rho_{22}$
\end{tabular}
\right)
$$
with $\rho_{21}=\overline\rho_{12}$ (the bar denoting complex conjugation) and
non-negative real eigen\-values $p_{1}$ and $p_{2}$ 
satisfying $p_{1}+p_{2}=1$.

The result of performing a measurement on the system in state $\rho$ is a
random variable $x$ taking values in a measure space 
$(\mathcal{X},\mathcal{A)}$ and with law of the form
$$
\Pr(x\in A)=\mathrm{tr}\{\rho M(A)\}
\thinspace ,
$$
where $M$ is a mapping from the $\sigma$-algebra $\mathcal{A}$ into the space
${\mathbb {SA} }_{+}(\mathcal{H})$ of non-negative self-adjoint operators on
$\mathcal{H}$ which satisfies $M(\mathcal{X})=\boldone$ (where 
$\boldone $\ is the identity operator) and
$$
\sum_{i=1}^{\infty}M(A_{i})=M(A)
$$
for any finite or countable sequence $\{A_{1},A_{2},\dots\}$ of disjoint
elements of $\mathcal{A}$ and $A=\cup_{i=1}^{\infty}A_{i}$ (the sum in the
formula being defined in the sense of weak convergence of operators). Such a
mapping $M$ is said to be a (generalised) \emph{measurement}. We 
shall also refer to $M$ as an \emph{operator-valued probability measure} or 
\emph{OProM} for short. In the literature the usual names and acronyms are 
probability operator-valued measure or
positive operator-valued measure (POM or POVM), and 
(nonorthogonal, generalised) resolution of the identity.

The most basic measurements, which are among the class of \emph{simple measurements}
defined in Section \ref{ss:meas}, 
have $\mathcal{X}$ a finite set of real numbers, with cardinality less
than or equal to the dimension of $\mathcal{H}$, $\mathcal{A}$ as the
$\sigma$-algebra of all subsets of $\mathcal{X}$,  
$M(\{x\})=\Pi_{[x]}$ for any atom $\{x\}$ of $\mathcal{A}$, 
the $\Pi_{[x]}$ being mutually orthogonal \emph{projection operators} with 
$\sum\Pi_{[x]} =\boldone $. We speak then of a 
\emph{projector-valued probability measure} or \emph{PProM}. 
The usual terminology in the literature
is a PVM or (orthogonal) resolution of the identity.
All the ingredients of such a simple measurement 
are encapsulated in the specification of a self-adjoint operator 
$Q$ on $\mathcal{H}$ with eigenvalues $x$ in $\mathcal{X}$ and eigenspaces which are
precisely those subspaces onto which the $\Pi_{[x]}$ project. 
The operator $Q=\sum x\Pi_{[x]}$ is called the \emph{observable}. 
Conversely, any self-adjoint operat\-or
on $\mathcal{H}$ can be given an interpretation as an observable. 
We denote the space of self-adjoint operators (observables) by 
${\mathbb {SA} }(\mathcal{H})$ and the set of states $\rho$ by 
$\mathcal{S}(\mathcal{H})$. The adjoint of an operator is indicated by 
an asterisk $\ast$.

Physics textbooks on quantum theory usually
take the concept of observables as a starting point.
In the infinite dimensional case, observables---self-adjoint operators,
not necessarily bounded---may have continuous spectrum instead of
discrete eigenvalues. But the one-to-one correspondence between PProM's 
and observables continues to hold. Any self-adjoint operat\-or
on $\mathcal{H}$ can be given an interpretation as an observable. 

Let $M$ be a measurement. We shall often assume that $M$ is dominated by a
$\sigma$-finite measure $\nu$ on $(\mathcal{X},\mathcal{A})$ and we shall
write $m(x)$ for the density of $M$ with respect to $\nu$. Thus
$$
M(A)=\int_{A}m(x)\nu(\mathrm d x)
\thinspace  .
$$
We can take $m(x)$ to be self-adjoint and nonnegative for all $x$.
(If $\mathcal{H}=\mathbb C ^{d}$, then $M(A)$ and $m(x)$ may be considered as
$d\times d$ matrices of complex numbers.) The law of $x$ is also dominated by $\nu$
and the probability density function of $x$ is
$$
p(x)=\mathrm{tr}\{\rho m(x)\}
\thinspace .
$$

The physical state may depend on an unknown parameter $\theta$, which runs
through some parameter space $\Theta$. In this case we denote the state by
$\rho(\theta)$. Then the law of the outcome $x$ of a measurement $M$ depends
on $\theta$ as well and we indicate this by writing $\mathrm P_{\theta}(A)$ or
$p(x;\theta)$ for the probability or the probability density, as the 
case may be. In particular,
\begin{equation}\label{density}
p(x;\theta)=\mathrm{tr}\{\rho(\theta)m(x)\}
\thinspace .
\end{equation}
It may also be relevant to stress the dependence on $M$ and we then write
$p(x;\theta;M)$, etc. We shall refer to the present kind of setting 
as a \emph{parametric quantum model} $(\boldrho,M)$ or $(\boldrho,m)$ with elements 
$\boldrho =(\theta\mapsto\rho(\theta))$ and $M$, or its density $m$.
It is also relevant to consider cases where the 
measurement $M$ depends on some unknown parameter, but we shall not 
discuss this possibility further in the present paper. When the
measurement $M$ is given, a problem of classical statistical inference 
results concerning the model (\ref{density}) for the distribution
of the outcome. However, it turns out 
that the model for the state $\theta\mapsto\rho(\theta)$ 
can be usefully studied independently of 
which measurement is made of the system (or in order to choose the 
best measurement) and then quantum analogues of many concepts from 
classical statistical inference become important.

OProM's specify the probabilistic law of the outcome of an actual measure\-ment
but do not say anything about the state of the physi\-cal system after the
measurement has been performed. The math\-ematical concept of 
\emph{quantum instrument} prescribes both the OProM for the measure\-ment and 
the posterior state.

The next three subsections discuss in more detail the concepts of states,
measurements (or OProM's), and quantum instruments.

\subsection{States}\label{ss:states}

As stated at the beginning of the section, the state of a quantum
system is represented by an \emph{operator} $\rho$ in 
$\mathcal{S}(\mathcal{H})$. It is often called the 
\emph{density matrix} or \emph{density operator} of the system.
We think of vectors $\psi$ in $\mathcal{H}$ as column vectors,
and will emphasise this by writing $| \psi \rangle$ (Dirac's \lq ket\rq\  notation).
The adjoint (complex conjugate and transpose) of $| \psi \rangle$ is a row
vector, which we denote by $\langle \psi |$ (Dirac's \lq bra\rq\ notation).

The simplest states, called \emph{pure states}, are the projectors 
of rank one, i.e.\ they are of the form 
$\rho=| \psi\rangle\langle\psi |$,
where $\psi$ is a unit vector in $\mathcal{H}$ (so 
$\langle\psi|\psi\rangle=1$), called the
\emph{state-vector} of the pure state $\rho$. If $\mathcal{H}$ 
has dimension $d$ then the set $\mathcal{S}_{1}(\mathcal{H})$ of pure states 
can be identified with the complex projective space $\mathbb C  P^{d-1}$. 
In particular, $\mathcal{S}_{1}(\mathbb C ^{2})$ can be identified 
with the sphere $S^{2}$, which is known in theoretical physics as the 
\textit{Poincar\'{e} sphere}, in quantum optics as the \textit{Bloch 
sphere}, and in complex analysis as the \textit{Riemann sphere}.

\begin{example}[Spin-half]\label{e:spin12}
Take $\mathcal{H} =\mathbb C ^{2}$, so that $\mathcal{H}$ has complex dimension 2, 
the space of general operators on $\mathcal{H}$ has real dimension 8,
and the space ${\mathbb {SA} }(\mathcal{H})$ of self-adjoint operators on 
$\mathcal{H}$ has real dimension 4. 

The space ${\mathbb {SA} }(\mathcal{H})$ is spanned by the identity matrix
$$
\boldone =\sigma_{0}=\left(
\begin{tabular}
[c]{cc}
$1$ & $0$\\
$0$ & $1$
\end{tabular}
\right)  
\thinspace ,
$$
together with the \emph{Pauli matrices}
$$
\sigma_{x}=\left(
\begin{tabular}
[c]{cc}
$0$ & $1$\\
$1$ & $0$
\end{tabular}
\right)  \quad\sigma_{y}=\left(
\begin{tabular}
[c]{cc}
$0$ & $-i$\\
$i$ & $0$
\end{tabular}
\right)  \quad\sigma_{z}=\left(
\begin{tabular}
[c]{cc}
$1$ & $0$\\
$0$ & $-1$
\end{tabular}
\right)  
\thinspace .
$$
Note that $\sigma_{x},\sigma_{y}$ and $\sigma_{z}$ satisfy the
\emph{commutativity relations}
\begin{align*}
\lbrack \sigma_{x},\sigma_{y}] ~&=~ 2i\sigma_{z}  \\
\lbrack \sigma_{y},\sigma_{z}] ~&=~ 2i\sigma_{x}  \\
\lbrack \sigma_{z},\sigma_{x}] ~&=~ 2i\sigma_{y} 
\end{align*}
where, for any operators $A$ and $B$, their commutator $[A,B]$ 
is defined as $AB - BA$; and note that
$$
\sigma_{x}^{2}=\sigma_{y}^{2}=\sigma_{z}^{2}=\boldone 
\thinspace .
$$

Any pure state has the form $|\psi\rangle\langle\psi|$ for some 
unit vector $|\psi\rangle$ in $\mathbb C ^{2}$. 
Up to a complex factor of modulus $1$ (the \emph{phase}, which does
not influence the state), we can write $|\psi\rangle$ as
$$
|\psi\rangle =
\left( 
\begin{matrix}
e^{-i\varphi/2}\cos(\vartheta/2) \\
e^{i\varphi/2}\sin (\vartheta/2)
\end{matrix}
\right)
\thinspace .
$$
The corresponding pure state is
$$
\rho=\left(
\begin{matrix}
\cos^{2}(\vartheta/2) & e^{- i\varphi}\cos(\vartheta/2)\sin(\vartheta/2)\\
e^{i\varphi}\cos(\vartheta/2)\sin(\vartheta/2) & 
\sin^{2}(\vartheta/2)
\end{matrix}
\right)  
\thinspace .
$$
A little algebra shows that $\rho$ can be written as 
$\rho=(\boldone +u_{x}\sigma
_{x}+u_{y}\sigma_{y}+u_{z}\sigma_{z})/2=
\frac{1}{2}(\boldone +\vec{u}\cdot\vec{\sigma})$, 
where $\vec{\sigma}=(\sigma_{x},\sigma_{y},\sigma_{z})$ are the three 
Pauli spin matrices and $\vec{u}=(u_{x},u_{y},u_{z})=\vec{u}(\vartheta,\varphi)$ 
is the point on $S^{2}$ with polar coordinates $(\vartheta,\varphi)$.
\hfill$\ \qedsymbol$\end{example}

\subsubsection{Mixing and Superposition}\label{sss:super}

There are two important ways of constructing new states from old. Firstly,
since the set of states is convex, new states can be obtained by 
mixing states $\rho_{1}, \dots, \rho_{m}$, i.e. taking convex combinations
\begin{equation}
p_{1} \rho_{1} + \dots + p_{m} \rho_{m} 
\thinspace , \label{mixed}
\end{equation}
where $p_1, \dots , p_m$ are real with $p_{i}\geq0$ and 
$p_{1}+ \dots+p_{m}=1$. 
If $\mathcal{H}$ is finite-dimensional then all states are of the 
form (\ref{mixed}) with the $\rho_{i}$ pure, so that $\mathcal{S}(\mathcal{H})$ 
is the convex hull of $\mathcal{S}_{1}(\mathcal{H})$:
in the infinite-dimensional case one needs infinite mixtures.
For this reason, states which are not pure are called \emph{mixed states}.
In particular, if $\mathcal{H} = \mathbb C ^2$ then the set of pure states is 
the Poincar\'{e} sphere, whereas the set of mixed states is the interior of 
the corresponding unit ball. 

If $\mathcal{H} = \mathbb C ^{d}$ then mixing the pure states
by the uniform probability measure on $\mathbb C  P^{d-1}$ gives a state
which is invariant under the action $\rho\mapsto U \rho U^{*}$ of 
$\mathrm{SU}(d)$, the group of special (determinant $+1$) unitary 
($U U^{*}=U^{*}U=\boldone $) matrices of order $d$;  
this is the unique such invariant state.

The other important way of constructing new states from old is by superposition. 
The \emph{super\-position principle} states that a complex linear
combination of state-vectors is also a physically possible state-vector. Let 
$|\psi _1 \rangle \langle \psi _1|$, \dots , 
$|\psi _m \rangle \langle \psi _m|$ be pure states on 
$\mathcal{H}$.
Then any state which can be written in the form
$\langle \psi|\psi \rangle ^{-1} |\psi \rangle \langle \psi |$, where
$$
\psi = w_{1}\psi _{1}+ \dots +w_{m}\psi _{m} 
$$ 
and $w_{1},\dots ,w_{m}$ are some complex numbers, is called a \emph{superposition} 
of the pure states with state-vectors $|\psi _{1} \rangle$, \dots , 
$|\psi _{m} \rangle $ (here the phases of the state-vectors
are relevant!).

The difference between superposition and mixing may be illustrated by 
a spin-half example: take $\langle\psi_{1}|=(1,0)$ and $\langle\psi_{2}|=(0,1)$. 
For the super\-position with $w_1 = w_2=1/\sqrt{2}$, we have
$$
\langle \psi|\psi \rangle ^{-1} |\psi \rangle \langle \psi | = 
{\textstyle{\frac12}}\left(
\begin{matrix}
1 & 1\\
1 & 1
\end{matrix}
\right)  
\thinspace ,
$$
whereas the mixed state 
$$
p_{1}|\psi_{1}\rangle\langle\psi_{1}|+p_{2}\ |\psi_{2}
\rangle\langle\psi_{2}|
={\textstyle{\frac12}}\left(
\begin{matrix}
p_{1} & 0\\
0 & p_{2}
\end{matrix}
\right)  
\thinspace 
$$
is different from the preceding superposition, whatever $p_{1}$ and $p_{2}=1-p_{1}$.
Taking $p_{1}=p_{2}=\frac12$,
if we measure the PProM defined by the two projectors
$|\psi_1 \rangle \langle \psi_1 |$ and $|\psi_2 \rangle \langle \psi_2 |$
and corresponding outcomes $+1$ and $-1$,
the two states are indistinguishable:
each gives probabilities of $1/2$ for the two outcomes. 
However, if we measure 
$\langle \psi|\psi \rangle ^{-1} |\psi \rangle \langle \psi |$ and
$|\psi \rangle ^{\perp} \langle \psi |^{\perp}$,
where $|\psi \rangle ^{\perp}$ denotes a unit vector in $\mathbb C ^2$ 
orthogonal to $|\psi \rangle$, then the second state again gives each outcome 
probability half, while the first state gives probabilities 1 and 0.

The possibility of taking complex superpositions of state-vectors to get 
new pure states corresponds to the wave-particle duality at the heart of
quantum mechanics (linear combinations of solutions to wave equations
are also solutions to wave equations). The new states obtained in 
this way will have distinctively different properties from the states out
of which they are constructed. On the other hand, taking mixtures of states
represents no more and no less than ordinary probabilistic mixtures:
with probability $p_i$ the system has been prepared in state $\rho_i$, 
for $i = 1, \dots , m$.
It is a fact that whatever physical predictions one makes about
a quantum system, they will depend on the $| \psi_i \rangle$ and 
on the $p_{i}$ or $w_i$ 
involved in mixed states or superpositions only through the 
corresponding matrix 
$\rho$. Since the representation of $\rho$ as a mixture of pure states
and the representation of a pure state as a superposition of other 
pure states are highly non-unique, we draw the conclusion that very 
different ways of preparing a quantum system, which result in the 
state $\rho$, cannot be distinguished from one another by any 
measurement whatsoever on the quantum system. 
This is a most remarkable feature of quantum mechanics, 
of absolutely non-classical physical nature.%
\marginpar{*}\endnote{O: book: contrast Neymann--Scott} 

\subsubsection{The Schr\"{o}dinger Equation}\label{sss:schroed}

Typically the state of a particle
undergoes an evolution with time under the influence of an external field. The
most basic type of evolution is that of an arbitrary initial state $\rho_{0}$
under the influence of a field with Hamiltonian $H$. This takes the form
$$
\rho_{t}=e^{tH/i\hbar}\rho_{0}e^{-tH/i\hbar}
\thinspace ,
$$
where $\rho_{t}$ denotes the state at time $t$, $\hbar$ is Planck's constant,
and $H$ is a self-adjoint operator on $\mathcal{H}$. If $\rho_{0}$ is a pure
state then $\rho_{t}$ is pure for all $t$ and we can choose unit vectors
$\psi_{t}$ such that $\rho_{t}=|\psi_{t}\rangle\langle\psi_{t}|$ and
\begin{equation}
\psi_{t}=e^{tH/i\hbar}\psi _0
\thinspace .\label{evoln}
\end{equation}
Equation (\ref{evoln}) is a solution of the celebrated 
\emph{Schr\"{o}dinger equation} $i\hbar(\mathrm d/\mathrm d t)\psi=H\psi$
or equivalently $i\hbar(\mathrm d/\mathrm d t)\rho=[H,\rho]$.

\subsubsection{Separability and Entanglement}\label{sss:entang}

When we study several quantum systems (with Hilbert spaces
$\mathcal{H}_1$, \dots, $\mathcal{H}_m$) interacting together, 
the natural model for the combined system has as its Hilbert space the 
tensor product $\mathcal{H}_1 \otimes \dots \otimes \mathcal{H}_m$.
Then a state such as $\rho_1 \otimes \dots \otimes \rho_m$ represents
`particle 1 in state $\rho_1$ and \dots and particle $m$ in state $\rho_m$'.
Suppose the states $\rho_{i}$ are pure with state-vectors 
$\psi_{i}$. Then the product state we have just defined is also
pure with state-vector $\psi_1 \otimes \dots \otimes \psi_m$.
According to the superposition principle, a complex superposition of
such state vectors is also a possible state-vector of the interacting
systems. 
Pure states which cannot be written in the product form 
$\rho _1 \otimes \dots \otimes \rho_m$ are called \emph{entangled}.
The same term is used for mixed states which cannot be written 
as a mixture of pure product states. A state which is not entangled, 
is called \emph{separable}.
The existence of entangled states is respons\-ible for 
extraordinary quantum phenomena,
which scientists are only just starting to harness (in quantum
communication, comput\-ation, teleportation, etc.; see Section 
\ref{quprob} for an introduction).

An important physical feature of unitary evolution in a tensor 
pro\-duct space is that, in general, it does not preserve 
non-entangledness of states. 
Suppose that the state $\rho _1 \otimes \rho _2$ evolves 
according to the Schr\"{o}dinger operator $U_{t}=e^{tH/i\hbar}$ 
on $\mathcal{H}_{1}\otimes \mathcal{H}_{2}$. 
In general, if $H$ does not have the special form 
$H_1 \otimes \boldone  _2 +  \boldone  _1 \otimes H_2$,
the corresponding state at any non-zero time is 
entangled. The notorious \emph{Schr\"{o}dinger Cat},
see Section \ref{ss:measprob}, is 
a consequence of this phenomenon of entanglement.
For an illustrative discussion of this see, 
for instance, \citet[{\ }Sect.\ 8.4.2]{isham95}. 

\subsubsection{Spin-$j$}\label{ss:spinj}

So far, our concrete examples have had a two-dimensional Hilbert space.
Quantum systems in which the Hilbert space $\mathcal{H}$ is 
\mbox{finite-dimensional} are sometimes called \textit{spin systems}.
A \textit{spin-$j$ system}, where $j$ is a positive half-integer,%
\marginpar{*}\endnote{O: book: explain why only half-integer} 
is one for
which the Hilbert space is $\mathbb C ^{2j+1}$. A physical interpretation of
a spin-$j$ system is in terms of a particle having spin angular momentum $j$.

An important class of spin-$j$ systems can be obtained from pure
spin-half systems as follows.
Let $| \psi \rangle$ be a state vector representing a spin-half
particle in a pure state $\rho$. Then the quantum system consisting of $n$ independent
particles, all prepared in this state, is  represented by the state vector
$\otimes ^{n} | \psi \rangle$ in $\otimes ^{n} \mathbb C ^{2}$.
Such state vectors lie in (and span) the subspace
$$
\odot ^{n} \mathbb C ^{2} =
\mathrm{span} \{ \otimes ^{n} | \psi \rangle : | \psi \rangle \in \mathbb
C ^{2} \}
$$
of $\otimes ^{n} \mathbb C ^{2}$.
The corresponding states have the form $\otimes ^{n} \rho$ and
are sometimes known as \emph{(angular momentum) coherent spin-$j$ 
states}.%
\marginpar{*}\endnote{R: book: Bloch sphere for coherent spin j} 

Let $\{ | \psi _0 \rangle ,| \psi _1 \rangle \}$ be any basis of $\mathbb C ^2$.
Put $j = n/2$ and, for $m = -j, \dots , j$, define $| m \rangle$ in
$\odot ^{n} \mathbb C ^{2}$ by
\begin{equation}
| m \rangle = 2^{-j} \Pi _{\odot} \left(
\sum_{k=0}^{n}
\left( \otimes ^{k}| \psi _0 \rangle \right) \otimes
\left( \otimes ^{n-k}| \psi _1 \rangle \right)
\right)
\thinspace ,
\label{mvector}
\end{equation}
where $\Pi _{\odot}$ denotes the orthogonal projection from
$\otimes ^{n} \mathbb C ^{2}$ to $\odot ^{n} \mathbb C ^{2}$.
The formula
$$
\otimes ^{2j} \left( \alpha | \psi _0 \rangle + \beta | \psi _1 \rangle \right) =Ê
\sum_{m=-j}^{j} \binom{ 2j }{ m} \alpha ^{j+m} \beta ^{j-m} | m \rangle
\qquad \alpha ,\beta \in \mathbb C
$$
(which can be obtained by binomial expansion) shows that
$\{ | m \rangle : m = -j, \dots , j \}$ spans
$\odot ^{n} \mathbb C ^{2}$. It is easy to check that this is a basis,
and so $\odot ^{n} \mathbb C ^{2}$ has dimension $2j+1$.

\begin{example}[Coherent Spin-$1$ states]
Take $j=2$. Then
$\{| \psi _0 \rangle  \otimes | \psi _0 \rangle,
(| \psi _0 \rangle \otimes | \psi _1 \rangle
+ | \psi _1 \rangle \otimes | \psi _0 \rangle )/\sqrt{2},
| \psi _1 \rangle \otimes | \psi _1 \rangle \}$ is a basis of
$\odot ^{2} \mathbb C ^{2}$.
Thus $\odot ^{2} \mathbb C ^{2}$ can be identified with
$\mathbb C ^{3}$, whereas $\otimes ^{2} \mathbb C ^{2}$ can be identified with
$\mathbb C ^{4}$. The subspace of $\otimes ^{2} \mathbb C ^{2}$
orthogonal to $\odot ^{2} \mathbb C ^{2}$ is spanned by
$(| \psi _0 \rangle \otimes | \psi _1 \rangle
- | \psi _1 \rangle \otimes | \psi _0 \rangle )/\sqrt{2}$.
The corresponding state is known as the \textit{singlet} or \textit{Bell}
state and helps to demonstrate non-classical properties of quantum
mechanics; see Section 8.2.

Spin-$1$ coherent states can be described in matrix terms as follows.
If $\rho = \frac{1}{2}(\boldone + u_{x}\sigma_{x}+u_{y}\sigma_{y}+u_{z}\sigma_{z})$
is a pure state on $\mathbb C ^{2}$ then
$u_{x}^{2} + u_{y}^{2} + u_{z}^{2} = 1$ and
$$
\rho \otimes \rhoÊ
=
\frac{1}{4} \left\{ \boldone + 2 (u_{x}S_{x}+u_{y}S_{y}+u_{z}S_{z})
+ u_{x}^{2} \sigma_{x} \odot \sigma_{x} + u_{y}^{2} \sigma_{y} \odot \sigma_{y}
+ u_{z}^{2} \sigma_{z} \odot \sigma_{z} \right\} \thinspace ,
$$
where, in terms of the basis
$\{ | - \frac{1}{2} \rangle , | 0  \rangle , | \frac{1}{2} \rangle \}$
of $\odot ^{2} \mathbb C ^{2}$,
$$
S_{x} = \frac{1}{\sqrt{2}} \left(
\begin{tabular}
[c]{ccc}
$0$ & $1$ & $0$\\
$1$ & $0$ & $1$\\
$0$ & $1$ & $0$
\end{tabular}
\right)  \quad
S_{y} = \frac{1}{\sqrt{2}} \left(
\begin{tabular}
[c]{ccc}
$0$ & $-i$ & $0$\\
$i$ & $0$  & $-i$\\
$0$ & $i$  & $0$
\end{tabular}
\right)  \quad
S_{z} = \frac{1}{\sqrt{2}} \left(
\begin{tabular}
[c]{ccc}
$1$ & $0$ & $0$\\
$0$ & $-1$ & $0$\\
$0$ & $0$ & $-1$
\end{tabular}
\right)Ê
$$
and
$$
\sigma_{x} \odot \sigma_{x} = \left(
\begin{tabular}
[c]{ccc}
$0$ & $0$ & $1$\\
$0$ & $1$ & $0$\\
$1$ & $0$ & $0$
\end{tabular}
\right)  \quad
\sigma_{y} \odot \sigma_{y} = \left(
\begin{tabular}
[c]{ccc}
$0$ & $0$ & $-1$\\
$0$ & $1$ & $0$\\
$-1$ & $0$ & $0$
\end{tabular}
\right)  \quad
\sigma_{z} \odot \sigma_{z} = \left(
\begin{tabular}
[c]{ccc}
$1$ & $0$  & $0$\\
$0$ & $-1$ & $0$\\
$0$ & $0$  & $1$
\end{tabular}
\right)  \thinspace .
$$
\hfill$\ \qedsymbol$\end{example}

\subsection{Measurements}\label{ss:meas}

Operator-valued probability measures, or OProM's, were introduced in the
beginning of the present section. We shall denote by 
$\mathrm{OProM}(\mathcal{X}, \mathcal{H})$ the set of  
OProM's on $\mathcal{X}$. 

As indicated earlier, a basic kind of operator-valued probability measures 
consists of those in which the operators $M(A)$ are orthogonal projections. 
Specifically, a \emph{projector-valued probability measure} (or 
\emph{PProM}, also called a \emph{simple measurement}) 
is an operator-valued probability measure $M$ such that
$$
M(A)=M(A)^{\ast}=M(A)^{2}\qquad A\in\mathcal{A}
\thinspace .
$$
We shall denote by 
$\mathrm{PProM}(\mathcal{X},\mathcal{H})$ the set of
PProM's on 
$\mathcal{X}$. As we noted, when the outcome space $\mathcal X$ is 
$\mathbb R $, the PProM's stand in one-to-one correspondence with the
self-adjoint operators on $\mathcal{H}$, which in this context
are also called observables. If one measures the observable $X$ on 
a quantum system in state $\rho$, it turns out that the expected
value of the outcome is given by the {\em trace rule}
\begin{equation}\label{tracerule}
\mathrm E(\mathrm{meas}(X;\rho))=\sum_{x} x \mathrm{tr}\{\rho\Pi_{[x]}\}=
\mathrm{tr}\{\rho\sum_{x}x\Pi_{[x]}\}=
\mathrm{tr}\{\rho X\}.
\end{equation}

\begin{example}[Spin-half, cont.]\label{e:spin12c1}
For any unit vector $\psi$ of $\mathbb C ^{2}$, the observable 
$2|\psi\rangle\langle\psi|-\boldone 
= |\psi\rangle\langle\psi|-|\psi^{\perp}\rangle\langle\psi^{\perp}|$ defines a PProM. 
It has eigenvalues $1$ and 
$-1$ and one-dimensional eigenspaces spanned by $\psi$ and 
$\psi^{\perp}$.
This operator measures the spin of the particle in the direction 
(on the Poincar\'e sphere) defined by $\psi$. We mentioned two of such
measurements in Section \ref{sss:super} on mixing and superposition.
\hfill$\ \qedsymbol$\end{example}

\begin{example}[Spin-half, cont.]\label{e:spin12c2}
In particular, with $\mathcal{X}=\{-1,1\}$, the specific\-ation
\begin{align*}
M(\{+1\}) ~&=~ {\textstyle{\frac12}}(\boldone +\sigma_{x})\\
M(\{-1\})  ~&=~ {\textstyle{\frac12}}(\boldone -\sigma_{x})
\end{align*}
defines an element of 
$\mathrm{PProM}(\mathcal{X},\mathbb C ^{2})$. It corresponds to
the observable $\sigma_{x}$: spin in the $x$-direction.
\hfill$\ \qedsymbol$\end{example}

We next discuss the notion of \emph{quantum 
random\-isation}
whereby adding an auxiliary quantum system to a system 
under study gives one further possibilities for probing the system of 
interest. This also connects to the important notion of \emph{realisation}:
representing generalised measurements by simple measurements on a 
quantum randomised extension.

Suppose given a Hilbert space $\mathcal{H}$, and a pair 
$(\mathcal{K},\rho_{a})$, where $\mathcal{K}$ is a Hilbert space and 
$\rho_{a}$ is a state on $\mathcal{K}$. 
Any measurement ${\widetilde{M}}$ in 
$\mathrm{OProM}(\mathcal{X},\mbox{$\mathcal{H}\otimes\mathcal{K}$})$ 
induces a measurement $M$ in $\mathrm{OProM}(\mathcal{X},\mathcal{H})$ which is 
determined by
\begin{equation}
\mathrm{tr}\left\{\rho M(A) \right\} = 
\mathrm{tr}\left\{ (\rho\otimes\rho_{a}) {\widetilde{M}} (A)\right\}  
\quad\rho\in\mathcal{S}(\mathcal{H}), 
\quad A \in\mathcal{A} 
\thinspace . \label{ancilla}
\end{equation}
The pair $(\mathcal{K},\rho_{a})$ is called an \emph{ancilla}. The following
theorem (Holevo's extension of Naimark's Theorem, see Appendix 
\ref{a:naimark}) states that any
measurement $M$ in $\mathrm{OProM}(\mathcal{X},\mathcal{H})$ is of the form
(\ref{ancilla}) for some ancilla $(\mathcal{K},\rho_{a})$ and some 
simple measurement ${\widetilde{M}}$ in 
$\mathrm{PProM}(\mathcal{X},\mathcal{H}\otimes
\mathcal{K})$. The triple $(\mathcal{K},\rho_{a},{\widetilde{M}})$ is called 
a \emph{realisation} of $M$ (the words \emph{extension} or 
\emph{dilation} are also used sometimes). 
Adding an ancilla before taking a simple measurement 
could be thought of as \emph{quantum randomisation}. 

\begin{thm}[\citealt{holevo82}]\label{t:holevo}
For every $M$ in
$\mathrm{OProM}(\mathcal{X},\mathcal{H})$, there is an ancilla 
$(\mathcal{K},\rho_{a})$ and an element ${\widetilde{M}}$ of 
$\mathrm{PProM}(\mathcal{X},\mathcal{H} \otimes\mathcal{K})$ which form a 
realisation of $M$.
\end{thm}

We use the term quantum randomisation, because of its analogy with the mathematical 
representation of randomisation in classical statistics, whereby one 
replaces the original probability space with a product space,
one of whose components is the original space of 
interest, while the other corresponds to an independent random 
experiment with probabilities under the control of the experimenter.
Just as randomisation in classical statistics is sometimes needed to 
solve optimisation problems of statistical decision theory, 
quantum randomisation sometimes allows for strictly better solutions 
than can be obtained without it.

Here is a simple spin-half example of an OProM which cannot be represented 
without \emph{quantum} randomisation. 

\begin{example}[The triad]
The triad, or Mercedes-Benz logo, has an outcome space consisting of just three 
outcomes:
let us call them $1$, $2$ and $3$. Let $\vec v_{i}$, $i=1,2,3$, denote three 
unit vectors in the same plane through the origin in $\mathrm R^{3}$, 
at angles of $120^{\circ}$ to one another. Then the matrices 
$M(\{i\})=\frac 13 (\mathbf 1 +\vec 
v_{i}\cdot\vec\sigma)$  define an OProM on the sample space 
$\{1,2,3\}$. It turns up as the optimal solution to 
the decision problem: suppose a spin-half system is generated in one of the 
three states $\rho_{i}=\frac 12 (\mathbf 1 -\vec 
v_{i}\cdot\vec\sigma)$, $i=1,2,3$, with equal probabilities. 
What decision rule gives the maximum probability of guessing the 
actual state correctly? There is no way to equal the success 
probability of this method, if one uses only simple measurements, even 
allowing for (classically) randomised procedures.
\hfill$\ \qedsymbol$\end{example}

Finally, we introduce some further terminology concerning measurements.
Given an OProM $M$ and a measurable function $T$ from its outcome space 
$\mathcal X$
to another space $\mathcal Y$, one can define a new measurement 
$M'=M\circ T^{{-1}}$ with outcome space $\mathcal Y$. It corresponds to 
restricting attention to the function $T$ of the outcome of the first measurement 
$M$. We call it a \emph{coarsening} of the original measurement, and 
conversely we say that $M$ is a \emph{refinement} of $M'$.

A measurement $M$ is called \emph{dominated} by a (real, sigma-finite) 
measure $\nu$ on the outcome space, if there exists a non-negative 
self-adjoint matrix-valued function $m(x)$, called the density of $M$, such that 
$M(B)=\int_{B}m(x)\nu(\mathrm d x)$ for all $B$. In the finite-dimensional case 
every measurement is dominated: take $\nu(B)=\mathrm{trace}(M(B))$.

To exemplify these notions, suppose for some dominated measurement $M$ one can write 
$m(x)=m_{1}(x)+m_{2}(x)$ for two non-negative self-adjoint 
matrix-valued functions $m_{1}$ and $m_{2}$. Then one can define a 
refinement $M'$ of $M$ as the measurement on the outcome space 
$\mathcal X'=\mathcal X\times\{1,2\}$ with density $m_{i}(x)$, $(x,i)\in 
\mathcal X'$, with respect to the 
product of $\nu$ with counting measure.

We described earlier how one can form product spaces from separate 
quantum systems, leading to notions of product states, separable 
states, and entangled states. Given an OProM $M$ on one component of a 
product space, one can naturally talk about `the same measurement' on the 
product system. It has components $M(B)\otimes\boldone$. Given 
measurements $M$ and $M'$ defined on the two components of a product 
system, one can define in a natural way the measurement `apply 
simultaneously $M$ and $M'$ to each component': its outcome space is 
the product of the two outcome spaces, and it is defined using obvious 
notation by $M\otimes M'(B\times B')=M(B)\otimes M'(B')$.

A measurement $M$ on a product space is called \emph{separable} if it 
has a density $m$ such that each $m(x)$ can be written as a positive 
linear combination of tensor products of non-negative components. It can 
then be thought of as a coarsening of a measurement with density $m'$
such that each $m'(y)$ is a product of non-negative 
components.

\subsection{Instruments}\label{ss:instru}

When a physical measurement is made on a quantum system, the system usually
changes state in some stochastic manner. Thus a complete description of 
the measurement specifies not just the probability distribution of the
outcome $x$ but also the new state of the system when the outcome is $x$. 
We shall refer to the states of the system before and after 
measurement as the \emph{prior state} and the 
\emph{posterior state}, and use the notation 
$\mathcal N$ 
to denote a particular mapping from prior states to probability 
distributions over outcomes, with a particular posterior 
state associated with each outcome and given prior state. 
Such mappings are called {\em instruments\/} 
(\citealt{davieslewis70, davies76}). 
Because of the basic rules of quantum mechanics, 
an instrument cannot be completely arbitrary but must satisfy certain 
constraints. We shall describe these constraints after we have 
introduced some further notation.

The word `instrument' is not very illuminating.  
The concept which we are 
trying to catch here is that of any interaction between a 
quantum system and the real world. The interaction will change the 
state of the quantum system, and cause changes in the real world. 
One can think of these changes as being information recorded 
in classical physical systems. Data stored on a CD-ROM or printed 
on paper is just one kind of classical physical information. A 
measurement, in the sense of a deliberately carried out experiment, 
is just one kind of interaction. The data which are available to 
an experimenter, after a measurement has been done, 
form only part of the totality of 
changes which have happened in the real world. So one can distinguish 
between what is somehow imprinted in the real world as a result of the 
interaction which takes place when the instrument is applied to the 
quantum system, and a coarsened or reduced version of this information, which 
is the outcome of the measurement as it is available to the experimenter.
What is relevant for the experimenter is the final state of the quantum 
system, conditioned on the data which he has available. This is 
typically different from the final state of the quantum system, 
conditioned on the final state of the real world.

In the following, the outcome of the instrument will refer to the 
data available to the experimenter, and the posterior state 
means the final state (possibly mixed) of the quantum system given this information only.

Consider an instrument $\mathcal N$ with outcomes $x$ in the measurable space
$(\mathcal X, \mathcal A)$. Let $\pi(\mathrm d x;\rho,\mathcal N)$ denote the
probability distribution of the outcome of the measurement, and let
$\sigma(x;\rho,\mathcal N)$ denote the posterior state when the prior 
state is $\rho$ and the outcome of the measurement is $x$. Now let $Y$
denote some observable on the quantum system and let $A\in\mathcal A$ 
denote a measurable set of outcomes. Suppose one `measures the 
instrument' on the state $\rho$, registers whether or not the outcome 
is in $A$, and subsequently measures the observable $Y$. Then the 
expected value of the indicator of the event `outcome is in $A$' times
the outcome of measuring the observable $Y$ is the number 
$\int_{A}\pi(\mathrm d x;\rho,\mathcal 
N)\mathrm{tr}\{(\sigma(x;\rho,\mathcal N) Y \}$, by using the
trace rule (\ref{tracerule}). Now it turns out that this number,
seen as a function of prior state $\rho$, measurable subset of outcomes $A$,
and observable $Y$, determines
$\mathcal N$ completely. By the interpretation of mixed states as probability 
mixtures, it follows that the expression is linear in $\rho$ and 
therefore  can be rewritten as $\mathrm{tr}\{\rho \mathcal N(A)[Y]\}$ 
where $\mathcal N(A)[Y]$,
for each event $A$ in the outcome space and each observable $Y$, 
is a uniquely defined (possibly unbounded) 
self-adjoint operator on $\mathcal H$. This linearity constraint restricts 
considerably the class of all possible $(\pi,\sigma)$.
One can show that $\mathcal N(A)[Y]$ must be countably additive in the
argument $A$, linear and positive in $Y$ (positive in the sense of 
mapping nonnegative operators to nonnegative operators), and 
normalised in the sense that $\mathcal N(\mathcal X)[\boldone ] = 
\boldone $. 

Thus, mathematically, an instrument $\mathcal N$ can be specified 
equally well by giving the probability distribution of the outcome of the measurement 
$\pi(\mathrm d x;\rho,\mathcal N)$, together with the posterior state
$\sigma(x;\rho,\mathcal N)$, as by giving an operator $\mathcal N(A)[Y]$
for each $A$ and $Y$.
The physical constraints imposed by quantum 
theory restrict the possible $(\pi,\sigma)$, and equivalently 
restrict the possible $\mathcal N(A)[Y]$.
The second specification is less direct but
more convenient from a theoretical point of view, since the physical 
constraints (additivity, linearity, positivity, normalization)
are much more simple to express in those terms. In a moment we 
indicate that, on further physical considerations,
the positivity condition should be strengthened to a 
condition called \emph{complete positivity}.

Following \citet{ozawa85}, we show how to recover $(\pi,\sigma)$ from
$\mathcal N$. 
The first step is to read off the measurement or OProM 
$M$ which is determined by the instrument $\mathcal N$, when we ignore the posterior 
state. This is given by the prescription
\begin{equation}
M(A)=\mathcal{N}(A)[\boldone ] 
\thinspace .
\label{instrmeas}
\end{equation}
The probability that the measurement of the state $\rho$ 
results in an outcome in $A$ is given by
$$
\pi(A;\rho,\mathcal N)=\mathrm{tr}\{\rho \mathcal{N}(A)[\boldone ]\}.
$$
If the system was in state $\rho$ just before the measurement then 
the state of the system after the measurement, given that the
measurement was observed to result in an outcome belonging to 
$A$, is determined as the solution $\sigma(A;\rho,\mathcal N)$ of the equation
$$
\mathrm{tr}\{\sigma(A;\rho,\mathcal N) Y\} =
\frac{\mathrm{tr}\{\rho \mathcal{N}(A)[Y]\}}
{\mathrm{tr}\{\rho \mathcal{N}(A)[\boldone ]\}}
\qquad Y \in \mathbb B (\mathcal{H}) 
$$
(provided that 
$\mathrm{tr}\{\rho\mathcal{N}(A)[\boldone ]\}>0$).
Finally, the family $\sigma(x;\rho,\mathcal N)$ of posterior states
is characterised (almost everywhere, with respect to $\pi$) by
$$
\mathrm{tr}\{\rho 
\mathcal{N}(A)[Y]\}=\int_{A}\mathrm{tr}\{\sigma(x;\rho,\mathcal N) Y\}
\pi(\mathrm d x;\rho,\mathcal N)
\qquad Y \in \mathbb B (\mathcal{H})
\qquad A \in \mathcal{A}
\thinspace .
$$

An extremely important class of quantum instruments
consists of those of the form 
\begin{equation}
\mathcal{N}(\mathrm d x)[Y] = 
\sum_{i} W_{i}(x)^* Y W_{i}(x) \nu (\mathrm d x)
\thinspace ,
\label{cpinstr}
\end{equation}
where $\nu$ is a $\sigma$-finite measure on $\mathcal{X}$
(and, without loss of generality, can be taken to be a probability 
measure), the index 
$i$ runs over some finite or countable set, and 
$W_{i}$ is a measurable function from 
$\mathcal{X}$ to $\mathbb B (\mathcal{H})$ such that
$$
\sum_{i} \int _{\mathcal{X}} W_{i}(x)  W_{i}(x)^* \nu (\mathrm d x) = 
\boldone 
\thinspace .
$$
For such quantum instruments, the posterior states are
$$
\sigma(x;\rho,\mathcal N) = \frac{\sum_{i}  W_{i}(x)^*  \rho  W_{i}(x) }
{\sum_{i} \mathrm{tr} \{\rho  W_{i}(x)  W_{i}(x)^* \} }
$$
and the distribution of the outcome is
$$
\pi(\mathrm d x;\rho,\mathcal N) 
    = \sum_{i} \mathrm{tr} \{\rho  W_{i}(x)  W_{i}(x)^* \} \nu (\mathrm d x).
$$
Such quantum instruments are almost generic,
in the sense that an instrument which satisfies the further 
physically motivated condition of \emph{complete positivity} 
can be represented as in (\ref{cpinstr}), except that the 
operators $W_{i}(x)$ need not be bounded (in which case the 
formulae we have given need to be interpreted with some 
care).

The mathematical definition of complete positivity is given in
Appendix \ref{aa:cp}. Its intuitive meaning is as follows. 
We can consider the instrument as acting not just on the 
system of interest $\mathcal H$ but also on a completely arbitrary 
system $\mathcal K$ somewhere else in the universe.
If the systems are independent, we can express the 
joint state as a tensor product, and the instrument acts on 
it by transforming the system of interest as we have 
already specified, while leaving the auxiliary system unchanged; the 
posterior joint state remains a product state. Now once we have 
specified how the extended instrument acts on product states, one can
calculate how it acts on {\em any} joint state, including entangled 
states, by using the linearity which is a basic feature of quantum 
physics. To be physically meaningful, this extended instrument
has to be positive, in the sense of mapping states 
(nonnegative matrices) to states (after all, the system we are 
studying may actually be in an entangled state with a system 
elsewhere). The mathematical statement of this 
physical property is called \emph{complete positivity}.

Formulae like (\ref{cpinstr}) are known in the 
physics literature as Kraus representations.
If we allow \emph{unbounded instruments} for which the self-adjoint 
operator $\mathcal N(A)[Y]$ is not necessarily bounded for all $A$ and $Y$, 
then the $W_{i}(x)$ need not be bounded either. In this case
posterior states may not be defined for each outcome of the measurement, but 
only for each measurable collection of outcomes of positive probability.
Allowing unbounded operators as well as bounded makes a 
difference only in infinite dimensional spaces, see Example
\ref{e:posinstr} in Appendix \ref{aa:cp}
Key references on instruments and complete positivity are 
\citet{stinespring55}, \citet{davieslewis70}, \citet{davies76}, 
\citet{kraus83}, \citet{ozawa85}, \citet{loubenets99,loubenets00}, 
and \citet{holevo01book}. 

\begin{example}[Simple Instruments]\label{e:simpinst}
Let $\{\Pi_{[x]}:x\in\mathcal X\}$ define a PProM on a 
finite-dimensional quantum system, corresponding to the
simple measurement of the observable $Q=\sum x \Pi_{[x]}$.
One can embed this measurement in many different instruments, i.e., 
the state could be transformed by the measurement in many different ways.
However the most simple description possible is obtained when one takes,
in (\ref{cpinstr}), $\nu$ to be counting 
measure on the finite set $\mathcal X$, the set of indices $i$ to 
contain a single element, and $W_{i}(x)=W(x)=\Pi_{[x]}$.
We call this particular instrument the corresponding \emph{simple} instrument.
If one applies it to a system in the pure state with state-vector 
$\psi$, and observes the outcome $x$, then the state of the system
remains pure but now has state vector $\Pi_{[x]} \psi/\|\Pi_{[x]} \psi\|$.
The probability of this event is precisely $\|\Pi_{[x]} \psi\|^{2}$.
When the state transforms in this way, one says that
von Neumann's or L\"uders' \emph{projection postulate} holds
for the measurement of the observable $Q$.

Two observables $Q$, $P$ are called \emph{compatible} if as operators 
they commute. For a Borel measurable function $f:\mathbb R  \rightarrow
\mathbb R $ and an observable $R$ with eigenvalues $r$ and 
eigenspaces the ranges of the projectors $\Pi_{[R=r]}$, the 
observable $f(R)$ is the operator $\sum f(r) \Pi_{[R=r]}$. A celebrated
result of von Neumann is that
observables $Q$ and $P$ are compatible if and only if they are both 
functions $f(R)$, $g(R)$ of a third observable $R$. 
Taking $R$ to have as coarse a 
collection of eigenspaces as possible, one can show that the results 
of the following three instruments are identical: 
the simple instrument for $Q$ followed by the simple instrument for $P$,
recording the values $q$ of $Q$ and $p$ of $P$; the simple instrument 
for $P$ followed by the simple instrument for $Q$, recording the values 
$q$ of $Q$ and $p$ of $P$; and the simple instrument for $R$, recording the
values $q=f(r)$ and $p=g(r)$ where $r$ is the observed value of $R$.

It follows that the probability distribution of the outcome of 
measurement of an observable $P$ is not altered when it is measured
(simply, jointly) together with any other compatible observables.
Note that the expected value of the outcome of a measurement of the 
observable $Q$ on a quantum system in state $\rho$ is 
$\mathrm{tr}\{\rho Q\}$, and the expected value of the real function 
$f$ of this outcome is $\mathrm{tr}\{\rho f(Q)\}$, identical
to the expectation of the outcome of a measurement 
of the observable $f(Q)$. We call this rule \emph{the 
law of the unconscious quantum physicist} since it is analogous to
the law of the unconscious statistician, according to which the 
expectation of a function $Y=f(X)$ of a random variable $X$
may be calculated by an integration (i) over the underlying probability 
space, (ii) over the outcome space of  $X$, (iii) over the outcome space 
of $Y$. 

A useful consequence of this calculus of functions of observables is
that the characteristic function of the distribution of a measurement 
of an observable $Q$ is equal to $\mathrm{tr}\{\rho e^{itQ}\}$. Since $Q$ 
is self-adjoint, $e^{itQ}$ is unitary and the trace may have a 
physical interpretation which aids its calculation.
\hfill$\ \qedsymbol$\end{example}

Further results of \citet{ozawa85} generalise the realisability of 
measurements (Naimark, Holevo theorems) 
to the realisability of an arbitrary completely positive 
instrument. Namely, after forming a compound system by taking the
tensor product with some ancilla, the instrument can be realised as a unitary 
(Schr\"odinger) evolution for some length of time,
followed by the action of a simple instrument 
(measurement of an observable, with state transition according to 
von Neumann's projection postulate). Therefore to say 
that the most general operation on a quantum system is a completely 
positive instrument comes down to saying: the only mechanisms
known in quantum mechanics are Schr\"odinger evolution, von Neumann 
measurement, and forming compound systems (entanglement).
Combining these ingredients in arbitrary ways, one
remains within the class of completely positive instruments;
moreover, anything in that class can be realised in this 
way.

Just as we introduced notions of coarsening and refinement for 
OProM's, and discussed OProM's on product systems, one can do the same
(and more) for instruments. The extra ingredient is \emph{composition}. 
Since the description of an 
instrument includes the state of the system after the measurement by 
the instrument, we are able to define mathematically the composition of 
two instuments, corresponding to the notion of applying first one 
instrument, and then the second, while registering the outcomes (data) produced 
at each of the two stages. The outcome space of the composition 
of two instruments is the product of the two respective outcome spaces.
A more complicated form of composition is possible, in which the 
second instrument is replaced by a family of instruments, indexed by 
possible outcomes of the first instrument. Informally: apply the first 
instrument, then choose a second instrument depending on the outcome 
of the first; keep the outcomes of both.
We do not write out the mathematical formalism for describing these rather 
natural concepts. 

For coarsening, we do write out some formal details, since we need later to 
refer to a specific result.
Let $\mathcal{N}$ denote an instrument on a Hilbert space $\mathcal{H}$
and with outcome space $(\mathcal{X},\mathcal{A})$ and let $\mathcal{N}
^{\prime }$ be a \emph{coarsening} 
of $\mathcal{N}$, i.e. $\mathcal{N}^{\prime }$
is an instrument on the same Hilbert space $\mathcal{H}$, with outcome space 
$(\mathcal{Y},\mathcal{B})$,\ and there is a mapping $T$ from $(\mathcal{X},
\mathcal{A})$ to $(\mathcal{Y},\mathcal{B})$\ such that 
$$
\mathcal{N}^{\prime}(B)[\cdot ]=\mathcal{N}(T^{-1}(B))[\cdot ]
$$
for all $B\in \mathcal{B}$.
This mathematical formalism defines the instrument 
corresponding to applying the instrument $\mathcal N$, registering the 
result of applying the function $T$ to the outcome $x$, and 
discarding $x$.
Because of this interpretation, one has the following relation between 
the posterior states $\sigma(x;\rho, \mathcal{N})$ and 
$\sigma(t;\rho,\mathcal{N}')$: 
\begin{equation}\label{coarsenedposterior}
\sigma (t;\rho, \mathcal{N}' )=\int_{T^{-1}(t)}\sigma(x;\rho, \mathcal{N} )
\pi(\mathrm{d }x|t;\rho, \mathcal{N}) ,
\end{equation}
where $\pi(\mathrm{d }x|t;\rho, \mathcal{N})$ is the conditional distribution 
of $x$ given $T(x)=t$ computed from 
$\pi(\mathrm{d }x;\rho, \mathcal{N})$.

An instrument defined on one component of a product 
system can be extended in a natural way (similar to that described in 
Section \ref{ss:meas} for measurements) to an instrument on
the product system. Conversely, it is of great 
interest whether instruments on a product system can in some way be
reduced to `separate instruments on the separate sub-systems'. There 
are two important notions in this context. 
The first (similar to the concept 
of separability of measurements) is the \emph{mathematical}
concept of separability of 
an instrument defined on a product system: this is that each 
$W_{i}(x)$ in some representation (\ref{cpinstr}) is a tensor product 
of separate matrices for each component. The second is the \emph{physical}
property which we shall call \emph{multilocality}: 
an instrument is called multilocal, if it can be represented as a 
coarsening of a
composition of separate instruments applied sequentially to separate components of 
the product system, where the choice of each instrument at each stage 
may depend on the outcomes of the instruments applied previously. 
Moreover, each component of the system may be measured several times 
(i.e., at different stages), and the choice of component measured at the 
$n$th stage may depend on the outcomes at previous stages.
One should think of the different components of the quantum system as 
being localised at different locations in space. At each location 
separately, anything quantum is allowed, but all communication between 
locations is classical. It is a theorem of \citet{bennettetal99a} that 
every multilocal instrument is separable, but that (surprisingly) not 
all separable instruments are multilocal. It is an open problem to 
find a physically meaningful characterisation of separability, and 
conversely to find a mathematically convenient characterisation of 
multilocality. (Note, our terminology is not standard: the word 
`unentangled' is used by some authors instead of separable, and 
`separable' instead of multilocal).

Not all joint measurements (by which we just 
mean instruments on product systems), are separable, let alone 
multilocal. Just as quantum randomised measurements can give strictly 
more powerful ways to probe the state of a quantum system than 
(combinations of) simple measurements and classical 
randomisation, so non-separable measurements can do strictly
better than separable
measurements at extracting information from product systems, even if 
a priori there is no interaction of any kind between the 
subsystems; this is a main conclusion of 
Section \ref{ss:asymptotic}.

\section{Parametric Quantum Models and Likelihood}\label{s:lik}

A measurement from a parametric quantum model $(\boldrho  ,m)$ 
results in an observation $x$ with density
$$
p(x;\theta)=\mathrm{tr}\{\rho(\theta)m(x)\}
$$
and log likelihood
$$
l(\theta)=\log\mathrm{tr}\{\rho(\theta)m(x)\}
\thinspace .
$$

For simplicity, let us suppose $\theta$ is one-dimensional.
For the calculation of log likelihood derivatives in the present setting it is
convenient to work with the \emph{symmetric logarithmic derivative}
or \emph{quantum score} of $\boldrho$, 
denoted by $\rho_{/\!\!/\theta}$. This is defined implicitly as 
the self-adjoint solution of the equation
\begin{equation}
\rho_{/\theta}=\rho\circ\rho_{/\!\!/\theta}
\thinspace , 
\label{SLD}
\end{equation}
where $\circ$ denotes the Jordan product, i.e.
$$
\rho\circ\rho_{/\!\!/\theta}=
{\textstyle{\frac12}}(\rho\rho_{/\!\!/\theta}+\rho_{/\!\!/\theta}\rho)
\thinspace ,
$$
$\rho_{/\theta}$ denoting the ordinary derivative of $\rho$ with respect to
$\theta$ (term by term differentiation in matrix representations of $\rho$).
(We shall often suppress the argument $\theta$ in quantities like 
$\rho$, $\rho_{/\theta}$, $\rho_{/\!\!/\theta}$, etc.) 
The quantum score
exists and is essentially unique subject only to mild conditions (for 
a discussion of this see, for example, \citealt{holevo82}).

The likelihood score $l_{/\theta}(\theta)=(\mathrm d/\mathrm d\theta)l(\theta)$ 
may be expressed in terms of the
quantum score $\rho_{/\!\!/\theta}(\theta)$ of 
$\rho(\theta)$ as
\begin{align*}
l_{/\theta} (\theta ) ~&=~ p(x;\theta)^{-1}
\mathrm{tr}\{\rho_{/\theta} (\theta ) m(x)\}  \\
~&=~ p(x;\theta)^{-1}{\textstyle{\frac12}}\mathrm{tr} \, 
\{(\rho (\theta ) \rho_{/\!\!/\theta} (\theta )
+ \rho_{/\!\!/\theta} (\theta ) \rho (\theta )) m(x)\} 
 \\
~&=~ p(x;\theta)^{-1}\Re  \, \mathrm{tr} \, 
\{\rho (\theta ) \rho_{/\!\!/\theta} (\theta ) m(x)\} 
\thinspace , 
\end{align*}
where we have used the fact that for any self-adjoint operators 
$P,Q,R$ on
$\mathcal{H}$ the trace operation satisfies 
$\mathrm{tr}\{PQR\}=\overline{\mathrm{tr}
\{RQP\}}$ and 
$\Re \, \mathrm{tr}\{Q\}=
\frac{1}{2} \mathrm{tr}\{Q+Q^{\ast}\}$.
It follows that
$$
\mathrm E_{\theta}[l_{/ \theta}(\theta )] = 
\mathrm{tr} \{ \rho (\theta ) \rho_{/\!\!/\theta} (\theta )\}
\thinspace .
$$
Thus, since the mean value of $l_{/ \theta}$ is $0$, we find that
\begin{equation}
\mathrm{tr}\{ \rho (\theta ) \rho_{/\!\!/\theta} (\theta )\} =0
\thinspace .\label{3.3}
\end{equation}
The expected (Fisher) information 
$i(\theta)=i(\theta;M)=\mathrm E_{\theta}[l_{/ \theta}(\theta )^{2}]$ may be written as
\begin{equation}
i(\theta;M)=\int p(x;\theta)^{-1}\left\{  
\Re \, \mathrm{tr}\{\rho (\theta ) 
\rho_{/\!\!/\theta} (\theta ) m(x)\}\right\}  ^{2}\nu(\mathrm d x)
\thinspace . \label{expinfo}
\end{equation}
It plays a key role in the quantum context, just as in classical
statistics, and is discussed in Section \ref{s:info}. In particular,
we will there discuss its relation with the expected or Fisher
\emph{quantum information}
\begin{equation}\label{Qinfo}
I(\theta)=\mathrm{tr} \{ \rho (\theta ) 
\rho_{/\!\!/\theta}(\theta)^{2} \}.
\end{equation}
The quantum score is a self-adjoint operator, and therefore may be
interpreted as an observable which one might measure on the quantum 
system.
What we have just seen is that the outcome of a simple measurement of
the quantum score has mean zero, and variance equal to the 
quantum Fisher information.

\section{Quantum Exponential and Quantum Transformation Models}
\label{s:qeqtm}

In traditional statistics, the two major classes of parametric models are the
exponential models (in which the log densities are affine functions of
appropriate parameters) and the transformation (or group) models (in which a
group acts in a consistent fashion on both the sample space and 
the parameter space); see \citet{barndorffnielsencox94}. The intersection
of these classes is the class of exponential transformation models, and its
members have a particularly nice structure. There are quantum analogues of
these classes, and they have useful properties.

\subsection{Quantum Exponential Models}\label{ss:qem}

A \emph{quantum exponential model} is a quantum statistical model for which
the states $\rho(\theta)$ can be represented in the form
$$
\rho(\theta)= e^{-\kappa(\theta)} 
e^{\frac12\overline{\gamma}^{r}(\theta) 
T_{r}^{*}}\rho
_{0}e^{\frac12\gamma^{r}(\theta) T_{r}} \qquad \theta \in \Theta 
\thinspace , 
$$
where $\gamma=(\gamma^{1},\ldots,\gamma^{k}): \Theta\rightarrow\mathbb C ^{k}$, 
$T_{1}, \dots,T_{k}$ are
operators on $\mathcal{H}$, $\rho_{0}$ is self-adjoint and non-negative
(but not necessarily a density matrix), the Einstein
summation convention (of summing over any index which appears as both a
subscript and a superscript) has been used, and $\kappa (\theta)$ is a 
log norming constant, given by
$$
\kappa (\theta) = \log \mathrm{tr} \{ 
e^{\frac12\overline{\gamma}^{r}(\theta) 
T_{r}^{*}
}\rho_{0} e^{\frac12\gamma^{r}(\theta)T_{r}}\} 
\thinspace .
$$

Three important special types of quantum exponential 
model are those in which $T_{1},\dots,T_{k}$ are bounded and self-adjoint, 
(and for the first type, $T_{0}$, $T_{1}$, \dots,$T_{k}$ all commute)
and the quantum states have the forms
\begin{align}
\rho({\theta}) ~&=~ e^{-\kappa(\theta)}
\exp\left\{ T_0 + \theta^{r}T_{r} \right\}  
\label{mech} \\
\rho({\theta}) ~&=~ e^{-\kappa(\theta)}
\exp\left\{ {\textstyle{\frac12}}\theta^{r}T_{r}\right\} \rho_{0} 
\exp\left\{ {\textstyle{\frac12}}\theta^{r}T_{r}\right\}  
\label{opsymm} \\
\rho({\theta}) ~&=~ \exp\left\{ -i{\textstyle{\frac12}}\theta^{r}T_{r}\right\}  
\rho_{0}
\exp\left\{  i{\textstyle{\frac12}}\theta^{r}T_{r}\right\} 
\thinspace , \label{unitary} 
\end{align}
respectively, where $\theta=(\theta^{1},\dots,\theta^{k}) \in 
\mathbb R ^{k}$ and $\rho_{0} \in \mathbb {SA} _{+}(\mathcal{H})$,
and the summation convention is in force.

We call these three types, the quantum exponential models of 
\emph{mechanical} type, \emph{symmetric} type, and \emph{unitary} type 
respectively. The mechanical type arises (at least, with $k=1$) in
quantum statistical mechanics as a state of statistical equilibrium,
see \citet[{\ }Sect.\ 2.4.2]{gardinerzoller00}.
The symmetric type has theoretical statistical 
significance, as we shall see, connected among other things
to the fact that the quantum 
score for this model is easy to compute explicitly. The unitary type 
has physical significance connected to the fact that it is also a 
transformation model (quantum transformation models are defined in 
the next subsection). The mechanical type is a special case of the
symmetric type when $T_{0}$, $T_{1}$, \dots,$T_{k}$ all commute.

In general, the statistical model obtained by applying a 
measurement to a quantum exponential model is not an 
exponential model (in the classical sense). However, 
for a quantum exponential model of the form (\ref{opsymm}) in which 
\begin{equation}
T_{j} = t_{j}(X) \quad j = 1, \dots, k \qquad 
\mbox{for some $X$ in $\mathbb {SA} (\mathcal{H})$}
\thinspace ,
\label{Tcommute}
\end{equation}
i.e., the $T_{j}$ commute, the statistical model 
obtained by applying the measurement $X$
is a full exponential model. Various pleasant properties of such 
quantum exponent\-ial models then follow from standard 
properties of the full exponent\-ial models. 

The classical Cram\'{e}r--Rao bound for the variance of an 
unbiased estimator $t$ of $\theta$ is 
\begin{equation}
\mathrm{Var}(t)\geq i(\theta;M)^{-1}
\thinspace . \label{classCRB}
\end{equation}
Combining (\ref{classCRB}) with \citeauthor{braunsteincaves94}'
(\citeyear{braunsteincaves94})
quantum information bound $i(\theta;M)\le I(\theta)$,
which we derive as (\ref{qinfoineq}) in Section \ref{ss:infoclass}, 
yields  \citeauthor{helstrom76}'s (\citeyear{helstrom76}) quantum 
Cram\'{e}r--Rao bound 
\begin{equation}
\mathrm{Var}(t)\geq I(\theta)^{-1}
\thinspace,  \label{QCRB} 
\end{equation}
whenever $t$ is an unbiased estimator based on a quantum measurement.
It is a classical result that,
under certain regularity conditions, the following are equivalent:
(i) equality holds in (\ref{classCRB}), (ii) the score is
an affine function of $t$, (iii) the model is exponential
with $t$ as canonical statistic 
(cf.\ pp.\ 254--255 of \citealt{coxhinkley74}).
This result has a quantum analogue, see Theorems \ref{t:new2} 
and \ref{t:peter} and Corollary \ref{c:peter} below,
which states that under certain regularity conditions, 
there is equivalence between (i) equality holds in 
(\ref{QCRB}) for some unbiased estimator $t$ based on some 
measurement $M$,
(ii) the symmetric quantum score is an affine function of
commuting $T_{1},\dots,T_{k}$, and (iii) the quantum model is a
quantum exponential model of type (\ref{opsymm}) where
$T_{1},\dots,T_{k}$ satisfy (\ref{Tcommute}).
The regularity 
conditions which we assume below are indubitably too strong: the 
result should be true under minimal smoothness assumptions.

\subsection{Quantum Transformation Models}\label{ss:qtm}

Consider a parametric quantum model $(\boldrho , M)$ 
consisting of a family
$\boldrho =\{ \rho (\theta ) : \theta \in \Theta \}$ of states and a 
measurement $M$ with outcome space 
$(\mathcal{X},\mathcal{A})$. Suppose there exists a group, $G$, with 
elements $g$, acting both on $\mathcal X$ and on $\Theta$ in such a 
way that the following consistency condition holds
\begin{equation}
\mathrm{tr} \{\rho(\theta) M(A)\} = \mathrm{tr} \{ \rho(g\theta)M(g^{-1}A)\} 
\label{QTM}
\end{equation}
for all $\theta$, $A$ and $g$. If, moreover, $G$ acts transitively 
on $\Theta$ we say that $(\boldrho ,M)$ is a quantum transformation 
model. In this case, the resulting statistical model for the outcome 
of a measurement of $M$, i.e. $(\mathcal X,\mathcal A, \mathcal P)$, 
where $\mathcal P=\mathrm{tr}\{\rho(\theta)M\}:\theta\in\Theta\}$, 
is a classical transformation model. 
Consequently, the Main Theorem for transform\-ation
models, see \citet[{\ }pp.\ 56--57]{barndorffnielsencox94} and references given
there, applies to $(\mathcal{X}, \mathcal{A}, \mathcal{P})$.

Of particular physical interest are situations where the actions of 
$G$ are such that
\begin{equation}
M(g^{-1}A) = U_{g}^{*} M(A) U_{g}   \qquad A \in\mathcal{A}, 
\label{equivmeas}
\end{equation}
\begin{equation}
\rho(g\theta)= U_{g}^{*} \rho(\theta) U_{g},
\label{equivstate}
\end{equation}
where the $U_{g}$ are unitary matrices satisfying
\begin{equation}
U_{gh} = w(g,h) U_{g} U_{h} \qquad g,h \in G 
\thinspace ,
\label{wgh}
\end{equation}
for some complex valued function $w$ with $|w(g,h)|=1$ for all $g$ 
and $h$. A mapping $g\mapsto U_{g}$ with the property (\ref{wgh}) is said
to constitute a \emph{projective unitary representation} of $G$ and a 
measurement $M$ satisfying (\ref{equivmeas}) is termed \emph{covariant} in 
the physical literature; \emph{equivariant} would be a more correct 
terminology. Under certain conditions, equivariant measurements are 
representable in the form 
$$
M(A) = \int_{ \{ g : g^{-1} x_{0} \in A \} } U_{g}^{*} R_{0} U_{g} 
\mu(\mathrm d g)
$$ 
for an invariant measure $\mu$ on $G$, a fixed
non-negative self-adjoint operator $R_{0}$ on $\mathcal H$ and some 
fixed point $x_{0}\in\mathcal X$.

\begin{example}[Equivariant measurements for spin-half]
Suppose both outcome space $\mathcal X$ and group $G$ are the unit 
circle $S ^{1}$. 
Let the Hilbert space $\mathcal H$ be $\mathbb C ^{2}$ and let
$S ^{1}$ act on $\mathcal H$ via the projective representation
$$
\phi \mapsto U_{\phi} = \left(
\begin{matrix}
e^{i\phi/2} & 0 \\
0 & e^{-i\phi/2}
\end{matrix}
\right)
\qquad \phi \in S ^{1}  \thinspace .
$$ 
Then by \citet[{\ }p.\ 175 with $j=\frac12$]{holevo82}
any equivariant $M$ has
$$
m(\phi)=\left(
\begin{matrix}
1 & a e^{i\phi} \\
\overline a e^{-i\phi} & 1
\end{matrix}
\right)
$$
with respect to the uniform distribution on $S ^{1}$,
for some $a$ with $|a|\le 1$.
\hfill$\ \qedsymbol$\end{example}

\begin{example}[Equivariant measurements for spin-$j$]\label{e:spinj}
The preceding example generalises to spin-$j$ coherent states.
Again, both the outcome space $\mathcal X$ and the group $G$ are the unit
circle $S ^{1}$. Now let the Hilbert space $\mathcal H$ be
$\odot ^{n} \mathbb C ^{2}$. Define the operator $J$ on $\mathcal H$ by
$$
J = \sum_{m=-j}^{j} m | m \rangle  \langle m |  \thinspace ,
$$
where $j=n/2$ and $| m \rangle$ is defined in (\ref{mvector}).
Then putting
$$
U_{\phi} = e^{i \phi J} \qquad \phi \in S ^{1}
$$
gives a projective representation of $S ^{1}$ on $\mathcal H$.
By \citet[{\ }p.\ 175]{holevo82} any equivariant measurement has density
$$
m(\phi)= e^{-i \phi J} R_{0} e^{i \phi J}
$$
with respect to the uniform distribution on $S ^{1}$,
for some positive operator $R_{0}$ satisfying
$$
\frac{1}{2 \pi} \int_{0}^{2 \pi} e^{-i \phi J} R_{0} e^{i \phi J} \mathrm{d} \phiÊ
= \boldone  \thinspace .
$$
\hfill$\ \qedsymbol$\end{example}

\subsection{Quantum Exponential Transformation Models}\label{ss:qetm}

A \emph{quantum exponential transformation model} is a quantum exponential
model which is also a quantum transformation model.
The pleasant properties of classical exponential 
transformation models \citep{barndorffnielsenetal82} are shared 
by a large class of quantum exponential transformation models of 
the form (\ref{opsymm}) which satisfy (\ref{Tcommute}). 
In particular, if $\mathcal{H}$ is finite-dimensional and the group 
acts transitively then there is a unique 
affine action of the group on $\mathbb R ^{k}$ such that 
$(t_{1},\dots,t_{k}):\mathcal{X}
\rightarrow\mathbb R ^{k}$ is equivariant.

\begin{example}[Spin-half: great circle model]\label{e:greatcircle}
Consider 
the spin-half model $\rho(\theta)=U \,
\frac12(\boldone+\cos\theta \sigma_{x}+\sin\theta\sigma_{y})\,U^{*}$ 
where $U$ is a fixed $2\times 2$ unitary matrix, and $\sigma_{x}$ 
and $\sigma_{y}$ are two of the Pauli spin matrices, while the 
parameter $\theta$ varies through $[0,2\pi)$; see Example 
\ref{e:spin12}. The matrix $U$ can 
always be written as $\exp(-i\phi\vec u\cdot\vec\sigma)$ for some real 
three-dimensional unit vector $\vec u$ and angle $\phi$. Considered 
as a curve on the Poincar\'e sphere, the model forms a great circle. 
If $U$ is the identity (or, equivalently, $\phi=0$) the curve just follows the 
line of the equator; the presence of $U$ corresponds to rotating the sphere carrying 
this curve about the direction $\vec u$ through an angle $\phi$.
Thus our model describes an arbitrary great circle on the Poincar\'e 
sphere, parametrised in a natural way. Since we can write 
$\rho(\theta)=UV_{\theta}U^{*}\rho(0)UV_{\theta}^{*}U^{*}$, where the unitary 
matrix $V_{\theta}$ corresponds to rotation of the Poincar\'e sphere by an 
angle $\theta$ about the $z$-axis, we can write this model as a 
unitary transformation model of the form (\ref{equivstate}). 
Together with any equivariant measurement, 
this model forms a quantum transformation model. 
The model is clearly also an exponential model of unitary type.
Perhaps surprisingly, it can be reparameterised so as also to be an 
exponential model of symmetric type. We leave the details (which 
depend on the algebraic properties of the Paul spin matrices) to the 
reader, but just point out that a one-parameter pure-state exponential 
model of symmetric type has to be of the form 
$\rho(\theta)=\exp(-\kappa(\theta))\exp(\frac12 \theta\vec 
u\cdot\vec\sigma)\frac12(\boldone+\vec v\cdot\vec \sigma)
\exp(\frac12 \theta\vec 
u\cdot\vec\sigma)$ for some real unit vectors $\vec u$ and $\vec v$, 
since every self-adjoint $2\times2$ matrix is an affine function of
a spin matrix $\vec u\cdot \vec\sigma$. Now write out the 
exponential of a matrix as its power series, and use the fact that the 
square of any spin matrix is the identity.

This example is due 
to \citet{fujiwaranagaoka95}.
\hfill$\ \qedsymbol$\end{example}

\section{Quantum Exhaustivity and Sufficiency}\label{s:exhaust}

This section introduces and relates some concepts connected to the 
classical notion of sufficiency.

\subsection{Quantum Exhaustivity}\label{ss:exhaust}

An important role is played by quantum instruments for which no 
inform\-ation on
the unknown parameter of a quantum parametric model of states can be obtained
from subsequent measurements on the given physical system. 

Recall that an instrument $\mathcal N$ is represented by a 
collection of observables $\mathcal{N}(A)[Y]$, defined in the 
following implicit fashion. For any particular $A$ and $Y$, 
the expectation of the outcome of measuring the observable 
$\mathcal{N}(A)[Y]$ on a system in state $\rho$, is the same as
the expectation of a function of the joint outcomes  
of \emph{first} applying the \emph{instrument} to a system in state 
$\rho$ and \emph{next} measuring the observable $Y$ on the posterior state:
namely, take 
the product of the indicator variable that the outcome of the instrument is in $A$,
with the outcome of the subsequent measurement of $Y$.
This collection of observables determines uniquely 
the probability distribution $\pi(\mathrm d x;\rho,\mathcal N)$ of the outcome 
of applying the instrument $\mathcal N$ to the state $\rho$, and the 
posterior state $\sigma(x;\rho,\mathcal N)$ given that the outcome is $x$.
They are related to the $\mathcal{N}(A)[Y]$ by the equality (which we 
just expressed in words)
$$
\mathrm{tr}\{\rho \mathcal{N}(A)[Y] \}=
\int_{A} \mathrm{tr}\{\sigma(x;\rho,\mathcal N) Y\}\pi(\mathrm d 
x;\rho,\mathcal N ).
$$

In the sequel we will drop 
the name of the instrument in the notation for $\pi$ and $\sigma$ and, 
when considering a 
parameterised family of prior states, replace the prior state 
$\rho(\theta)$ by the parameter value $\theta$: thus $\pi(\mathrm d 
x;\theta)$ denotes the probability distribution of the outcome, and
$\sigma(x;\theta)$ denotes the posterior state.

\begin{defn}[Exhaustive instruments]
A quantum instrument $\mathcal{N}$ 
is \emph{exhaustive} for a para\-meterised set 
$\boldrho : \Theta\rightarrow\mathcal{S}(\mathcal{H})$ of states if
for all $\theta$ in $\Theta$ and for $\pi(\cdot ;\theta)$-almost all $x$, 
$\sigma(x;\theta)$ does not depend on $\theta$. 
\end{defn}

Thus the posterior states obtained from exhaustive quantum instruments are
completely determined by the result of the measurement and do not depend on 
$\theta$. 

A useful strong form of exhaustivity is defined as follows.

\begin{defn}[Completely exhaustive instruments]
A quantum instrument $\mathcal{N}$ is \emph{completely exhaustive} if it is 
exhaustive for all parameterised sets of states. 
\end{defn}

Recall that any completely positive instrument---in other words, 
virtually any 
physically realisable instrument---has the form (\ref{cpinstr}) 
of $\mathcal{N}(A)[Y]$, given by
\begin{equation}
\mathcal{N}(\mathrm d x)[Y] = 
\sum_{i}\mathrm{tr}\{ W_{i}(x)^* Y W_{i}(x)\} \nu (\mathrm d x)
\end{equation}
with posterior states
$$
\sigma(x;\rho) = \frac{\sum_{i}  W_{i}(x)^*  \rho  W_{i}(x) }
{\sum_{i} \mathrm{tr} \{\rho  W_{i}(x)  W_{i}(x)^* \} }
$$
and outcome distributed as
$$
\pi(\mathrm d x;\rho) 
    = \sum_{i} \mathrm{tr} \{\rho  W_{i}(x)  W_{i}(x)^* \} \nu (\mathrm d x).
$$
The following Proposition (which is a slight generalisation of a result of 
\citealt{wiseman99}) shows one way of constructing completely exhaustive 
completely positive quantum instruments.

\begin{prop}
Let the quantum instrument $\mathcal{N}$ be as above, with
$W_{i}(x)$ of the form
\begin{equation}
W_{i}(x) = | \psi_{x} \rangle\langle\phi_{i,x} | 
\thinspace ,
\label{Wcomplete}
\end{equation}
for some functions $(i,x) \mapsto\phi_{i,x}$ and $x \mapsto\psi_x$.
Then $\mathcal{N}$ is completely exhaustive. 
\end{prop}

\begin{prf}
By inspection we find that the posterior state is
$$
\sigma(x;\rho) = \frac{\sum_{i}  |\phi_{i,x}\rangle  \langle 
\phi_{i,x} | }
{\sum_{i} \langle 
\phi_{i,x} | \phi_{i,x}\rangle },
$$
which does not depend on the prior state $\rho$.
\hfill$\ \qedsymbol$\end{prf}

\subsection{Quantum Sufficiency}\label{ss:suff}

Suppose the measurement $M^{\prime }=M\circ T^{-1}$ is a coarsening of
the measurement $M$. In this situation we say that 
$M^{\prime }$\ is \emph{(classically) sufficient} for $M$ with respect to a 
family of
states $\boldrho =\{\rho (\theta ):\theta \in \Theta \}$ on $\mathcal{H}$ 
if the mapping $T$ is sufficient for the identity mapping on $(\mathcal{X},
\mathcal{A})$ with respect to the family $\{P(\cdot ;\theta ;M):\theta \in
\Theta \}$ of probability measures on $(\mathcal{X},\mathcal{A})$ induced by 
$M$ and $\boldrho$ (that is, 
$P(\cdot ;\theta ;M)=\mathrm{tr}\{M(\cdot)\rho (\theta )\}$).

As a further step towards a definition of quantum sufficiency, we introduce a
concept of inferential equivalence of parametric models of states.

\begin{defn}[Inferential equivalence]
Two parametric families of
states $\boldrho  =\{\rho (\theta ):\theta \in \Theta \}$ and 
$\boldsigma =\{\sigma (\theta ):\theta \in \Theta \}$ on Hilbert spaces 
$\mathcal{H}$ and $\mathcal{K}$\ are said to be 
\emph{inferentially equivalent} if for every measurement $M$ on $\mathcal{H}$
there exists a measurement $N$ on $\mathcal{K}$\ such that for all $\theta
\in \Theta $ 
\begin{equation}\label{infequiv}
\mathrm{tr}\{M(\cdot )\rho (\theta )\}
= \mathrm{tr}\{N(\cdot )\sigma (\theta )\}
\end{equation}
and vice versa.
(Note that, implicitly, the outcome spaces of $M$ and $N$ are assumed to be
identical.)
\end{defn}

In other words, $\boldrho$ and $\boldsigma$ are equivalent if
and only if they give rise to the same class of possible classical models
for inference on the unknown parameter.

\begin{example}[Two identical spin-half particles vs.\ one coherent
spin-\\
one particle]
Let $\mathbf{\rho} =\{\rho (\theta ):\theta \in \Theta \}$
be a parametric family of coherent spin-$1$ states;
see Section \ref{ss:spinj} above. Then the associated Hilbert space
$\mathcal{H}$ is $\mathbb C ^{2}\otimes \mathbb C ^{2}$.
Recall that the state vectors of coherent spin-$1$ states lie in the
subspace $\mathcal{K} = \mathbb C ^{2}\odot \mathbb C ^{2}$ of
$\mathbb C ^{2}\otimes \mathbb C ^{2}$.
Define the parametric family $\mathbf{\sigma}  = \{\sigma (\theta ):\theta \in \Theta \}$ by
$\sigma (\theta ) =  \Pi _{\odot} \rho (\theta ) \iota$, where $\Pi _{\odot}$
and $\iota $ are the orthogonal projection from $\mathbb C ^{2}\otimes \mathbb C ^{2}$
to $\mathbb C ^{2}\odot \mathbb C ^{2}$ and the inclusion of
$\mathcal{K}$ in $\mathcal{H}$, respectively.
Given a measurement $M$ on
$\mathcal{H}$, we can define a measurement $N$ on
$\mathcal{K}$ by $N(\cdot ) =  \Pi _{\odot} M(\cdot ) \iota$.
Similarly,  given a measurement $N$ on
$\mathcal{K}$, we can define a measurement $M$ on
$\mathcal{H}$ by $M(\cdot ) =  \iota N(\cdot ) \Pi _{\odot}$.
It is simple to verify that (\ref{infequiv}) is satisfied, and so $\mathbf{\rho}$ and
$\mathbf{\sigma}$ are inferentially equivalent.
\hfill$\ \qedsymbol$\end{example}

\begin{rmk} It is of interest to find characterisations of
inferential equivalence. This is a nontrivial problem, even in the case
where the Hilbert spaces $\mathcal{H}$\ and $\mathcal{K}$ are the same.
\end{rmk}

Next, let $\mathcal{N}$ denote an instrument on a Hilbert space $\mathcal{H}$
and with outcome space $(\mathcal{X},\mathcal{A})$ and let $\mathcal{N}
^{\prime }=N\circ T^{-1}$ be a coarsening
of $\mathcal{N}$ with outcome space 
$(\mathcal{Y},\mathcal{B})$, generated by a mapping $T$ from $(\mathcal{X},
\mathcal{A})$ to $(\mathcal{Y},\mathcal{B})$.
According to (\ref{coarsenedposterior}) in Section \ref{ss:instru}, 
the posterior states for the two instruments are related by
$$
\sigma (t;\theta, \mathcal{N}' )=\int_{T^{-1}(t)}\sigma(x;\theta, \mathcal{N} )
\pi(\mathrm{d }x|t;\theta, \mathcal{N}) ,
$$
where $\pi(\mathrm{d }x|t;\theta, \mathcal{N})$ is the conditional distribution 
of $x$ given $T(x)=t$ computed from 
$\pi(\mathrm{d }x;\theta, \mathcal{N})$.

\begin{defn}[Quantum sufficiency of instruments] Let 
$\mathcal{N}^{\prime }$ be a coarsening of an instrument $\mathcal{N}$
by $T : (\mathcal{X}, \mathcal{A}) 
\rightarrow (\mathcal{Y}, \mathcal{B})$. Then 
$\mathcal{N}^{\prime }$ is said to be \emph{quantum sufficient}\ with respect
to a family of states $\{\rho (\theta ):\theta \in \Theta \}$\ if

\begin{enumerate}
\item[(i)]  the measurement $M^{\prime }(\cdot )=\mathcal{N}^{\prime }(
\cdot)[\boldone ]$ is sufficient for the measurement $M(\cdot )=
\mathcal{N}(\cdot)[\boldone ]$, with respect to the family 
$\{\rho (\theta ):\theta \in \Theta \}$

\item[(ii)] for any $x\in \mathcal{X}$, 
the posterior families $\{\sigma(x; \theta, \mathcal{N}):\theta \in \Theta
\} $ and $\{\sigma(T(x);\theta, \mathcal{N}' ):\theta \in \Theta \}$ are inferentially
equivalent.
\end{enumerate}
\end{defn}

\subsection{Exhaustivity, Sufficiency, Ancillarity and Separability}

In the theory of classical statistical inference, many important
concepts (such as sufficiency, ancillarity and cuts) can be expressed
in terms of the decomposition by a measurable function
$T : (\mathcal{X}, \mathcal{A}) \rightarrow (\mathcal{Y},
\mathcal{B})$
of each probability
distribution on $(\mathcal{X}, \mathcal{A})$ into the corresponding
marginal
distribution of $T(x)$
and the family of conditional distributions of $x$ given $T(x)$.
In quantum statistics there are analogous concepts based on the
decomposition
\begin{equation}
\rho \mapsto (\pi ( \cdot ; \rho, \mathcal{N}), \sigma (\cdot ; \rho
, \mathcal{N}))
\label{qdecomp}
\end{equation}
by a quantum instrument $\mathcal{N}$ of each state $\rho$
into a measurement and a family of posterior states; see Section 2.3.

The classical concept of a cut encompasses those of sufficiency and
ancillarity and is therefore more basic.
A measurable function $T$ is a
{\it cut} for a set $\mathcal{P}$ of probability distributions on
$\mathcal{X}$ if for all $p_1$ and $p_2$ in
$\mathcal{P}$, the distribution on $\mathcal{X}$ obtained by
combining the marginal distribution of $T(x)$ given by $p_1$
with the family of conditional distributions of $x$ given $T(x)$
given by $p_2$ is also in $\mathcal{P}$; see, e.g.\
p.\ 38 of \citealt{barndorffnielsencox94}.
Recent results on cuts for exponential models can be found in
\citep{barndorffnielsenkoudou95}, which also gives references
to the useful role which cuts have played in graphical models.
A generalisation to {\it local cuts} has become important
in econometrics \citep{christensenkiefer94, christensenkiefer00}.
Replacing the decomposition into marginal and conditional
distributions in the definition of a cut by the decomposition
(\ref{qdecomp}) yields the following quantum analogue.

\begin{defn}[Quantum cuts]
A quantum instrument $\mathcal{N}$ is said to be a \emph{quantum cut}
for a family $\boldrho$ of states if for all
$\rho_1$ and $\rho_2$ in $\boldrho$, there is a $\rho_3$ in
$\boldrho$ such that
\begin{align*}
\pi ( \cdot ; \rho_3, \mathcal{N}) ~&=~
\pi ( \cdot ; \rho_1, \mathcal{N})  \\
\sigma (\cdot ; \rho_3 , \mathcal{N}) ~&=~
\sigma (\cdot ; \rho_2 , \mathcal{N}) . 
\end{align*}
\end{defn}
 ÊÊ
Thus, if $\mathcal{N}$ is a quantum cut for a family
$\boldrho = \{ \rho( \theta ) : \theta  \in \Theta \}$
with $\rho$ a one-to-one function then
$\Theta$ has the product form $\Theta = \Psi \times \Phi$ and furthermore
$\sigma (\cdot ; \rho( \theta ), \mathcal{N})$ depends on $\theta$
only through $\psi$, and $\pi (\cdot ; \rho( \theta ),\mathcal{N})$
depends on $\theta$ only through $\phi$.

\begin{example}[Simple quantum cuts]
Let $\{ \Pi _{[x]} : x \in \mathcal{X} \}$ be a PProM on a
finite-dimensional quantum system. Suppose that sets $\Psi$ and $\Psi$
are
given, together with collections of functions (indexed by $x$ in
$\mathcal{X}$)
$f_x :\Phi \rightarrow [0,1]$ andÊ
$M_x :\Psi  \rightarrow \mathcal{S}(\mathcal{H})$ which satisfy
\begin{align*}
\sum _{x  \in \mathcal{X}}  f_{x} (\phi  ) ~&=~ 1 &
\qquad &\phi  \in \Phi  \\
M_x (\psi  ) ~&=~ \Pi _{[x]} M_x  (\psi  ) \Pi _{[x]}Ê&
\qquad &\psi  \in \Psi . 
\end{align*}
Then we can define a family of states
$\{ \rho (\psi , \phi ) : \psi \in \Psi , \phi \in \Phi \}$
by
$$
\rho (\psi , \phi )
= \sum _{x  \in \mathcal{X}}  f_x (\phi  ) M_x (\psi  )
\qquad (\psi , \phi ) \in \Psi \times \Phi .
$$
As indicated in Example \ref{e:simpinst},
$\{ \Pi _{[x]} : x \in \mathcal{X} \}$ gives
rise to a simple quantum instrument $\mathcal{N}$, defined by
$$
\mathcal{N} ( \{ x \} ) [Y] = \Pi _{[x]} Y \Pi _{[x]} .
$$
A straightforward calculation using the orthogonality of the
projectionsÊÊÊÊÊÊÊÊÊÊÊÊÊÊÊ
$ \Pi _{[x]}$ shows that
\begin{align*}
\sigma ( x ; \rho (\psi , \phi ) , \mathcal{N}) ~&=~ M_x (\psi  )
 \\
\pi ( x ; \rho (\psi , \phi ) , \mathcal{N}) ~&=~  f_x (\phi  ) , 
\end{align*}
and so $\mathcal{N}$ is a quantum cut for $\mathbf{\rho}$.
\hfill$\ \qedsymbol$\end{example}

Since a quantum instrument is exhaustive for a parameterised set
$\boldrho = \{ \rho( \theta ) : \theta  \in \Theta \}$ of states
if the family $\sigma (\cdot ; \rho (\theta ) , \mathcal{N})$ of
posterior states does not depend on $\theta$, exhaustive quantum
instruments are quantum cuts of a special kind. They can be regarded
as quantum analogues of sufficient statistics.
At the other extreme are the quantum instruments for which the
measurements $\pi (\cdot ; \rho (\theta ) , \mathcal{N})$ do not depend
on $\theta$. These can be regarded as quantum analogues of ancillary
statistics.

Unlike exhaustivity, the concept of quantum sufficiency involves not
only a quantum instrument but also a coarsening. The definition of
quantum sufficiency can be extended to the following version
involving parameters of interest.

\begin{defn}[Quantum sufficiency for interest parameters]
Let $\boldrho = \{ \rho (\theta ) : \theta \in  \Theta \}$ be a
family of states and let $\psi : \Theta \rightarrow \Psi$ map $\Theta$
to the space $\Psi$ of interest parameters.
A coarsening $\mathcal{N}'$ of an instrument $\mathcal{N}$ by a
mapping $T$ is said to be \emph{quantum sufficient} for $\psi$ on
$\boldrho$ if
\begin{enumerate}
\item[(i)] the measurement $\mathcal{N}'( \cdot)[\mathbf{1}]$ is
sufficient for $\mathcal{N}( \cdot)[\mathbf{1}]$ with respect to the
family $\boldrho$,ÊÊÊ
\item[(ii)] for all $\theta_1$ and $\theta_2$ with $\psi (\theta_1)
= \psi (\theta_2)$ and for all $x$ in $\mathcal{X}$, the sets
$\sigma (x; \rho (\theta _1), \mathcal{N})$ and
$\sigma (T(x); \rho (\theta _2), \mathcal{N}')$ of posterior states
are
inferentially equivalent.
\end{enumerate}
\end{defn}
 ÊÊ
Consideration of the likelihood function obtained by applying a
measurement to a parameterised set of states suggest that the
following weakening of the concept of inferential equivalence may be useful.

\begin{defn}[Weak likelihood equivalence]
Two parametric families of states
$\boldrho = \{ \rho (\theta ) : \theta \in \Theta \}$ andÊ
$\boldsigma = \{ \sigma (\theta ) : \theta \in \Theta \}$ on
Hilbert spaces
$\mathcal{H}$ and $\mathcal{K}$ respectively are said to be
\emph{weakly likelihood equivalent} if for every measurement $M$ on
$\mathcal{H}$ 
there is a measurement $N$ on $\mathcal{K}$ with the same outcome 
space, 
such that
$$
\frac{ {\rm tr} \{ M(\mathrm{d} x)  \rho (\theta ) \} }
{ {\rm tr} \{ M(\mathrm{d} x)  \rho (\theta ')\} }
= \frac{ {\rm tr} \{ N(\mathrm{d} x) \sigma (\theta )\} }
{ {\rm tr} \{ N(\mathrm{d} x) \sigma (\theta ')\} }
\qquad  \theta , \theta ' \in \Theta
$$
(whenever these ratios are defined) and vice versa.
\end{defn}

Thus the likelihood function of the statistical model obtained by
applying $M$ to $\boldrho$ is equivalent to that obtained by
applying $N$ to $\boldsigma$, 
for the same outcome of 
each instrument.

Consideration of the distribution of the likelihood ratio leads to 
the following definition.

\begin{defn}[Strong likelihood equivalence]
Two parametric families of states
$\boldrho = \{ \rho (\theta ) : \theta \in \Theta \}$ andÊ
$\boldsigma = \{ \sigma (\theta ) : \theta \in \Theta \}$ on
Hilbert spaces
$\mathcal{H}$ and $\mathcal{K}$ respectively are said to be
\emph{strongly likelihood equivalent} if for every measurement $M$ on
$\mathcal{H}$ with outcome space $\mathcal{X}$ 
there is a measurement $N$ on $\mathcal{K}$ with some outcome space 
$\mathcal{Y}$ such that the likelihood ratios 
$$
\frac{{\rm tr} \{ M(\mathrm{d} x)  \rho (\theta ) \}}
{{\rm tr} \{ M(\mathrm{d} x)  \rho (\theta ')\}}
\qquad  \rm{and} \qquad  \frac{{\rm tr} \{ N(\mathrm{d} y) \sigma (\theta )\}}
{{\rm tr} \{ N(\mathrm{d} y) \sigma (\theta ')\}}
$$
have the same distribution for all $\theta , \theta '$ in $\Theta$,
and vice versa.
\end{defn}

The precise connection between likelihood equi\-valence and
inferential equivalence is not yet known but the following
conjecture appears reasonable.

\vskip 12pt
\noindent
{\bf Conjecture.}
Two quantum models are strongly likelihood equivalent if and only if they are
inferentially equivalent
up to quantum randomisation.

\section{Quantum and Classical Fisher Information}\label{s:info}

In Section \ref{s:lik} we showed how to express the Fisher information 
in the outcome of a measurement in terms of the quantum score. 
In this section we discuss quantum analogues of Fisher 
information and their relation to the classical concepts.

\subsection{Definition and First Properties}\label{ss:infodef}

Differentiating (\ref{3.3}) with respect to $\theta$, writing $\rho
_{/\!\!/\theta/\theta}$ for the derivative of the symmetric 
logarithmic derivative
$\rho_{/\!\!/\theta}$ of $\rho$, and using the defining 
equation (\ref{SLD}) for
$\rho_{/\!\!/\theta}$ and the fact that $\rho$ and 
$\rho_{/\!\!/\theta}$ are
self-adjoint, we obtain
\begin{align*}
0 ~&=~ \Re \, \mathrm{tr}
\{\rho_{/\theta}(\theta) \rho_{/\!\!/\theta}(\theta)
+ \rho (\theta) \rho_{/\!\!/\theta/\theta}(\theta)\}  \\
~&=~ \Re \, \mathrm{tr}\left\{  
{\textstyle{\frac12}}\Bigl(\rho (\theta) \rho_{/\!\!/\theta}(\theta)
+\rho_{/\!\!/\theta}(\theta) \rho(\theta)\Bigr) \rho_{/\!\!/\theta}(\theta)
\right\}  
+\Re  \,
\mathrm{tr}\{\rho (\theta) \rho_{/\!\!/\theta/\theta}(\theta)\}  \\
~&=~ I(\theta)- \mathrm{tr} (\rho (\theta) J(\theta))
\thinspace , 
\end{align*}
where
$$
I(\theta)=\mathrm{tr} \left\{ \rho (\theta ) \rho_{/\!\!/\theta}(\theta)^{2} \right\} 
$$
is the \emph{expected} (or \emph{Fisher}) \emph{quantum information},
already mentioned in Sections \ref{s:lik} and \ref{s:qeqtm}, and
$$
J(\theta)=-\rho_{/\!\!/\theta/\theta} (\theta)
\thinspace , 
$$
which we shall call the \emph{observable quantum information}. 
Thus
$$
I(\theta)=\mathrm{tr} \left\{ \rho (\theta) J(\theta) \right\}
\label{I=EJ}
\thinspace ,
$$
which is a quantum analogue of the classical relation 
$i(\theta)=\mathrm E_{\theta} [j(\theta)]$ between expected and observed 
information 
(where $j(\theta) = - l_{/ \theta / \theta}(\theta)$). 
Note that $J(\theta)$ is an observable, just as $j(\theta)$ is a random 
variable.

Neither $I(\theta)$ nor $J(\theta)$ depends on the choice of measurement,
whereas $i(\theta)=i(\theta;M)$ does depend on the measurement $M$. 

For parametric quantum models of states of the form
$$
\boldrho :\theta \mapsto 
\rho_{1}(\theta)\otimes\dots \otimes \rho_{n}(\theta)
$$
(which model `independent particles'), the associated 
expected quantum information satisfies
$$
I_{\rho_{1}\otimes\dots\otimes\rho_{n}}(\theta) = 
\sum_{i=1}^{n}I_{\rho_{i}}(\theta)
\thinspace , 
$$
which is analogous to the additivity property of Fisher information.
In particular, for parametric quantum models of states of the form
\begin{equation}
\boldrho :\theta\mapsto\rho(\theta)\otimes\dots\otimes\rho(\theta)
\label{iidmodel}
\end{equation}
(which model $n$ \lq independent and identical particles\rq ),
the associated expected quantum information $I_{n}$ satisfies
\begin{equation}
I_{n}(\theta)=nI(\theta)
\thinspace , 
\label{Iaddind}
\end{equation}
where $I(\theta)$ denotes the expected quantum information for 
a single measure\-ment of the same type. 

In the case of a multivariate parameter $\theta$, the 
\emph{expected quantum
inform\-ation} matrix $I(\theta)$ is defined in terms of the 
quantum scores by
\begin{equation}
I(\theta)_{jk} = {\textstyle{\frac12}} \mathrm{tr} \left\{ 
\rho_{/\!\!/\theta_{j}}(\theta) \rho (\theta) 
\rho_{/\!\!/\theta_{k}}(\theta) 
+ \rho_{/\!\!/\theta_{k}}(\theta) \rho (\theta)
\rho_{/\!\!/\theta_{j}}(\theta)
\right\}
\thinspace .
\label{matrixI}
\end{equation}

\subsection{Relation to Classical Expected Information}
\label{ss:infoclass}

Suppose that $\theta$ is one-dimensional.
There is an important relationship between expected quantum
inform\-ation $I(\theta)$ and classical expected information
$i(\theta;M)$, due to 
\citet{braunsteincaves94}, namely that for any measurement $M$ with density 
$m$ with respect to a $\sigma$-finite measure $\nu$ on $\mathcal{X}$,
\begin{equation}
i(\theta;M)\leq I(\theta)
\thinspace , \label{qinfoineq}
\end{equation}
with equality if and only if, for $\nu$-almost all $x$,
\begin{equation}
m(x)^{1/2}\rho_{/\!\!/\theta}(\theta) \rho (\theta)^{1/2} 
= r(x) m(x)^{1 /2} \rho (\theta) ^{1/2}
\thinspace ,
\label{QCREcond2}
\end{equation}
for some real number $r(x)$. For a proof see Appendix \ref{a:bc}.

For each $\theta$, there are measurements which attain the bound in the
quantum information inequality (\ref{qinfoineq}). For instance, we can choose
$M$ such that each $m(x)$ is a projection onto an eigenspace of the quantum
score $\rho _{/\!\!/ \theta} (\theta)$. Note that this attaining measurement 
may depend on $\theta$.

\begin{example}[Information for spin-half]\label{e:spin12qi}
Consider a spin-half particle in the pure state
$\rho=\rho(\eta,\theta)=|\psi(\eta,\theta)\rangle\langle\psi(\eta,\theta)|$ 
given by
$$
|\psi(\eta,\theta)\rangle=\left(
\begin{matrix}
e^{-i\theta/2}\cos(\eta/2)\\
e^{i\theta/2}\sin(\eta/2)
\end{matrix}
\right)  
\thinspace .
$$
As we saw in Example \ref{e:spin12} (where we wrote $(\eta,\vartheta)$
for $(\eta,\theta)$), $\rho$ can be written as 
$\rho=(\boldone +u_{x}\sigma
_{x}+u_{y}\sigma_{y}+u_{z}\sigma_{z})/2=
\frac{1}{2}(\boldone +\vec{u}\cdot\vec{\sigma})$, 
where $\vec{\sigma}=(\sigma_{x},\sigma_{y},\sigma_{z})$ are the 
three Pauli spin matrices and 
$\vec{u}=(u_{x},u_{y},u_{z})=\vec{u}(\eta,\theta)$ is
the point on the Poincar\'e sphere $S^{2}$ with polar coordinates
$(\eta,\theta)$. Suppose that the colatitude $\eta$ is known
and exclude the degenerate cases $\eta=0$ or $\eta=\pi$; the 
longitude $\theta$ is the unknown parameter.

Since all the $\rho(\theta)$ are pure, one can show that
$\rho_{/\!\!/\theta}(\theta) =2\rho_{/\theta}(\theta)
=\vec{u}_{/\theta}(\theta)\cdot\vec{\sigma} =
\sin(\eta)\,\vec{u}(\pi/2,\theta+\pi/2)\cdot \vec{\sigma}$.
Using the properties of the Pauli matrices, one finds that 
the quantum information is 
\[
I(\theta)=\mathrm{tr}\{\rho (\theta) \rho_{/\!\!/\theta}(\theta)^{2}\} =
\sin^{2}\eta.
\]
Following \citet{barndorffnielsengill00}, 
we now state a condition that a measurement must satisfy in order for 
it to achieve this information.

It follows from (\ref{QCREcond2}) that, for a pure 
spin-half state $\rho=|\psi\rangle\langle \psi|$, a necessary and sufficient
condition for a measure\-ment to achieve the information bound is: 
for $\nu$-almost all $x$, $m(x)$ is proportional to
a one-dimensional projector $|\xi(x)\rangle\langle\xi(x)| $ satisfying
$$
\langle\xi|2\rangle\langle2|a\rangle ~ = ~ r(x)\langle\xi|1\rangle
\thinspace ,
$$
where $r(x)$ is real, $|1\rangle=|\psi\rangle$, $|2\rangle=|\psi\rangle ^{\perp}$ 
($|\psi \rangle ^{\perp}$ being a unit vector in $\mathbb C ^2$
orthogonal to $|\psi \rangle$)
and $|a\rangle=2|{\psi}\rangle_{/ \theta}$. It can be seen that geometrically 
this means that $|\xi(x)\rangle$ corresponds to a point 
on $S^{2}$ in the plane
spanned by $\vec{u}(\theta)$ and $\vec{u}_{/\theta}(\theta)$.

If $\eta\ne\pi/2$, this is for each value of $\theta$ a different 
plane, and all these planes intersect in the origin only. Thus no
single measurement $M$ can satisfy $I(\theta ) = i(\theta ; M)$
for all $\theta$. On the other hand, if
$\eta=\pi/2$, so that the states $\rho(\theta)$ lie on a
great circle in the Poincar\'{e} sphere, then the planes defined for
each $\theta$ are all the same. In this case {\em any}
measurement $M$ with all components proportional to projector 
matrices for directions in the plane $\eta=\pi/2$ satisfies 
$I(\theta ) = i(\theta ; M)$ for all $\theta \in \Theta$. In particular,
\emph{any simple measurement in that plane} has this property. 

More generally, a smooth one-parameter model of a spin-half pure 
state with everywhere positive quantum information admits a uniformly 
attaining measurement, i.e. such that $I(\theta ) = i(\theta ; M)$ for 
all $\theta \in \Theta$, if and only if the model is a great circle on 
the Poincar\'e sphere. This is actually a quantum exponential 
transformation model, see Example \ref{e:greatcircle}.
\hfill$\ \qedsymbol$\end{example}

When the state $\rho$ is strictly positive, and under further 
nondegeneracy conditions, essentially the only way to achieve the bound
(\ref{qinfoineq}) is through measuring the quantum score. 
In the discussion below we first keep the value of $\theta$ fixed.
Since any nonnegative self-adjoint matrix can be written as a 
sum of rank-one matrices (using its eigenvalue-eigenvector 
decomposition), it follows that any dominated measurement can be refined 
to a measurement for which each $m(x)$ is 
of rank $1$, thus $m(x)=r(x)|\xi(x)\rangle\langle\xi(x)| $
for some real $r(x)$ and state-vector $|\xi(x)\rangle$,
see the end of Section \ref{ss:meas}.
If one measurement is the refinement of another, then the 
distributions of the outcomes are related in the same way. Therefore,
under refinement of a measurement, Fisher expected 
information cannot decrease. Therefore if any measurement achieves
(\ref{qinfoineq}), there is also a measurement with rank $1$ 
components achieving the bound. Consider such a measurement.
Suppose that $\rho>\boldzero$ and that all the 
eigenvalues of $\rho_{/\!\!/\theta}$ are different. The
condition $m(x)^{1/2}\rho_{/\!\!/\theta}\rho^{1/2} 
=r(x) m(x)^{1 /2} \rho ^{1/2}$ is then 
equivalent to $|\xi(x)\rangle\langle \xi(x) |\rho_{/\!\!/\theta}
=r(x)|\xi(x)\rangle\langle \xi(x) |$, which states that
$\xi(x)$ is an eigenvector of $\rho_{/\!\!/\theta}$. Since we must 
have $m(x)\mu(\mathrm d x)=\boldone$, it follows that all 
eigenvectors of $\rho_{/\!\!/\theta}$ occur in this way in components $m(x)$ of 
$M$. The measurement can therefore be reduced or coarsened
(the opposite of refined)
to a simple measurement of the quantum score, and the reduction (at the 
level of the outcome) is sufficient.

Suppose now the state $\rho(\theta)$ is strictly positive for all
$\theta$, and that the quantum score has distinct eigenvalues for at 
least one value of $\theta$.
Suppose a single measurement exists attaining (\ref{qinfoineq}) 
uniformly in $\theta$. Any refinement of this measurement therefore 
also achieves the bound uniformly, in particular, the refinement to 
components which are all proportional to projectors onto orthogonal
one-dimensional eigenspaces of the quantum score at the value of 
$\theta$ where the eigenvalues are distinct. Therefore the 
eigenvectors of the quantum score at this value of $\theta$ are 
eigenvectors at all other values of $\theta$. Therefore there is a 
self-adjoint operator $X$ with distinct eigenvalues such that
$\rho_{/\!\!/\theta}(\theta)=f(X;\theta)$ for each $\theta$.
Fix $\theta_{0}$ and 
let $F(X;\theta)=\int_{\theta_{0}}^{\theta}f(X;\theta)\mathrm d\theta$.
Let $\rho_{0}=\rho(\theta_{0})$.
If we consider the defining equation (\ref{SLD}) as a differential 
equation for $\rho(\theta)$ given the quantum score, and with initial 
condition $\rho(\theta_{0})=\rho_{0}$, we see that a
solution is $\rho(\theta)=\exp\{\frac12 
F(X;\theta)\}\rho_{0}\exp\{\frac12 F(X;\theta)\}$.
Under smoothness conditions the solution is unique.
Rewriting the form of this solution, we come to the following theorem:

\begin{thm}[Uniform attainability of quantum information bound]\label{t:new}
Suppose that the state is everywhere positive, the quantum 
score has distinct eigenvalues for some value of $\theta$, and is 
smooth. Suppose that a measurement $M$ exists with $i(\theta;M)=I(\theta)$ 
for all $\theta$, thus attaining the Braunstein--Caves information bound 
(\ref{qinfoineq})
uniformly in $\theta$. Then there is an observable $X$ such that a simple 
measurement of $X$ also achieves the bound uniformly, and the model 
is of the form 
\begin{equation}
    \rho(\theta)=c(\theta)\exp\{{\textstyle \frac12} F(X;\theta)\}\rho_{0}\exp\{
    {\textstyle \frac12}
    F(X;\theta)\}
\label{new}
\end{equation}
for a function $F$, indexed by $\theta$, of an observable $X$ where 
$c(\theta)=1/\mathrm{tr}\{\rho_{0}\exp(F(X;\theta))\}$, $\rho_{/\!\!/ 
\theta}(\theta)=f(X;\theta)-\mathrm{tr}\{\rho(\theta) f(X;\theta)\}$, and
$f(X;\theta)=F_{/ \theta}(X;\theta)$.
Conversely, for a model of this form, a measurement of $X$ 
achieves the bound uniformly.
\end{thm}

\begin{rmk}[Spin-half case]
For spin-half, if the information is positive then the 
quantum score has distinct eigenvalues, since the outcome of a 
measurement of the quantum score always equals one of the eigenvalues,
has mean zero, and positive variance.\hfill$\qedsymbol$
\end{rmk}

\begin{thm}[Uniform attainability of quantum Cram\'er--Rao bound]\label{t:new2}
Suppose the positivity and nondegeneracy conditions of the previous 
theorem are satisfied, and suppose that for the outcome of some measurement 
$M$ a statistic $t$ exists which is 
for all $\theta$ an unbiased estimator of $\theta$ achieving Helstrom's
quantum Cram\'er--Rao bound (\ref{QCRB}), $\mathrm{Var}(t) = I(\theta)^{-1}$.
Then the model is actually a quantum exponential model of symmetric 
type (\ref{opsymm}), 
\begin{equation}
    \rho(\theta)=c(\theta)\exp\{{\textstyle{\frac12}} \theta T \} \rho_{0}\exp\{
    {\textstyle{\frac12}}
    \theta T \}
\label{new2}
\end{equation}
for some observable $T$, and simple measurement of $T$ is equivalent to
the coarsening of $M$ according to $t$.
\end{thm}
\begin{prf}
The coarsening of the measurement $M'=M\circ t^{-1}$ corresponding to $t$ also 
achieves the quantum information bound (\ref{qinfoineq}) uniformly,
$i(\theta;M')= I(\theta)$. Apply Theorem \ref{t:new} to this 
measurement and we discover that the model is of the form (\ref{new}),
while (if necessary refining the measurement to have rank one 
components) $t$ can be considered as a function of the outcome of a 
measurement of the observable $X$, and it achieves the classical 
Cram\'er--Rao bound for unbiased estimators of $\theta$ based on this
outcome. Now the density of the outcome
(with respect to counting measure on the eigenvalues of 
$X$) is found to be 
$c(\theta)\exp(F(x;\theta))\mathrm{tr}\{\rho_{0}\Pi_{[X=x]}\}$.
Hence, up to addition of functions of $\theta$ or $x$ alone,
$F(x;\theta)$ is of the form $\theta t(x)$.
\hfill$\ \qedsymbol$\end{prf}

Example \ref{e:spin12qi} concerned pure spin-half models given by circles of 
constant latitude on the Poincar\'{e} sphere.
Taking the product of $n$ identical copies of such a model produces a spin-$j$ 
model, with $j=n/2$,
parameterised by a circle.
It follows from the discussion at the end of Example \ref{e:spin12qi}, 
(\ref{Iaddind}) and the additivity of Fisher information that if
such a spin-$j$ model is given by a great
circle then there is a measurement $M$ such that equality holds in
(\ref{qinfoineq}).

The basic inequality (\ref{qinfoineq}) holds also when the 
dimension of $\theta$ is greater than one. 
In that case, the quantum information matrix $I(\theta)$ is 
defined in (\ref{matrixI}) and the Fisher information matrix 
$i(\theta;M)$ is defined by
$$
i_{rs}(\theta;M) = \mathrm E_{\theta}[l_{r}(\theta)l_{s}(\theta)]
\thinspace , 
$$
where $l_{r}$ denotes $l_{/\theta^{r}}$ etc. 
Then (\ref{qinfoineq}) holds in the sense that 
$I(\theta) -  i(\theta;M)$ is positive semi-definite. The inequality 
is sharp in the sense that $I(\theta)$ is the smallest matrix 
dominating all $i(\theta;M)$. However it is typically not attainable,
let alone uniformly attainable.

Theorem \ref{t:new} can be generalised to the case of a vector parameter. 
This also leads to a generalisation of Theorem \ref{t:new2}, which is the 
content of Corollary \ref{c:peter} below.
First we give a lemma.

\begin{lemma}\label{l:peter}
Let $\boldrho : \Theta \rightarrow \mathcal{S}(\mathcal{H})$ 
be a twice differentiable parametric quantum model.
Then
$$
( \rho_{/\!\!/j}  \rho_{/\!\!/i}  - \rho_{/\!\!/i} \rho_{/\!\!/j}  ) 
\rho +
\rho 
( \rho_{/\!\!/i}  \rho_{/\!\!/j}  - \rho_{/\!\!/j} \rho_{/\!\!/i}  ) 
= 2
(\rho_{/\!\!/i/j} - \rho_{/\!\!/j/i}) \circ \rho,
$$
where $\rho_{/\!\!/\theta} = (\rho_{/\!\!/1}  , \dots , \rho_{/\!\!/k} )$ denotes the
symmetric quantum score and $\circ$ denotes the
Jordan product.
\end{lemma}

\begin{prf}
By definition of $\rho_{/\!\!/\theta}$, we have
$$
4 \rho _{/i} = 2 \left( \rho_{/\!\!/i}   \rho  + \rho \rho_{/\!\!/i}  \right) .
$$
Differentiating this gives
\begin{align*}
4 \rho _{/ij} ~&=~
2 \left( \rho_{/\!\!/i/j}  \rho + \rho_{/\!\!/i}  \rho _{/j} + \rho _j
\rho_{/\!\!/i}  + \rho \rho_{/\!\!/i/j}  \right)  \\
~&=~ 2 \left( \rho_{/\!\!/i/j}  \rho + \rho \rho_{/\!\!/i/j}  \right)
+ \rho_{/\!\!/i}  \rho \rho_{/\!\!/j}  + \rho_{/\!\!/i}  \rho_{/\!\!/j}  \rho
+ \rho \rho_{/\!\!/j}  \rho_{/\!\!/i}  + \rho_{/\!\!/j}  \rho 
\rho_{/\!\!/i}  . 
\end{align*}
Since $\rho _{/ij} = \rho _{/ji}$, this leads to
$$
\left( \rho_{/\!\!/j}  \rho_{/\!\!/i}  - \rho_{/\!\!/i} \rho_{/\!\!/j}  \right) \rho +
\rho \left( \rho_{/\!\!/i}  \rho_{/\!\!/j}  - \rho_{/\!\!/j} 
\rho_{/\!\!/i}  \right)
=
2 \left\{
\left( \rho_{/\!\!/i/j}  - \rho_{/\!\!/j/i}  \right) \rho
+ \rho \left( \rho_{/\!\!/i/j}  - \rho_{/\!\!/j/i}  \right) \right\} .
$$
\hfill$\ \qedsymbol$\end{prf}

\begin{thm}\label{t:peter}
Let $\boldrho : \Theta \rightarrow \mathcal{S}(\mathcal{H})$ be a 
twice differentiable parametric quantum model.
If
\begin{enumerate}
\item[(i)]      there is a measurement $M$ with $i(\theta;M)=I(\theta)$ 
for all $\theta$, 
\item[(ii)]     $\rho (\theta ) > \boldzero$ for all $\theta$,
\item[(iii)]     $\Theta$ is simply connected
\end{enumerate}
then, for any $\theta _0$ in $\Theta$, there are an observable $X$ and a
function $F$ (possibly depending on $\theta _0$) such that
$$
\rho (\theta )  = \exp \left\{ {\textstyle \frac12} F(X; \theta) \right\} 
\rho (\theta _0)
\exp \left\{ {\textstyle \frac12} F(X; \theta) \right\} .
$$
\end{thm}

\begin{prf}
Since $i(\theta ; M) = I(\theta )$, it follows from equation (\ref{QCREcond2})  
and (iii) that there are real-valued functions
$r_1, \dots , r_{{\rm dim} \Theta}$ on $\mathcal{X} \times \Theta$ such that 
$$
m(x)\rho_{/\!\!/i}(\theta)  = r_{i}(x, \theta) m(x) ,
$$
for all $\theta$ in $\Theta$ and $\nu$-almost all $x$.
Then
$$
m(x)\rho_{/\!\!/i}(\theta) \rho_{/\!\!/j}(\theta ') = 
r_{i}(x, \theta) r_{j}(x, \theta ') m(x) 
= m(x)\rho_{/\!\!/j}(\theta ') \rho_{/\!\!/i}(\theta ) ,
$$
for all $\theta , \theta '$ in $\Theta$ and $1 \le i,j \le \mbox{dim $\Theta$}$. 
Integration over $\mathcal{X}$ shows 
that $\rho_{/\!\!/i}(\theta)$ and $\rho_{/\!\!/j}(\theta ')$ commute.
By von Neumann's Theorem, there is an operator $X$ and real-valued functions
$f_1, \dots , f_{{\rm dim} \Theta}$ on $\mathbb R \times \Theta$ such that
\begin{equation}
\rho_{/\!\!/i}  (\theta) =  f_i (X; \theta) .
\label{lami=fi}
\end{equation}
Using condition (iii), and the fact that $\rho_{/\!\!/i} $ and $\rho_{/\!\!/j} $
commute, it follows from the Lemma that
$\rho_{/\!\!/i/j} =  \rho_{/\!\!/j/i} $. By condition (iv),
(\ref{lami=fi}) can be integrated to give a function $F$ such that
$$
\rho_{/\!\!/i}  (\theta) =  F_{/i} (X; \theta) .
$$
The result follows by uniqueness of solutions of differential
equations.
\hfill$\ \qedsymbol$\end{prf}

\begin{cor}\label{c:peter}If under the conditions of Theorem 
\ref{t:peter} there exists an unbiased estimator $t$ of $\theta$ 
based on the measurement $M$ achieving (\ref{QCRB}), then the model
is a quantum exponential family
of symmetric type (\ref{opsymm}) with commuting $T_{r}$.
\end{cor}

Versions of these results have been known for some time; 
see \citet{young75}, \citet{fujiwaranagaoka95}, \citet{amarinagaoka00};
compare especially our Corollary \ref{c:peter} to 
\citet[{\ }Theorem 7.6]{amarinagaoka00},
and our Theorem \ref{t:peter} to parts (I) to (IV) of the subsequent 
outlined proof in  \citet{amarinagaoka00}.
Unfortunately the precise regularity conditions and detailed proofs 
seem to be available only in some earlier publications in Japanese.
Note that we have obtained the same conclusions, by a different proof, 
in the spin-half pure state case, Example \ref{e:spin12qi}. This 
indicates that a more general result is possible without the 
hypothesis of positivity of 
the state.

The symmetric logarithmic derivative is not the unique quantum 
analogue of the classical statistical concept of score.
Other analogues include the right, left and balanced 
derivatives obtained by suitable variants of (\ref{SLD}).
Each of these gives a quantum information inequality 
and a quantum Cram\'er--Rao bound analogous to 
(\ref{qinfoineq}) and (\ref{QCRB}). See \citet{belavkin76}.
There is no general relationship between the 
various quantum information inequalities when the dimension of 
$\theta$ is greater than one.

In the next subsection we discuss the issue of asymptotic 
attainability of these and similar bounds.

\subsection{Asymptotic Information Bounds}\label{ss:asymptotic}

In classical statistics, the Cram\'er--Rao bound is attainable 
uniformly in the unknown parameter only under rather special circumstances.
On the other hand, the restriction to unbiased estimators is hardly 
made in practice and indeed is difficult to defend. 
However, we have a richly developed asymptotic theory 
which states that in large samples certain estimators (e.g., the 
maximum likelihood estimator) are approximately unbiased and 
approximately normally distributed with variance attaining the 
Cram\'er--Rao bound. Moreover, no estimator can do better, in various 
precise mathematical senses (the H\'ajek--LeCam asymptotic local minimax 
theorem and convolution theorem, for instance). Recent work by
\citet{gillmassar00}, surveyed in \citet{gill01}, makes a first 
attempt to carry over these ideas to quantum statistics.
Similar results have been obtained, 
interestingly, with quite different 
methods, in a series of papers, by \citet{young75}, 
\citet{fujiwaranagaoka95}, \citet{hayashi97}, and 
\citet{hayashimatsumoto98}.
Another very recent approach, using large deviation theory rather than 
central limit theory, is given by \citet{keylwerner01}.
The aim of \citet{gillmassar00} was to answer a question first posed by 
\citet{pereswootters91}: do joint measurements on a product of 
identical quantum systems contain more information about the common state of 
the subsystems, than separate measurements? The question was first 
answered---in the affirmative---in a rather specific form, by 
\citet{massarpopescu95}: they considered for the most part just $n=2$ 
copies of a spin-half pure state, in a Bayesian setting with
a special loss function and prior distribution. Work of 
\citet{barndorffnielsengill00} showed that this advantage of joint 
over separate measurements disappears, for the spin-half pure state 
example, 
as $n\to\infty$.%
\marginpar{*}\endnote{R: book: Massar antipodal spins example.} 

The approach of \citet{gillmassar00} 
is firstly to delineate more precisely the class of attainable
information matrices $i_{n}(\theta;M)$ based on arbitrary (or special 
classes) of measurements on the model (\ref{iidmodel}) of $n$ 
identical particles each in the same state $\rho(\theta)$. Next, 
using the van Trees inequality,
a Bayesian version of the Cram\'er--Rao 
inequality, see \citet{gilllevit95}, bounds on $i_{n}(\theta;M)$
are converted into bounds on the asymptotic scaled mean quadratic 
error matrix of regular estimators of $\theta$. Thirdly, one 
constructs measurements and estimators which achieve 
those bounds asymptotically. The first step yields the following theorem.

\begin{thm}[Gill--Massar information bound] In the model (\ref{iidmodel}),
one has
\begin{equation}
    \mathrm{tr}\{I(\theta) ^{-1}i_{n}(\theta;M)/n\}~\le ~
    \mathrm{dim}(\mathcal H)-1
\label{gmbound}
\end{equation}
in any of the following cases: (i) $\mathrm{dim}(\theta)=1$
and $\mathrm{dim}(\mathcal H)=2$, (ii) 
$\rho$ is a pure state, (iii) the measurement $M$ is separable.
\end{thm}
Case (i) follows from the earlier information inequality 
(\ref{qinfoineq}) from which follows, without any further conditions,
$\mathrm{tr}\{I(\theta) ^{-1}i_{n}(\theta;M)/n\}\le
    \mathrm{dim}(\theta)$.
The class of separable measurements, see Section \ref{ss:meas},
includes all \emph{multilocal} instruments, i.e., instruments which are
composed of a sequence of instruments acting on separate particles,
see Section \ref{ss:instru}. 
Thus it is 
allowed that the measurement made on particle $2$ depends on the 
outcome of the measurement on particle $1$, and even that after these 
two measurements, yet another measurement, depending on the results 
so far, is made on the first particle in its new state, etc.

In the spin-half case the bound (\ref{gmbound}) is achievable in the 
sense that for any matrix $K$ such that 
$\mathrm{tr}\{I(\theta) ^{-1}K\}\le 1$, there exists a measurement $M$
on one particle, generally depending on $\theta$, such that
$i(\theta;M)=K$. The measurement is 
a randomised choice of several simple measurements of spin, one 
spin direction for each component of $\theta$.

Application of the van Trees inequality gives the following 
asymptotic bound:

\begin{thm}[Asymptotic information bound] In the model (\ref{iidmodel}),
let $V(\theta)$ denote the limiting scaled mean quadratic error 
matrix of a 
regular sequence of estimators $\widehat\theta_{n}$ 
based on a sequence of measurements $M_{n}$ on $n$ particles; i.e.,
$V^{ij}(\theta)=\lim_{n\to\infty}
n \mathrm E_{\theta}\{(\widehat\theta_{n}^{i}-\theta^{i})
               (\widehat\theta_{n}^{j}-\theta^{j})\}$. Then $V$ satisfies 
the inequality
\begin{equation}
    \mathrm{tr}\{I(\theta) ^{-1}V(\theta)^{-1}\}~\le~
    \mathrm{dim}(\mathcal H)-1
\end{equation}
in any of the following cases: (i) $\mathrm{dim}(\theta)=1$
and $\mathrm{dim}(\mathcal H)=2$, (ii) 
$\rho$ is a pure state, (iii) the measurements $M_{n}$ are separable.
\end{thm}
A \emph{regular estimator sequence} is one for which the mean quadratic error 
matrices converge uniformly in $\theta$ to a continuous limit. It is
also possible to give a version of the theorem in terms of 
convergence in distribution, H\'ajek-regularity and $V$ the mean quadratic error 
matrix of the limiting distribution, rather than the limit of the 
mean quadratic error.

In the spin-half case, this bound is also asymptically achievable, in 
the sense that for any continuous matrix function $W(\theta)$ such that
$\mathrm{tr}\{I(\theta) ^{-1}W(\theta)^{-1}\}\le 1$ there exists a sequence of
separable measurements $M_{n}$ with asymptotic scaled mean quadratic 
error matrix equal to $W$. This result is proved by consideration of 
a rather natural two-stage measurement procedure. Firstly, on a small
(asymptotically vanishing) proportion of the particles, carry out 
arbitrary measurements allowing consistent estimation of $\theta$,
resulting in a preliminary estimate $\widetilde\theta$.
Then on each of the remaining particles, carry out the measurement 
$\widetilde M$ (on each separate particle) which is optimal in the sense that 
$i(\widetilde\theta;\widetilde M)=K=W(\widetilde\theta)^{-1}$.
Estimate $\theta$ by maximum likelihood estimation, conditional on 
the value of $\widetilde\theta$, on the outcomes obtained in the 
second stage. For large $n$, since $\widetilde\theta$ will then be 
close to the true value of $\theta$, the measurement $\widetilde M$ 
will have Fisher information $i(\theta;\widetilde M)$ close to that of
the `optimal' measurement on one particle with Fisher information
$i(\theta,M)=W(\theta)^{-1}$. By the usual properties of maximum 
likelihood estimators, it will therefore have scaled mean quadratic 
error close to $W(\theta)$. 
These measurements are not just 
separable, but multilocal, and within that class, adaptive and 
sequential with each new subsystem being measured
only once.

In the spin-half case we have therefore a complete asymptotic 
efficiency theory in any of the three cases (i) a one-dimensional 
parameter, (ii) a pure state, (iii) separable measurements. By 
`complete' we mean that it is precisely known what is the set of all 
attainable limiting scaled mean quadratic error matrices. This 
collection is described in terms of the quantum information matrix for 
one particle.  What is interesting is that when none of these three 
conditions hold, greater asymptotic precision is possible. For 
instance, \citet{gillmassar00} exhibit a measurement%
\marginpar{*}\endnote{R: book: Gill--Massar example} 
of two spin-half particles which, for a completely unknown mixed state
(a three-parameter model), has about $50\%$ larger total Fisher 
information (for certain parameter values) than any separable 
measurement on two particles. Therefore if one has a large number $n$ 
of particles, one has about $25\%$ better precision when using the 
maximum likelihood estimator applied to the outcomes of this 
measurement on $n/2$ pairs of particles, than any separable 
measurement whatsoever on all $n$. It is not known whether taking 
triples, quadruples, etc., allows even greater increases of precision.
It would be valuable to delineate precisely the set all attainable Fisher 
information matrices when non-separable measurements are allowed on 
each number of particles.

A similar instance of this phenomenon was called 
\emph{non-locality without entangle\-ment} by \citet{bennettetal99a}. One 
could say that though the $n$ particles are not in an entangled state, 
one needs an `entangled measurement', presumably brought about by 
bringing the particles into interaction with one another 
(unitary evolution starting from the product state) before 
measurement, in order to extract maximal information about their 
state. The word `non-locality' refers to the possibility that the $n$ 
particles could be widely separated and brought into interaction 
through other entangled particles; see Section \ref{quprob} for 
further examples of this kind in the context of optimal information 
transmission and in teleportation.

\section{Infinite Dimensional Space}\label{s:further}

So far our examples have concerned spin-half systems, for which the
dimension of the Hilbert space $\mathcal{H}$ is $2$, and occasionally 
spin-$j$ systems (dimension $2j+1$). In this section
we give a survey of an important infinite dimensional example.
The finite dimensional cases led us to parametric quantum statistical 
models. If the system has an infinite-dimensional
Hilbert space, non- and semi-parametric quantum statistical
models make an entrance.
So far, they have been little studied from the point of view of
modern mathematical statistics, despite their significance in experimental
quantum physics, especially quantum optics.

\subsection{Harmonic Oscillator}\label{ss:harmo}

In this subsection we summarise some useful basic theory, and in
the next we consider a basic statistical problem.

The simple harmonic oscillator is the basic model for the motion
of a quantum particle in a quadratic potential well on the real line.
Precisely the same mathematical structure describes oscillations
of a single mode of an electromagnetic field (a single frequency
in one direction in space). A useful orthonormal basis
in the latter situation is given by the state-vectors
of the pure states representing zero, one, two, \ldots 
photons. We denote these state-vectors by
$|0 \rangle , |1 \rangle , |2 \rangle , \ldots$.
This basis is called the \emph{number basis}.
For the simple harmonic oscillator, the pure state with
state-vector $|m \rangle$ is a state of definite
energy $1/2 + m$ units, $m = 0, 1, 2, \dots$.
A pure state with state-vector $|\psi \rangle=\sum c_{m}|m \rangle$,
where $\sum |c_{m}|^{2}=1$,  is a complex superposition of these 
states. A mixed state $\rho$
is a probability mixture over pure states $|\psi \rangle \langle \psi|$
with state-vectors $|\psi\rangle$.

Some key operators in this context, together with their common names, are 
\begin{equation}
\begin{aligned} 
A^+ | n \rangle &= \sqrt {n+1}\, | {n+1} \rangle  & \quad&\mbox{Creation} \\
A^- | n \rangle &= \sqrt {n}\, | {n-1} \rangle  & \quad&\mbox{Annihilation} \\
N | n \rangle &= n \,| n \rangle & \quad&\mbox{Number} \\
Q &=  (A^- + A^+)/\sqrt 2 & \quad&\mbox{Position} \\
P &= \frac 1 {i}(A^- - A^+)/\sqrt 2 & \quad&\mbox{Momentum} \\
X_\phi &= (\cos\phi)\,Q + (\sin \phi)\,P & \quad&\mbox{Quadrature at phase $\phi$}
\thinspace .
\end{aligned}
\end{equation}
One should observe that
\begin{equation}
\begin{aligned}
N = A^+ A^- &= A^- A^+ -\boldone  = \frac12(Q^2+P^2-\boldone ) \\
 \mbox{$ [Q,P]$} &= i \boldone 
\thinspace .
\end{aligned}
\end{equation}

In the simple quantum harmonic oscillator, the state of a part\-icle
evolves under the Hamiltonian $H=\frac12(Q^2+P^2)=N+\frac12\boldone $; 
thus the state-vector of a pure state satisfies
$|\psi(t) \rangle =e^{-i Ht}|\psi(0) \rangle$, and an arbitrary
state evolves as $\rho(t)
= e^{-i Ht}\rho(0)e^{i Ht}$.
The operators $Q$ and $P$ correspond to the position
(on the real line) and the momentum of the particle. 
Indeed, the spectral decompositions
of these two operators yield the PProM's of measurements of
position and momentum respect\-ively.  It turns out that
for a complex number $z=r e^{i\phi}$ and the cor\-responding operator
(called the \emph{Weyl operator})
$W_{z}=\exp(i r X_\phi)$, we have
$e^{i\theta N}W_{z}e^{-i\theta N}=W_{e^{i\theta}z}$, 
or in terms of the operator $X_{\phi}$, 
$e^{i\theta N}e^{it X_{\phi}}e^{-i\theta N}=e^{it X_{\phi+\theta}}$.
These relations become especially powerful when we note a short cut
to the computation of the probability distribut\-ion of the measurement
of the PProM corresponding to an observable $X$: 
it is the probability distribution with characteristic
function $\mathrm{tr}\{\rho e^{itX}\}$. 
Combining these facts, we see that the distribution of the
outcome of a measurement of
position $Q$ on the particle at time $t$ is the same as that
of $X_t$ at time $0$. In particular, with $t=\pi/2$, measuring $P$
at time $0$ has the same distribution as measuring $Q$ at time $\pi/2$.
For future reference, define $F=e^{-i(\pi/2)N}$ and note the
relation $FP=QF$.

We mention for later reference that the Weyl operators form
a projective unitary representation of the translation group
on the real plane, since these are unitary operators with
$W_{z}W_{z'}=w(z,z')W_{z+z'}$ for a certain complex function $w$
of modulus $1$, cf.\ (\ref{wgh}).

In order to derive the probability distributions of outcomes of
measurements of the observables defined above, it is useful to 
consider a particular concrete re\-presentation of the 
abstract Hilbert space $\mathcal H$ as $L^2_{\mathbb C }(\mathbb R )$, 
that is, the space
of complex-valued, Borel measurable, square integrable functions on
the real line. The basis vectors $| n \rangle$ will be represented by
normalised Hermite polynomials times the square root 
of the normal density with mean zero and 
variance half.
The observables $Q$ and $P$ become rather easy to describe in this
representation. At the same time, algebraic results
from the theory of representations of groups provide further
relations between the observables $X_\phi$, $N$, $Q$ and $P$.

Let us define the Hermite polynomials $H_n(x)$, $n=0,1,2\dots$, by
\begin{equation}
H_n(x)=e^{x^2}(-1)^n\frac {\mathrm d^n}{\mathrm d x^n} e^{-x^2}.
\end{equation}
It follows that $H_n(x)$ is an $n$'th order polynomial 
with leading term $(2x)^n$. These polynomials can also be defined
starting from the simple poly\-nomials
$(2x)^n$, $n=0,1,2,\dots$ by Gram--Schmidt 
orthogonalisation with respect to the normal density with mean $0$ 
and variance $1/2$,
$n(x)=(1/\sqrt \pi) \exp(-x^2)$. 
Now if $X$ is normal with mean zero and variance half, 
then $\mathrm E(H_n(X)^2)=2^n n!$. 
Normalising, we
obtain the following ortho\-normal sequence $u_n$
in the space $L^2_{\mathbb C }(\mathbb R )$:
\begin{equation}
u_n(x)=\sqrt{\frac{n(x)}{2^n n!}}H_n(x)
\thinspace .
\end{equation}
This sequence is not only orthonormal but complete---it forms a basis
of $L^2_{\mathbb C }(\mathbb R )$. The functions $u_n$ satisfy
the following recursion relations
\begin{align*}
\sqrt 2\, x u_n(x) ~&=~ \sqrt{n+1}\, u_{n+1}(x)+\sqrt n\, u_{n-1}(x)  \\
\frac d{d x} u_n(x) ~&=~  \sqrt 2 \sqrt {n}\, u_{n-1}(x) - x  u_{n}(x)
\thinspace .  
\end{align*}
This shows us that under the equivalence defined
by $| n \rangle \longleftrightarrow u_n$, one has the following correspondences
\begin{equation}
\begin{aligned}
Q= (A^- + A^+)/\sqrt 2 ~&\longleftrightarrow ~ x  \\
P= \frac 1 i (A^- - A^+)/\sqrt 2 ~&\longleftrightarrow ~ \frac 1 i \frac 
{\mathrm d} {\mathrm d x}  \\
2N+\boldone =Q^2+P^2 ~&\longleftrightarrow ~ \Bigl(x^2-\frac {\mathrm d^2} 
{\mathrm d x^2}\Bigr) 
\thinspace ,
\end{aligned}
\end{equation}
where, on the first line, by `$x$' we mean the operator of multiplication 
of a function of $x$ by $x$ to obtain a new function.
In this representation the operator $Q$ has \lq diagonal\rq\ form, 
corresponding to the PProM with element $\Pi(B)$, $B$ a Borel set of the real 
line, being the operator `multiply by the indicator function $1_{B}$'.
Thus for a pure state with state-vector $|\psi\rangle$ in $\mathcal H$ 
represented by the wave-function $x\mapsto\psi(x)$ in 
$L^2_{\mathbb C }(\mathbb R )$, 
the probability that a measurement of $Q$ takes a value in $B$ is
equal to $\|1_B \psi\|^2=\int_B|\psi(x)|^2 \mathrm d x$, so that the outcome
of the measurement has probability density $|\psi(x)|^2$.
Moreover,
\begin{equation}
\frac{1}{\sqrt{2\pi}}\int_{-\infty}^\infty e^{-itx} u_n(x) \mathrm d x = (-i)^n u_n(t)
\thinspace .
\end{equation}
By expanding an arbitrary wave function $\psi$ as a series
of coefficients times $u_{n}$, one sees from this that
the operator $F=e^{-i(\pi/2)N}=(-i)^{N}$ is nothing 
else than the Fourier transform, and its adjoint $F^{*}$ is the inverse
Fourier transform.%
\marginpar{*}\endnote{R: book: further explanation Fourier} 
The relation $FP=QF$ between $Q$ and $P$ involving $F$ tells us that the 
probability distribution of a measurement of momentum $P$ on a 
particle in the pure state 
with state-vector $|\psi \rangle$ has density equal to the absolute value 
of the square of the Fourier transform of the wave function $\psi(x)$.%
\marginpar{*}\endnote{R: book: further expl density momentum meas} 
Measurement of $Q$ is further studied in Example \ref{e:Qinstr}in Appendix 
\ref{aa:cp}.

More generally, for the observable $X_\phi$ and considering mixed
states instead of pure, from
$e^{i\phi N}e^{it Q}e^{-i\phi N}=e^{it X_{\phi}}$
one may derive the following expression for the probability
density of a measurement of $X_{\phi}$ on a system in state $\rho$:
\begin{equation}\label{dxphimixed}
    p_{\rho}(x;\phi)=\sum_m\sum_{m'}  \rho_{m,m'} e^{i(m-m')\phi} u_m(x)u_{m'}(x)
\thinspace , 
\end{equation}
where $\rho_{m,m'}= \langle m|\rho|m' \rangle$. The sense in which this
double infinite sum converges is rather delicate; however, if only 
finitely many matrix elements $\rho_{m,m'}$ are non-zero, the formula
makes sense as it stands.

\subsection{Quantum Tomography}\label{ss:tomo}

In this subsection we discuss a statistical problem,
called for historical reasons \emph{quantum tomography}, concerning
the observables introduced in the previous subsection.
Some key references are the book \citet{leonhardt97} and the survey 
papers \citet{dariano97a,dariano97b}, though there has been much 
further progress since then.
In its simplest form, the problem of quantum tomography is:
given in\-dependent observations of measurements of 
the quadrature at phase $\phi$, $X_{\phi}$,
with $\phi$ drawn repeatedly at random from the uniform distribution
on $[0,2\pi]$, reconstruct the state $\rho$. 
In statistical terms,
we wish to do nonparametric estimation of $\rho$ from $n$ 
independent and identically distributed observations $(\phi_{i},x_{i})$,
with $\phi_{i}$ as just described and $x_{i}$ from the density 
(\ref{dxphimixed}) with $\phi=\phi_{i}$.  In quantum optics, measuring a single
mode of an electromagnetic field in what is called a quantum homodyne
experi\-ment, this would be the appropriate model with perfect
photo\-detectors. In practice, independent Gaussian noise should be
added.%
\marginpar{*}\endnote{R: book: picture of homodyne meas}

Recalling that the probability density of a measurement of 
$X_{\phi}$ has 
$\mathrm{tr}\{ \rho e^{i t(\cos\phi  Q+\sin\phi  P)}\}$
as its characteristic function, we note that
{\em if} $Q$ and $P$ were actually commuting operators (they
are not!) then the joint characteristic function of a
measurement of the two simul\-taneously would have been the function
$\mathrm{tr}\{ \rho e^{i(s Q+ t P)}\}$ of the two variables $(s,t)$.

Now the latter may not be the bivariate characteristic function
of a joint probability density, but it is the characteristic
function of a certain function called the Wigner function.
This function $W_{\rho}(q,p)$ is known to characterise $\rho$.
It is a real-valued function, integrating to $1$ over the whole
plane, but generally taking negative as well as positive
values. The relation between the characteristic function of
a measurement of $X_{\phi}$ and the characteristic function
of the Wigner function which we have just described, shows that
the probability density of a measurement of $X_{\phi}$ can be
computed from the Wigner function by treating it as a joint
probability density of two random variables $\widetilde Q,\widetilde P$
and computing from
this density the marginal density of the linear combination
$\cos\phi \, \widetilde Q+\sin\phi \, \widetilde P$.  
Now this computation is nothing else
than a computation of the Radon transform of $W_{\rho}(q,p)$: its projection
onto the line $(\cos\phi)q+(\sin\phi)p=0$. This transform is
well known from computer-aided tomography, when for instance the
data from which an X-ray image must be computed is the collection
of one-dimensional images obtained by projecting onto all
poss\-ible directions. Thus from the collection of all densities
$p_{\rho}(x;\phi)$ of measurements of $X_{\phi}$, one could in principle
compute the Wigner function $W_{\rho}(q,p)$ by inverse Radon transform,
from which one can compute other represent\-ations of $\rho$ by 
further appropriate transformations. In particular, a double 
infinite integral over $(p,q)$ of the product of the Wigner function with
an appropriate kernel results in $\rho$ in the `position'
representation, i.e., as the kernel of an integral transform 
mapping $L^{2}$ into $L^{2}$. Not all states can be
so represented, but at least all can be approximated in this
way. A further double infinite integral over $(x,x')$ of another
kernel results in $\rho$ in the `number' representation, i.e.,
the elements $\rho_{m,m'}$.

The basic idea of quantum tomography is to carry out
this sequence of mathematical transformations
on an empirical version of the density $p_{\rho}(x;\phi)$
obtained by some combination of smoothing and binning of
the observations $(\phi_{i},x_{i})$.
This theoretical possibility was discovered by \citet{vogelrisken89},
and first carried out experimentally by M.G. Raymer and colleagues in
path-breaking experiments in the early 1990's, see \citet{smitheyetal93}.
Despite the
enthusiasm with which the initial results were received, the method
has a large number of drawbacks. To begin with, it depends on
some choices of smoothing parameters and/or binning intervals,
and later, during the succession of integral transforms, on truncations
of infinite integrals among other numerical approximations. It has
been discovered that these `smoothings' tend to destroy precisely
the interesting `quantum' features of the functions being 
reconstructed. The final result suffers from both bias and variance,
neither of which can be evaluated easily. Inverting the Radon 
transform is an ill-posed inverse problem and the whole procedure needs
massive numbers of observations before it works reasonably well.

In the mid 1990's G.M. D'Ariano and his coworkers in Pavia have discovered
a fascinating method to short-cut this approach, see 
\citet{dariano97a,dariano97b}.
Using the fact that
that the Weyl operators introduced above form an irreducible 
projective representation of the translation group on $\mathbb R ^{2}$,
they derived an elegant `tomographic formula' expressing the
mean of any operator $A$ (not necessarily self-adjoint), 
i.e., $\mbox{tr}(\rho A)$, as the integral of
a function (depending on the choice of
$A$) of $x$ and $\phi$, multiplied by $p_{\rho}(x;\phi)$, with
respect to Lebesgue measure on $\mathbb R \times[0,2\pi]$.
In particular, if we take the operator $A$ to be $|m'\rangle \langle m|$
for given $(m,m')$, we have hereby expressed $\rho_{m,m'}$
as the mean value of a certain function, indexed by $(m,m')$,
of the observations $(\phi_{i},x_{i})$, as long as the phases
$\phi_{i}$ are chosen uniformly at random.

The  key relation of their approach is the identity
\begin{equation}
    A=\pi^{-1}\int_{\mathbb C }\mbox{tr}(AW_{z})W_{\overline 
    z}\mathrm d z
\thinspace ,
\end{equation}
which can be derived (and generalised)
with the theory of group representations. From this follows
\begin{equation}
   \mbox{tr}(\rho A)=\pi^{-1}\int_{\mathbb C }\mbox{tr}(AW_{z})
   \mbox{tr}(\rho W_{\overline z})\mathrm d z
\thinspace .
\end{equation}
The left hand side is the mean value of interest. The first
\lq trace\rq\ on the right hand side is a known function of the
operator of interest $A$ and the variable $z$. In the second
\lq trace\rq\ on the right hand side, after expressing $z=re^{i\phi}$
in polar coordinates, we recognise the characteristic function
evaluated at the argument $r$ of the probability density of
our observations $p_{\rho}(x;\phi)$. Writing the characteristic
function as the integral over $x$ of $e^{irx}$ times this density, 
transforming the integral over $z$ into integrals over $r$ and $\phi$,
and reordering the three resulting integrals,
we can rewrite the right hand side as
$$\int_{x=-\infty}^{\infty}\int_{\phi=0}^{2\pi}\left[\int_{r=0}^{\infty}
K_{A}(r,x,\phi) \mathrm d r \right] p_{\rho}(x;\phi) \mathrm d x \mathrm 
d\phi/(2\pi).$$

The innermost integral can sometimes be evaluated analytically, 
other\-wise numerically; but in either case we have succeeded in
our aim of rewriting means of operators of interest as means of
known kernel functions of our observations. In the case 
$A=|m'\rangle \langle m|$, of interest for reconstructing $\rho_{m,m'}$,
the kernel turns out to be bounded and hence we obtain 
unbiased estimators of the $\rho_{m,m'}$ with variance 
equal to $1/n$ times some bounded quantities.

Still this approach has its drawbacks. The required kernel function,
in the case of reconstructing the density in the
number representation, is highly oscillatory and even though 
everything is bounded, still huge numbers of observations are needed
to get informative estimates. Also, the unbiased
estimators constructed in this way are not unique and one may wonder whether
better choices of kernels can be found.  
However, the approach does open a window of
opportunity for further mathematical study of the mapping from 
$p_{\rho}(x;\phi)$ to  $\rho_{m,m'}$ which could be a vital tool
for developing the most recent approach, which we now outline briefly.

As we made clear, the statistical estimation problem seems related
to the problems of nonparametric curve estimation, or more 
precisely, estimation of a parameter lying in an infinite 
dimensional space.  Modern experience with such problems has 
developed an arsenal of methods, of which penalised and
sieved likelihood, and nonparametric Bayesian methods, hold
much promise as `universal' approaches leading to opt\-imal methods.
In the present context, sieved maximum likelihood is very natural,
since truncation of the Hilbert space in the number bas\-is leads to
finite dimensional parametric models which can in principle be tackled
by maximum likelihood. One can hope that, from a study of the balance
between truncation error (bias) and variance, it would be possible to 
derive data-driven methods to estimate $\rho$ optimally with 
respect to a user-specified loss function.  So far, only the 
initial steps in this research programme have been taken; in recent
work \citet{banaszeketal00} and \citet{parisetal01} have shown that maximum likelihood 
estimation of the parameters in the density (\ref{dxphimixed})
is numerically feasible, after the number basis 
$\{ |m \rangle : m = 0,1, \dots \}$ is truncated at (e.g.) $m = 15$ or $m = 20$.
This means estimation of about $400$ real parameters constrained to produce a 
density matrix. 
Numerical optim\-isation was used after a reparameterisation by
writing $\rho=T T^*$ as the product of an upper-triangular
matrix and its adjoint, so that only one constraint (trace $1$)
needs to be incorporated. We think that it is a major open problem
to work out the asymptotic theory of this method, taking account
of data-driven truncation, and possibly alleviating the problem
of such a large parameter-space by using Bayesian methods.
The method should be tuned to the estimation of various
functionals of $\rho$ of interest, and should provide standard
errors or confidence intervals.

The quantum statistical model introduced above is that of optical 
homodyne measurements.
There is also an elegant mathematical model for another 
experimental set-up called heterodyne measurement. In this case the 
measurement is a generalised measurement or OProM, and it can be 
\emph{realised} by taking the product of the Hilbert space of 
the system of interest with another infinite dimensional system, in 
its ground state. Write $Q'$, $P'$ for the position and momentum 
operators on the ancillary system. It turns out that $P+P'$ and 
$Q-Q'$ commute, and therefore could in principle be measured 
simultaneously. A joint measurement of the two is a realisation of a 
heterodyne measurement. As an OProM it is invariant under the rotation 
group (corresponding to the phase changes $\phi$ of the homodyne 
measurement) 
and under a certain parametric model for the state, called 
the Gaussian or coherent state of the harmonic oscillator,
possesses some decision-theoretic optimality 
properties because of this, see \citet{holevo82}. The pair now form a 
quantum transformation model in the sense of Section \ref{ss:qtm}.

The field of quantum tomography is rapidly developing, with some of 
the latest (not yet published) results from the Pavia group of G.M. 
d'Ariano being \emph{quantum holographic} methods to estimate 
not an unknown state,
but an unknown transformation of a state (i.e., a completely positive
instrument with trivial outcome 
space).

\section{From Quantum Probability to Quantum Statistics}\label{quprob}

A recurring theme in this section is the relation between 
classical and quantum probability and statistics. This has been a matter of 
heated controversy ever since the discovery of quantum mechanics. 
It has mathematical, physical, and philosophical 
ingredients and much confusion, if not controversy, has been generated by 
problems of interdisciplinary communication between mathematicians, 
physicists, philosophers and more recently statisticians. 
Authorities from both physics and mathematics, 
perhaps starting with \citet{feynman51},
have promoted vigorously the standpoint that `quantum probability' 
is something very different from `classical probability'.
Most recently, in two papers on Bell's inequality (which we discuss in 
Section \ref{ss:bell}) \citet{accardiregoli00a,accardiregoli00b}, 
state ``the real origin of 
the Bell's inequality is the assumption of the applicability of 
classical (Kolmogorovian) probability to quantum mechanics'' which 
can only be interpreted as a categorical statement that classical probability 
is \emph{not} applicable to quantum mechanics.
\citet{malleyhornstein93} conclude from the perceived conflict between 
classical and quantum probability that 
`quantum statistics' should be set apart from classical statistics.

We disagree.
In our opinion, though fascinating mathematical facts and physical 
phenomena lie at the root of these statements, cultural preconceptions
have also played a role. Statistical problems from 
quantum mechanics fall definitely in the framework of classical statistics 
and the claimed distinctions have retarded the adoption of 
statistical science in physics. The phenomenon of quantum entanglement
in fact has far-reaching technological implications, which are easy to 
grasp in terms of classical probability; their development will surely 
involve statistics too.

In the first subsection we discuss, from a mathematical point of view, the 
distinction between classical and quantum probability. Next, we consider 
physical implications of the probabilistic predictions of quantum mechanics 
through the celebrated example of the \citet{bell64} inequalities and
the \citet{aspectetal82a,aspectetal82b} experiment.
We appraise the `classical versus quantum' question in the 
light of those implications. Finally we review a number of controversial 
issues in the foundations of quantum physics (locality, realism, the 
measurement problem) and sketch the basics of quantum teleportation,
emphasizing that emerging quantum technology (entanglement-assisted 
communication, quantum computation, quantum holography and 
tomography of instruments) aims to capitalise on precisely 
those features of quantum mechanics which in the past have often been seen
as paradoxical theoretical nuisances.

\subsection{Classical versus Quantum 
Probability}\label{ss:classicalvsquantum}

Our stance is that the predictions which quantum mechanics makes of the 
real world are stochastic in nature. A quantum physical model of a 
particular phenomenon allows one to compute probabilities of all possible 
outcomes of all possible measurements of the quantum system. The word 
`probability' means here: relative frequency in many independent repetitions.
The word `measurement' is meant in the broad sense of: macroscopic results of
interactions of the quantum system under study with the outside world.
These predictions depend on a summary of the state of the quantum system.
The word `state' might suggest some fundamental property of a particular 
collection of particles, but for our purposes all we need to understand 
under the word is: a convenient mathematical encapsulation of
the information needed to make any such predictions. 
Some physicists argue that it is meaningless to talk
of the state of a particular particle, one can only talk of the state of a large 
collection of particles prepared in identical circumstances;
this is called a \emph{statistical ensemble}. Others take the point of view 
that when one talks about the state of a 
particular quantum system one is really talking about a property of the 
mechanism which generated that system. 
Given that quantum mechanics predicts only probabilities, as far
as real-world predictions are concerned
the distinction between on the one hand a property of an 
ensemble of particles or of a procedure to prepare particles, and on 
the other hand a property of one particular particle,
is a matter of semantics.  However, if one would like to understand quantum 
mechanics by somehow finding a more classical (intuitive) physical theory in the 
background which would explain the observed phenomena, this becomes an 
important issue. It is also an issue for cosmology, when there is 
only one closed quantum system under study: the universe.

It follows from our standpoint that `quantum statistics' is, for us, classical 
statistical inference about unknown parameters in models for data arising 
from measurements on a quantum system. However, just as in biostatistics,
geostatistics, etc., etc., many of these statistical problems have a common 
structure and it pays to study the core ideas and common 
features in detail. As we have 
seen, this leads to the introduction of mathematical objects such as 
quantum score, quantum expected information, quantum exponential family, 
quantum transformation model, and so on; the names are deliberately chosen 
because of analogy and connections with the existing notions from classical 
statistics.

Already at the level of probability (i.e., before statistical considerations 
arise) one can see analogies between the mathematics of quantum states and 
observables on the one hand, and classical probability measures and random variables 
on the other. This analogy is very strong and indeed mathematically very 
fruitful (also very fruitful for mathematical physics). Note that 
collections of both random variables 
and operators can be endowed with algebraic structure (sums, products, 
\dots). It is a fact that from an abstract point of view a basic structure 
in probability theory---a collection of random variables $X$ on a countably 
generated probability space, together with their expectations $\int X 
\mathrm d P$
under a given probability measure $P$---can be represented by a 
(commuting)
subset of the set of self-adjoint operators $Q$ on a separable Hilbert space 
together with the expectations $\mathrm{tr}\{\rho Q\}$ computed using the trace 
rule under a given state $\rho$. Thus: a \emph{basic} structure in classical 
probability theory is isomorphic to a \emph{special case} of a basic structure in 
quantum probability. `Quantum probability', or `noncommutative 
probability theory' is the name of the branch of mathematics which studies 
the mathematical structure of states and observables in quantum mechanics.
From this mathematical point of view, one may justly 
claim that classical probability is a special case 
of quantum probability. The claim does entail, however, a rather narrow view 
of classical probability. Moreover, many probabilists will feel that
abandoning commutativity is throwing away the baby with the 
bathwater, since this broader mathematical structure has no analogue of the
sample outcome $\omega$, and hence no opportunity for a
probabilist's beloved probabilistic arguments.
We discuss Quantum Probability further in Section \ref{ss:qstochproc} under the 
heading of Quantum Stochastic Processes.

\subsection{Bell, Aspect, et al.}\label{ss:bell}

We now discuss some physical predictions of quantum mechanics of a most%
\marginpar{*}\endnote{R: book: ref Mermin et al sequel to Bell} 
striking `nonclassical' nature. Many authors have taken this as 
a defect of classical probability theory and there have been proposals to 
abandon classical probability in favour of alternative theories (negative, 
complex or $p$-adic probabilities; nonmeasurable events; noncommutative 
probability; \dots) in order to `resolve the paradox'. However in our 
opinion, the phenomena are real and the defect, if any, lies in 
believing that quantum phenomena do not contradict classical \emph{physical}
thinking. This opinion is supported by the recent development of 
(potential) technology which acknowledges the extraordinary nature of the 
predictions and exploits the discovered phenomena (teleportation, 
entanglement-assisted communication, and so on). In other words, one should 
not try to explain away the strange features of quantum mechanics as some 
kind of defect of classical probabilistic thinking, but one should use 
classical probabilistic thinking to pinpoint these features.

Consider two spin-half particles, for which the customary state space 
is $\mathcal H= \mathbb C ^2 \otimes \mathbb C ^2$. Let $|0\rangle$ and
$|1\rangle$ denote the orthonormal basis of $\mathbb C ^2$ corresponding to 
`spin up' and `spin down', thus two eigenvectors of the Pauli spin matrix
$\sigma_z$. We write $|ij\rangle$ as an abbreviation for 
$|i\rangle\otimes|j\rangle$, defining four elements of an orthonormal basis
of our $\mathcal H$. 

For $\vec u$ in $S^2$, let 
$\sigma_{\vec u}=u_x\sigma_x +u_y\sigma_y +u_z\sigma_z$, the observable 
`spin in the direction $\vec u$' for one spin-half particle. It has 
eigenvalues $\pm 1$ and its eigenvectors are the state-vectors 
$\psi(\pm \vec u)$ corresponding 
to the directions $\pm\vec u$ in $S^2$. The appropriate model for measurement
of spin in direction $\vec u$ on the first particle and spin in the 
direction $\vec v$ on the second particle is a joint simple measurement of 
the two compatible observables $\sigma_{\vec u}\otimes\boldone $ and
$\boldone \otimes\sigma_{\vec v}$ (see Example \ref{e:simpinst}). The possible outcomes $\pm 1, \pm 1$
correspond to the one-dimensional subspaces spanned by the four orthogonal
vectors $\psi(\pm \vec u)\otimes \psi(\pm\vec v)$.

Now if the state of the system is a tensor product $\rho_1\otimes\rho_2$
of separate states of each particle, then one can directly show that the
outcomes for particle $1$ and particle $2$ are independent, and distributed
as separate measurements on the separate particles, as one would hope.
If the joint state is a mixture of product states, then the outcomes will
be distributed as a mixture of independent outcomes. For an entangled 
state, the outcomes can be even more heavily dependent.

Consider the entangled pure state with state vector
$\{|10\rangle-|01\rangle\}/\sqrt 2$. This state is often called the 
\emph{singlet} or \emph{Bell} state. Straightforward calculations, 
see for instance \cite{barndorffnielsenetal02}, show that for
this state the two spin measurements have the following joint distribution:
the marginal distribution of each spin measurement is Bernoulli($\frac12$),
the probability that the two outcomes are equal (both $+1$ or both $-1$)
is $\frac12(1-\vec u\cdot\vec v)$. In particular, if the two measurements 
are taken in the \emph{same} direction, then the two outcomes
are different with 
probability $1$; in the opposite direction, the two outcomes
are always the same; in orthogonal directions the probability of equality 
is $\frac12$ so, taking account of the marginal distributions, the two 
outcomes are independent.

The singlet state is an appropriate description 
for the spins of two spin-half particles 
produced simultaneously in some nuclear scattering or decay processes
where a total spin of $0$ is conserved. The two particles have exactly
opposite spin, which seems reasonable. The two particles are 
together in a pure state, which is also reasonable if the process involved 
was a Schr\"odinger evolution starting from a pure state. The model also 
exhibits a rotational invariance. These are all good reasons to expect 
the model to be not just a hypothetical possibility but a real possibility
(and indeed, it is).

Fix a special choice of two possible different values of $\vec u$
and two possible different values of $\vec v$. Let us suppose that all 
four directions are in the same great circle on $S^2$ and let $\vec u_1$
and $\vec u_2$ be in the directions $0^\circ$ and $120^\circ$, let
$\vec v_1$ and $\vec v_2$ be in the directions $180^\circ$ and $60^\circ$.
Since $\cos(60^\circ)=\frac12$ we see that: when the directions are the pair
$(0^{\circ},180^{\circ})$ then the probability the
two spins are found to be equal is $1$; but when the 
directions are any of the three pairs $(0^{\circ},60^{\circ})$ or 
$(120^{\circ},180^{\circ})$ or $(120^{\circ},60^{\circ})$
the two spins are found to be equal with probability $\frac14$. 
Is this surprising?

Consider an experiment where pairs of particles are generated in the 
singlet state, and then made to travel to two far-apart locations, at each of
which spin is measured in one of the two directions just specified.
Suppose the experiment is repeated many times, with random and 
independent choice of the two directions for measurement at each of
the two locations. We have just computed the probabilities of all 
possible outcomes under each of the four possible combinations of 
directions.

Let us try to simulate the predicted statistics of the 
experiment using classical objects. To be very 
concrete, consider two people who try to simulate two spin-half particles.
They start in a room together but then leave by different doors.
Outside the room they are separately told a direction,
$\vec u_1$ or $\vec u_2$ for person $1$, $\vec v_1$ or $\vec v_2$ for 
person $2$, and asked to choose an outcome `$+1$' or `$-1$'. 
They are not allowed to communicate 
any more once they have left the room. Moreover the directions will be 
chosen independently and randomly. The whole procedure will be repeated 
many many times and their aim is to simulate the quantum probabilities
stated above. The two persons obviously will need randomisation in order
to imitate the randomness of spin-half particles. We allow them to
toss dice or coins, in any way they like, and to do this together in the
room before leaving. They can simulate in this way any degree of dependence 
or independence they like. Let us call the outcome of their randomisation 
process $\omega$. Their strategy will then be two pairs of functions of
$\omega$, with values $\pm 1$, which determine the answers each person 
would give when confronted with each of his two directions on leaving the 
room, when the randomisation produces the outcome $\omega$. 

This whole set-up defines four Bernoulli $\pm 1$-valued
random variables, let us call them
$X_1$, $X_2$, $Y_1$, $Y_2$; the $X$ variables for person $1$ and the $Y$
variables for person $2$. The four must be such that any pair $X_i,Y_j$
has the same joint distribution as the result of measuring spins in the 
directions $\vec u_i$ and $\vec v_j$. Now it is easy to check that since 
these four variables are binary,  $X_1\ne Y_2$ and $Y_2\ne X_2$
and $X_2\ne Y_1$ implies $X_1\ne Y_1$ (just fill in $+1$, $-1$, $+1$, $-1$
for $X_1$, $Y_2$, $X_2$, $Y_1$ in order; or alternatively $-1$, $+1$, 
$-1$, $+1$.) Conversely, therefore, $X_1=Y_1$ implies $X_1=Y_2$ or $Y_2=X_2$
or $X_2=Y_1$. Therefore we have 
$$
\mathrm P(X_1=Y_1)\le 
\mathrm P(X_1=Y_2)+\mathrm P(Y_2=X_2)+
\mathrm P(X_2=Y_1).
$$
But the four probabilities we are trying to simulate are
$1$, $\frac14$, $\frac14$, $\frac14$ and it is not true that 
$1\le\frac14+\frac14+\frac14$.
Therefore it is not possible to simulate with classical means (people or
computers or other classical physical systems) the predicted outcomes of 
measurements of two spin-half particles!

The inequality we have just derived is due to \citet{bell64} who contrasted
it with the prediction of quantum mechanics in order to prove the failure, a 
priori, of any
attempt through the introduction of \emph{hidden variables}
to explain the randomness of outcomes 
of measurements of quantum systems through `mere statistical variation' in 
not directly observed and uncontrollable
(hence hidden) properties of the quantum systems or 
measurement devices. He assumed that any physically meaningful hidden 
variables model would satisfy the physically reasonable property of 
\emph{locality}, that is to say, the outcome of a measurement on one 
particle in one location should not depend on the measurement being carried 
out simultanously on the other particle in another distant location.
Inspection of the argument we have given shows that Bell's inequality 
is not due to our slavish adherence to classical probability, but 
simply through the assumption that \emph{the outcome of a measurement on 
one particle should not depend on} which \emph{measurement is being made on the 
other particle}. This is reason enough for some authors, for instance
\citet{maudlin94}, to conclude that Bell's argument shows that the 
predictions of quantum mechanics violate \emph{locality}; he goes on to 
study the possible conflicts with relativity theory and concludes that 
there is no conflict in the sense that this phenomenon does not violate the 
requirements that cause and effect should not spread faster than the speed 
of light, and there is not a conflict with the basic relativistic
(Minkowski) invariance property. Thus 
quantum mechanics lives in uneasy but peaceful coexistence with relativity 
theory.

All this would be purely academic were it not the case that the model we have
just described truly is appropriate in certain physical situations and the 
predictions of quantum theory have been experimentally verified; first by
Alain Aspect and his coworkers in a celebrated experiment 
(reported in \citealt{aspectetal82a,aspectetal82b}) in Orsay, Paris,
where polarisation of pairs of entangled photons emitted from an excited 
caesium atom was measured with 
polarisation filters several metres apart; the 
orientation of the filters being fixed independently and randomly
after the photons had been emitted from the source and before they arrived 
at the polarisation filter. (Polarisation of photons has a very similar 
mathematical description to spin of spin-half particles, except that all 
angles need to be halved: entangled photons have equal behaviour at 
polarisation filters oriented $90^\circ$ to one another.) More recently, 
the experiment has been done on the glass fibre network of Swiss telecom 
with the two filters being $10$ km apart on different shores of Lake Geneva.

Our conclusion is that quantum mechanics makes extraordinary physical 
predictions, predictions which are properly stated and interpreted in the 
language of classical probability. Technological implications of these 
predictions are only just beginning to be explored.  One proposal is 
\emph{entanglement-assisted communication},
see \citet{bennettetal99b, bennettetal01, holevo01c}.
Suppose A would like to send a message to B by 
encoding the message in the states of a sequence of spin-half particles 
transmitted one by one from A to B. At the receiving end B carries out 
measurements on the received particles on the basis of which he infers the 
message. Obviously the results will be random, especially if the 
communication channel suffers from noise, of classical or quantum nature.
Using the theory of instruments one can describe mathematically all 
physically possible communication channels and all physically possible 
decoding (measurement) schemes, and compute analogously to classical 
information theory the maximum rate of transmission of information through 
the channel. Suppose now A and B allow themselves a further resource for
communication. In between A and B a third person C is located, and he sends 
A and B simultaneously pairs of entangled spin half particles, in step 
with the transmission of particles from A to B. `Obviously' there is no way 
these particles can be used to transmit information from A to B. They come 
from a different source altogether and are created in a fixed and known 
state. Yet it turns out that if A uses one part of the entangled pair in his
encoding step with each particle he transmits, and B uses the other part 
of the pair in his decoding step, the rate of transmission can be 
\emph{doubled}.

These extraordinary results show that it would be foolish to `explain 
away' the phenomenon discovered by Bell by turning to some exotic 
probability theory (though many authors have done precisely this!).
On the contrary, the mathematics---using classical probability---shows 
that strange things are going on and indeed it seems likely that one will 
be able to harness them in future technology.

\subsection{Teleportation}

As an example we show how the singlet state of a pair of spin-half 
particles, supposed to be in two distant locations,
can be used to transmit a third spin-half state from one location to the 
other. This scheme was invented by \citet{bennettetal93} and experimentally 
carried out by A. Zeilinger's group in Innsbruck, see 
\citet{bouwmeesteretal97}. 
For a recent survey including references to the results of other 
experimental groups see \citet{bouwmeesteretal01}.
The method illustrates how quantum technology (e.g., 
computation) will combine the basic ingredients of simple measurements, 
unitary evolution, and entanglement (product systems).
The state being teleported is supposed to be completely unknown. This means 
that any attempt to measure it, and then teleport it by communicating in a 
classical way the results of measurement, cannot succeed, since the 
outcomes will be random, do not determine the initial state, and the 
initial state will have been destroyed by the measurement. The 
\emph{no-cloning theorem} of
\citet{wootterszurek82}, \citet{dieks82}
shows that there is no instrument which can 
transform a state $\rho$ together with an ancillary quantum system
into two identical copies $\rho\otimes\rho$.

Consider a single spin-half particle in the pure state with state-vector 
$\alpha|1\rangle+\beta|0\rangle$. It is brought into interaction with a 
pair of particles in the singlet state so that the whole system is in the 
pure state with state-vector, after multiplication of the tensor product,
and up to a factor $1/\sqrt 2$,
$\alpha|110\rangle-\alpha|101\rangle+\beta|010\rangle-\beta|001\rangle$.
The three particles are here written in the sequence: particle to be 
teleported, first entangled particle at the source location, second 
entangled particle at the destination location.
Now we introduce the following four orthogonal state-vectors for the two 
particles at the source location, neglecting another constant factor $1/\sqrt 2$, 
$\Phi_1=|10\rangle-|01\rangle$, $\Phi_2=|10\rangle+|01\rangle$,
$\Psi_1=|11\rangle+|00\rangle$, $\Psi_2=|11\rangle-|00\rangle$,
and we note that our three particles together are in a pure state 
with state-vector which may be
written (up to yet another factor, $1/\sqrt 4$)
$\Psi_1\otimes(\alpha|0\rangle-\beta|1\rangle)+
\Psi_2\otimes(\alpha|0\rangle+\beta|1\rangle)+
\Phi_1\otimes(-\alpha|1\rangle-\beta|0\rangle)+
\Phi_2\otimes(-\alpha|1\rangle+\beta|0\rangle)$.
So far nothing has happened at all: we have simply rewritten the 
state-vector of the three particles as a superposition of four 
state-vectors, each lying in one of four orthogonal two-dimensional
subspaces of $\mathbb C ^2 \otimes \mathbb C ^2 \otimes \mathbb C ^2$:
namely the subspaces
$\Phi_{1}\otimes\mathbb C ^2$,
$\Phi_{2}\otimes\mathbb C ^2$, $\Psi_{1}\otimes\mathbb C ^2$ and
$\Psi_{2}\otimes\mathbb C ^2$.

To these four subspaces corresponds a simple instrument. 
It only involves the two particles at the source
location and hence may be carried out by
the person at that location.
He obtains one of four different outcomes, each with probability 
$\frac14$, so he learns nothing about the particle to be teleported.
However, conditional on the outcome of his measurement, 
the particle at the destination is in one 
of the four pure states with state-vectors $\alpha|0\rangle-\beta|1\rangle$,
$\alpha|0\rangle+\beta|1\rangle$, $-\alpha|1\rangle-\beta|0\rangle$,
$-\alpha|1\rangle+\beta|0\rangle$. The mixture with equal probabilities of 
these four states is the completely mixed state $\rho=\frac12 \boldone $,
so nothing has happened at the destination: the state of the second part of 
the entangled pair still is in its original (marginal) state.
But once the outcome of the measurement at the source is transmitted to 
the destination (two bits of information, transmitted by classical means),
the receiver is able by means of one of four unitary transformations to 
transform the resulting pure state into the state with state-vector 
$\alpha|0\rangle+\beta|1\rangle$: teleportation is succesful.
Neither source nor destination learn anything 
at all about the particle being transmitted by this procedure. If 
the state being teleported was a mixture, then decomposing it into pure 
components which are teleported independently and perfectly shows that the 
final destination state is the \emph{same} mixture. In short, by 
transmitting two classical bits of information we are able to copy a point 
in the unit ball (specified by three real numbers) from A to B, without learning 
anything about the point at all in the process.

\subsection{The Measurement Problem}\label{ss:measprob}

We summarise here the problem raised by Schr\"odinger's cat, and
survey briefly some responses.  Consider a spin-half particle in the 
pure state with state-vector $\alpha|0\rangle+\beta|1\rangle$, where
$|\alpha |^{2}+|\beta |^{2}=1$. Suppose a measurement is made of the 
PProM with elements $\{|0\rangle\langle 0|,|1\rangle \langle 1 | 
\}$, resulting in the outcomes $0$ and $1$ with probabilities 
$|\alpha |^{2},|\beta |^{2}$. Next to the measurement device is a cage
containing a cat and a closed bottle of poison.  If the outcome
is $1$, an apparatus automatically releases the poison and the cat dies.
Otherwise, it lives. We suppose this whole system is enclosed in a 
large container and isolated from the rest of the universe. 

Now the contents of that container are themselves just one large quantum 
system, and presumably it evolves unitarily under some Hamiltonian.
If $\alpha=0$, the final situation involves a dead cat. Let us denote
its state-vector then by $|\text{dead}\rangle$. If $\beta=0$ then the
final state of the cat has state-vector $|\text{alive}\rangle$.
So by linearity, in general the final state of the cat has 
state-vector  $\alpha|\text{alive}\rangle+\beta|\text{dead}\rangle$.
How would the cat experience being in this state?

When the container is opened and we look in, presumably a measurement
does take place of the state of the cat, and at that moment (and only 
at that moment) it collapses into one of the two states 
with state-vectors $|\text{alive}\rangle$, $|\text{dead}\rangle$
with the probabilities $|\alpha |^{2},|\beta |^{2}$.
Recently, a number of experiments have been done which are purported to 
produce Schr\"odinger cats, in the sense of quantum superpositions of 
macroscopically distinct physical states of physical systems. 
For instance, \citet{mooijetal99} report on an experiment in which an 
electronic 
current involving of the order of a billion ($10^{9}$) 
electrons flows in a superposition 
of clockwise and anticlockwise directions around a supercooled 
alumuminium ring of a few micrometers in diameter (a thousand times 
larger than a typical molecular dimension). See \citet{gill01b} for a 
discussion of this experiment and of the role of quantum statistics in 
confirming its 
success.

The situation is made more complicated when another person, known
in the literature as Wigner's friend, is included in the system.
He is in a room together with the container and at some point
looks in the container. Only later does he report his findings to us. 

This weird story accentuates some strange features of quantum mechanics.
We told it as if `the state' of a quantum system is something 
with physical reality, as it were, `engraved' in the particles 
constituting the system. This idea leads us to suppose states exist
which are very hard to imagine, and never observed in the real world.
We see that the `collapse of the state-vector', supposed to occur 
when a measurement takes place, seems to contradict the fact that 
measurement devices are physical systems themselves, and the device 
and the system being measured should evolve unitarily, not suddenly
jump randomly from one state to another. We see that the dividing line 
between quantum system and the outside world is completely arbitrary,
yet plays a central role in the theory (separating deterministic
unitary evolution from random state-collapse).

Many different standpoints can be taken on these issues. The most 
extreme are those of the empiricist (or instrumentalist, or 
pragmatician) on the one hand, and the realist
(who is actually an idealist) on the other. The empiricist does not 
believe in some kind of physical reality behind observed facts. He is 
interested only in making correct predictions about observable 
features of the world. For this person the only problem in our story 
is that the dividing line between quantum system and classical 
environment is somewhat arbitrary. If different descriptions 
lead to different prescriptions, there is a problem with the 
mathematical model. Below we present a simplified version of a 
consistency argument, which aims to show that there is no 
conflict between the two ingredients of quantum theory, and
no inconsistency when the Heisenberg divide between quantum system and 
outside world may be placed at several different places.

Very similar considerations as those used in the consistency argument 
are also often used to argue 
that the von Neumann (random) collapse of the wave function can be
\emph{derived} from (deterministic) Schr\"odinger evolution. However 
we are inclined to believe that such claims are incomplete. If one 
believes that the state of things in the world is described by 
wave-functions, one still has a problem in relating wave-functions to 
physical properties of real objects. This problem is supposedly 
addressed by Everett's many worlds theory, van Fraassen's modal 
interpretation, and Griffiths' and Omn\`{e}s' theory of 
consistent histories, among others. We find none of these 
attempts to make von Neumann redundant 
very convincing. However, the realist who wants the wave-function to 
be actually there in reality, and who believes that the true dynamics 
of physical systems is according to Schr\"odinger's equation alone, 
is forced in this direction. For cosmologists, wanting to model the 
whole universe without external observer, there seems to be a 
problem, since quantum randomness is a key part of modern theories of 
the origin of the universe.

The alternative for the realist is to extend or alter Schr\"odinger's 
dynamics in order to introduce a random element, which should make no 
difference to small quantum systems but should `simulate' the von 
Neumann collapse, on big ones. Two fairly well explored variants of 
this idea are Bohm's hidden variables model, and the `continuous 
spontaneous localisation' model of Ghirardi, Rimini and Weber. Most 
physicists are unhappy about these theories, since their claim to 
legitimacy is essentially that they \emph{reproduce} unitary evolution 
and wave-function collapse in the two extreme situations where these should 
hold; `in between' the physics is too difficult to make predictions, 
let alone test them by experiment. Thus the models do not seem to have 
new, testable consequences, while they include variables which 
determine the outcome of measurement, hence must be non-local.

Now we turn to the consistency argument, which aims to show that there
is no contradiction between Schr\"odinger evolution and von Neumann 
collapse, in the sense that placing the dividing line between quantum
system and outside world at different levels does not lead to different
conclusions (at least, for an observer who is always in the outside 
world). This particular version was communicated to us by Franz Merkl.

Consider a spin-half particle which passes through the magnetic field
of a Stern-Gerlach apparatus and then, if its spin is `up', hits a
photographic plate where a chain reaction produces a visible spot.
If the spin is `down' suppose the particle is lost. (This is a bit
simpler than allowing the spin-down particle to hit the photographic
plate at a different position: we have to model the
interaction only in the spin-up case). We will call the photographic plate
the detector. If the particle starts in the
state $\alpha|0\rangle+\beta|1\rangle$, where
where $|0\rangle$ and $|1\rangle$ represent spin-up and
spin-down respectively, and the coefficients $\alpha$ and $\beta$
satisfy $|\alpha|^2+|\beta|^2=1$, we get to see the spot with 
probability $|\alpha|^2$. Now the consistency problem arises
because we could just as well have considered particle plus 
photographic plate as one large quantum system evolving jointly
under some Hamiltonian for some length of time. If the detector
started off in some pure state, then the final joint state of
the joint system is another pure state, and no random jump to one
of two possible final states has taken place. Let us however admit
that the large system of the photographic plate involves many, many
particles, and repetition of the experiment with the whole system in
an identical pure state is physically meaningless to consider. At
each repetition there are myriads of tiny differences. 
Therefore physically relevant predictions are only obtained when
we use a mixed state as input for the macroscopic system.
To make the mathematics even more simple, we will suppose that what
varies from instance to instance is the length of time of the 
interaction. Let $|\psi\rangle$ be the state-vector of the 
detector, before the interaction starts. The joint
system starts in the pure state with state-vector 
$(\alpha|0\rangle+\beta|1\rangle)\otimes|\psi\rangle$.
Now the Hamiltonian of the interaction between particle and 
detector must be of the form 
$|0\rangle\langle 0| \otimes H$ where $H$ acts on the huge Hilbert
space of the detector, since there is a change to the
detector if the particle starts in the spin-up state,
but not at all if the particle starts in the spin-down state.
Let the length of time of the interaction be $\tau$.
Then the final state of the joint system after the interaction
is the pure state with state-vector
$\alpha|0\rangle\otimes e^{-i H \tau /\hbar}|\psi \rangle
+\beta|1\rangle\otimes |\psi\rangle $.
The corresponding density-matrix can be written out, partitioned
according to the first component of the joint system, as
\begin{equation*}
\left(
\begin{matrix}
|\alpha|^2 
e^{-i H \tau /\hbar}|\psi \rangle \langle \psi | e^{i H \tau /\hbar}
&
\alpha\overline\beta e^{-i H \tau /\hbar}|\psi \rangle \langle \psi |
\\
\overline\alpha \beta |\psi \rangle \langle \psi | e^{i H \tau /\hbar}
&
|\beta|^2  |\psi \rangle \langle \psi |
\end{matrix}
\right)
\end{equation*}
Now suppose we replace $H\tau$ by $H\tau+I\epsilon$ where $I$ is
the identity matrix. The idea here is that $H\tau$ must in some sense
be large, since it produces a macroscopic change in a large
quantum system. Thus this is a tiny perturbation
of the interaction if $\epsilon$ is small, but on the other hand,
since $\hbar$ is so tiny, $\epsilon/\hbar$ can still be very large.
As we vary $\epsilon$ smoothly over some small interval,
$\epsilon/\hbar$ varies smoothly over a huge range of values, and
therefore the fractional part of $\epsilon/(2\pi\hbar)$ is 
close to uniformly distributed over the interval $[0,1]$.
Consequently, the factor $e^{-i \epsilon / \hbar}$ is close to uniformly
distributed over the unit circle. Now after we have made this
perturbation to the interaction, the density matrix of the joint
state is
\begin{equation*}
\left(
\begin{matrix}
|\alpha|^2 
e^{-i H \tau /\hbar}|\psi \rangle \langle \psi | e^{i H \tau /\hbar}
&
e^{-i \epsilon / \hbar}
\alpha\overline\beta e^{-i H \tau /\hbar}|\psi \rangle \langle \psi |
\\
e^{i \epsilon / \hbar}
\overline\alpha \beta |\psi \rangle \langle \psi | e^{i H \tau /\hbar}
&
|\beta|^2  |\psi \rangle \langle \psi |
\end{matrix}
\right).
\end{equation*}
On averaging over $\epsilon$, the off-diagonal factors disappear and
we find the density matrix
\begin{equation*}
\left(
\begin{matrix}
|\alpha|^2 
e^{-i H \tau /\hbar}|\psi \rangle \langle \psi | e^{i H \tau /\hbar}
&
0
\\
0
&
|\beta|^2  |\psi \rangle \langle \psi |
\end{matrix}
\right).
\end{equation*}
This is the density matrix of the joint system which with probability
$|\alpha|^2$ is in the pure state with state-vector
$|0\rangle\otimes e^{-i H \tau /\hbar}|\psi \rangle$ and with
probability $|\beta|^2$ is in the pure state with state-vector
$|1\rangle\otimes |\psi \rangle$. In other words, either a
spin-up particle and a detector which indicates a particle was
detected, or a spin-down particle and a detector which indicates
no particle was detected.

This argument is simple and one can criticise it in many ways.
One would prefer to put the initial randomness into the many
particles making up the detector, rather than into the interaction,
and it should not have such a special form. 
But this is not a problem. Much more realistic models can
be worked through which lead to the same qualitative conclusion:
allowing variability in the initial conditions of the macroscopic
measuring device, of a most innocuous kind, allows random phase 
factors such as $e^{-i \epsilon / \hbar}$ to wipe out off-diagonal
terms in a large density matrix, so that all future predictions
of the joint system are the same as if a random jump had occured
during the initial interaction to one of two macroscopically distinct 
states.

In conclusion, it seems that as long as one is interested in 
using quantum mechanics only to predict what happens in a small part of 
the universe, and takes the randomness of quantum mechanics as 
intrinsic, not something which should be explained in a deterministic 
way, there are no logical inconsistencies in the theory. The state 
vector or state matrix of a quantum system should not be thought of as 
having an objective reality, somehow `engraved' in the physical nature 
of a single instance of some quantum system, but is rather a 
characteristic of the preparation of the quantum system which, at least 
conceptually if not actually, could be repeated many times. Thus a 
statistical description goes in, and a statistical description comes 
out. The working quantum physicist even makes do without the von Neumann collapse 
of a quantum system, on measurement, since realistic quantum mechanical 
modelling of the quantum system under study \emph{together} with the 
macroscopic measurement device allows one to introduce statistical 
variation in the initial state of the measurement device of the kind 
we have just described, and this leads irrevocably, it seems, to 
density matrices which are diagonal in the bases expressing 
macroscopically distinguishable states. In other words, 
unitary evolution alone, starting from the mixed initial state of 
quantum system plus measuring environment, is enough to determine the 
correct probability distribution over macroscopically distinguishable, 
thus `real world', outcomes. The working quantum physicist is also 
well aware that the Hamiltonians he uses are only `effective 
Hamiltonians' relative to some energy cut-off, which in turn 
corresponds to some approximation of a much larger state space by a 
smaller one. So the concerns of 
workers in the foundations of physics, worried about whether 
`the state vector of the universe' evolves in a unitary, deterministic 
way, or a random, non-unitary way, could turn out in the long run to 
be as purely academic as those of medieval theologians trying to calculate how many 
angels could dance on the head of a pin, since sooner or later 
physicists will learn that quantum mechanics was itself only a 
limiting case of a better theory, as happened to Newtonian mechanics 
before. If we think about it carefully, we realise that the reality of 
basic concepts of classical physics is as illusory as that of basic 
concepts of modern 
physics.%
\marginpar{*}\endnote{R: book: Fuchs \& Peres;  Mandel's ghost, bomb testing 
[Penrose, Science]}

\section{Some Further Topics}\label{s:conclude}

\subsection{Quantum stochastic processes}\label{ss:qstochproc}

Since its inception in the early 1980's, through pioneering work of 
Hudson and  Parthasarathy,
quantum---or noncommutative---probability has 
grown into a mature and sophisticated mathematical field. 
The criticsm which we levelled at the philosphical standpoint of its 
protagonists in Section  \ref{ss:classicalvsquantum} does nothing to 
reduce the mathematical and physical results which have been achieved;
see, for instance, \citet{accardietal97}.
An excellent introduction to
the field has been given by \citet{biane95} and a more comprehensive account is
available from the hand of \citet{meyer93}, see also \citet{parthasarathy92}.
A new journal \textit{Infinite Dimensional Analysis, Quantum Probability and Related
Topics}, now in its fourth year, is home to many of the more recent
developments. Here we shall summarise briefly some aspects of quantum
stochastic processes, under several subheadings.

\paragraph{Quantum optics}

Quantum optics is one of the currently most active and exciting fields of
quantum physics, particularly from the viewpoint of the present paper. Laser
cooling, on which we comment separately below, is, or may be viewed as, one
of the areas in this field. Here we discuss briefly the Markov quantum
(optical) master equation (MQME) and its quantum stochastic differential
equation (QSDE) counterparts.

The Markovian quantum master equation provides an (approximate) description
of a wide range of quantum system evolutions. The MQME is of the form 
\begin{equation*}
\dot{\rho}(t)=L(t)\rho (t),
\end{equation*}
where $L(t)$\ is a linear operator. In order for this equation to have a
solution\ such that $\rho (t)$ is a density operator for each $t$, $L(t)$\
must be of the \emph{Lindblad form} (\citealt{lindblad76}) 
\begin{equation}
L\rho =-\frac{i}{\hbar }[H,\rho ]+\sum_{k}\Bigl(A_{k}\rho A_{k}^{\ast }-
\frac12\rho A_{k}^{\ast }A_{k}-\frac12 A_{k}^{\ast }A_{k}\rho \Bigr) \thinspace ,
\label{Lindblad}
\end{equation}
where $H$\ is some Hermitian operator and the $A_{k}$ are (bounded)
operators. To each such operator there exists a variety of QSDE's for a
process $\psi (t)$ with values in $\mathcal{H}$\ such that, writing $\hat{
\rho}(t)=|\psi (t)\rangle \langle \psi (t)|/\langle \psi (t)|\psi (t)\rangle$, 
we have $\mathrm{E}[\hat{\rho}(t)]=\rho (t)$. See, for instance, 
\citet{moelmercastin96}, \citet{wiseman96} and 
\citet[{\ }Chap.\ 5]{gardinerzoller00}.

Interestingly, the same Markov quantum master equation has turned up 
in the Ghirardi-Rimini-Weber `continuous spontaneous localisation' 
approach to the measurement problem, whereby unitary Schr\"odinger 
evolution is replaced by a stochastic differential equation, which is 
able to mimic, according to the circumstances, both purely unitary 
evolution of a closed quantum system, and the von Neumann collapse of the wave 
function of a quantum system interacting with a large (measuring) 
environment.

To illustrate how equation (\ref{Lindblad}) can be numerically 
calculated by simulating many times a QSDE in what is called the 
quantum Monte Carlo approach, we consider the simplest case, 
when the index $k$ just takes a single value
and can therefore be omitted. Moreover, absorb the constant $\hbar$ 
into the Hamiltonian $H$. We show that the evolution is identical to 
the mean evolution of the following stochastic process
for an unnormalised state vector $\psi$: the deterministic but 
non-Hamiltonian evolution
$$
\dot\psi~=~-iH\psi-\frac12 A^{*}A\psi
$$
interupted by \emph{collapses}
$$
\psi~\to~A\psi
$$
with stochastic intensity
$$
I~=~\|A\psi\|^{2}/\|\psi\|^{2}.
$$
Introducing a counting process $N$ with intensity $I$ one can combine 
these equations into one QSDE of jump type,
$$
\mathrm d \psi~=~(-iH\psi-\frac12 A^{*}A\psi)\mathrm d t 
+(A\psi-\psi)\mathrm d N.
$$
Define $\widetilde \rho=\psi\psi^{*}$, the unnormalised random density 
matrix corresponding to the stochastic evolution, and 
$\widehat\rho=\widetilde\rho/\mathrm{tr}\,\widetilde\rho$. Note that 
$I=\mathrm{tr}(A\widetilde\rho A^{*})/\mathrm{tr}\,\widetilde\rho=
\mathrm{tr}(A\widehat\rho A^{*})/\mathrm{tr}\,\widehat\rho$.
Since $\mathrm d \widetilde\rho=\mathrm d \psi.\psi^{*}+
\psi.\mathrm d \psi^{*}$ and 
$\dot{\psi^{*}}=i\psi^{*}H-\frac12\psi^{*}A\psi$, the smooth part of 
the evolution can be rewritten as
$$\dot{\widetilde\rho}~=~-i[H,\widetilde\rho]-\frac 12 
(A^{*}A\widetilde\rho+\widetilde\rho A^{*}A).
$$
Taking the trace, we find on the smooth part $\mathrm 
d(\mathrm{tr}\,\widetilde\rho)=-\mathrm{tr}(A\widetilde\rho A^{*})$.
Together, this yields
\begin{align*}
     \frac{\mathrm d \widehat 
     \rho}{\mathrm d t}~&=~\frac{1}{\mathrm{tr}\,\widetilde 
     \rho}\frac{\mathrm d\widetilde\rho}{\mathrm d t} -
     \frac{\widetilde \rho}{(\mathrm{tr}\,\widetilde\rho 
     )^{2}}\frac{\mathrm 
     d(\mathrm{tr}\,\widetilde\rho)}{\mathrm d t}\nonumber
\\
     &=~-i[H,\widehat\rho]-\frac12(A^{*}A\widehat\rho+\widehat\rho 
     A^{*}A) +I\widehat\rho.\nonumber
\end{align*}
For the jump part, define $N(t)$ to be the number of jumps in the time 
interval $(0,t]$. Then at a jump time we can write
\begin{align*}
    \mathrm d \widehat\rho~&=~\Bigl( \frac{A\widehat\rho_{-} A^{*}}
    {\mathrm{tr}(A\widehat\rho_{-} 
    A^{*})}-\widehat\rho_{-}\Bigr)\mathrm d N\nonumber\\
    &=~\Bigl( \frac{A\widehat\rho_{-} A^{*}}
    {I(t)}-\widehat\rho_{-}\Bigr)(\mathrm d N -I\mathrm d t)
    +(A\widehat\rho_{-} A^{*}-I \widehat\rho_{-})\mathrm d t.\nonumber
\end{align*}
Together this gives, at all time points,
\begin{align*}
    \mathrm d \widehat\rho~=~
    \Bigl( -i[H &,\widehat\rho]-\frac12(A^{*}A\widehat\rho+\widehat\rho 
     A^{*}A) + A\widehat\rho A^{*}\Bigr)\mathrm d t\nonumber\\
     &+\Bigl( \frac{A\widehat\rho_{-} A^{*}}
    {I(t)}-\widehat\rho_{-}\Bigr)(\mathrm d N -I\mathrm d t).\nonumber
\end{align*}
Taking the expectation throughout, the martingale part (the second line) of this 
equation disappears, and $\widehat\rho$ in the first line is replaced by its expected 
value which we call $\rho$. The resulting nonstochastic differential 
equation for $\rho$ is precisely (\ref{Lindblad}). Moreover since 
$\widehat\rho$ was by construction a random density matrix (nonnegative, 
self-adjoint and trace one) we see that the solution $\rho$ of 
(\ref{Lindblad}),
being the expected value of a density matrix,
is also a density matrix; something which is not obvious from (\ref{Lindblad}).

\begin{example}[Quantum Monte Carlo for spin-half]
Consider a two dimensional quantum system and choose a basis such 
that $H=E_{1}\ketbra 1 1 +E_{2}\ketbra 2 2$, for real numbers $E_{1}$ 
and $E_{2}$. These are the two energy levels of the Hamiltonian. 
Suppose $A$ is diagonal in this basis with $A\ket 2=\alpha\ket 1$ and 
$A\ket 1=0$ (the zero vector), where $\alpha$ is real. This is the 
model for the energy of a  two-level atom which, on the spontaneous emission of a 
photon to its environment, can decay from its excited state to its 
ground state. Consider the evolution of an unnormalised state $\psi=c_{1}\ket 
1+c_{2}\ket 2$, where $c_{1}$ and  $c_{2}$ are complex functions of 
time. One discovers, since $H$ and $A^{*}A$
are simultaneously diagonalizable, that the smooth part of the 
evolution decouples as $\dot c_{1}=(-iE_{1}-\frac12\alpha^{2})c_{1}$,
$\dot c_{2}=(-iE_{2})c_{1}$. Thus starting in state $\ket 1$ or in 
state $\ket 2$, we stay there, as long as no collapse occurs. If we 
are in state $\ket 2$ collapse has intensity $0$. However in state 
$\ket 1$ there is a constant intensity $\alpha^{2}$ of collapse to 
state $\ket 1$. Thus starting in state $\ket 1$, the QSDE predicts an 
exponential waiting time of collapse to $\ket 2$ with rate $\alpha^{2}$.
The reader may like to compute the probability distribution of the 
time to collapse to state $\ket 2$, starting from an arbitary pure 
state $\psi=\alpha\ket 1+\beta\ket 2$.
\hfill$\ \qedsymbol$\end{example}

As we remarked above, the same Lindblad equation can be represented 
as the mean evolution of a whole range of QSDE's, of jump type,
diffusion type, and mixed type. Consider the same Lindblad 
equation as we were discussing above (no summation over $k$, drop 
$\hbar$). For an arbitrary real number $\mu$ define two matrices 
$D_{\pm}=(\mu\boldone \pm A)/\sqrt 2$. Then the original Lindblad 
equation can be rewritten again in Lindblad form, with two different 
values of $k$, and the corresponding $A_{k}$ being $D_{+}$ and $D_{-}$.
This has a Quantum Monte Carlo representation of a smooth evolution
$\dot\psi=(-iH-\frac12 D_{+}{D_{+}}^{*}-\frac12 D_{-}{D_{-}}^{*})\psi$, 
interupted by collapses $\psi\to D_{\pm}\psi$ 
with intensities $\|D_{\pm}\psi\|^{2}/\|\psi\|^{2}$. The total 
intensity of jumps can be calculated as 
$\mu^{2}+\|A\psi\|^{2}/\|\psi\|^{2}$. As $\mu\to\infty$ the rate of 
jumping increases without limit, but the relative change in the 
state at each jump becomes smaller and smaller. In the limit 
(after normalising suitably) one obtains a diffusion representation
$$
\mathrm d\phi~=~\Biggl(-iH\phi+\frac12\Bigl(\phi^{*}A\phi.A-\frac12 A^{*}A -\frac 
12 \phi^{*}A^{*}\phi . \phi^{*}A\phi \Bigr)\phi \Biggr)\mathrm d t 
+\frac12 \Bigl(2A-\phi^{*}A\phi
-\phi^{*}A^{*}\phi\Bigr)\mathrm d W
$$
where $W$ is of course a standard Wiener 
process.

\paragraph{Laser cooling}

The paper by \citet{moelmercastin96} on Monte Carlo techniques, for
calculating expectation values for dissipative quantum systems, has been
instrumental in particular in the context of laser cooling. Laser cooling is
a topic of great current in interest in physics, both from the theoretical
point of view and in terms of experimental advances opening up possibilities
of studying many basic quantum phenomena, for instance Bose--Einstein
condensation.

For a full understanding of the posssibility of subrecoil cooling, leading
physicists were led to develop theoretical results that from the viewpoint
of probability belong to renewal theory and add interesting new results and
problems to that theory. For an introduction to this, see 
\citet{barndorffnielsenbenth01}. 
A comprehensive account is given in \citet*{bardouetal01}. 
\citet*{barndorffnielsenbenthjensen00a,barndorffnielsenbenthjensen00b}
present some extensions to the setting of 
(classical) Markov processes.

\paragraph{Quantum infinite divisibility and L\'{e}vy processes}

Several types of quantum analogues of infinite divisibility and L\'{e}vy
processes have recently been introduced. Two belong to free probability and
are mentioned below. Infinitely divisible instruments and associated
instrumental processes with independent increments are discussed in 
\citet{holevo01}. See also \citet[{\ }Chap.\ 7]{meyer93}, 
\citet{barchiellipaganoni96},
and \citet*{albeverioetal01}.

\paragraph{Free probability and random matrices}

The subject area of free probability evolves around the concept of free
independence, also termed freeness. The latter was originally introduced by
Voiculescu in the mid 1980's in a study of free-group von Neumann factors
but was shortly afterwards realised to be naturally connected to the
limiting properties of products of large and independent 
self-adjoint random matrices (of complex numbers).
More specifically, suppose that $X_{i}^{(n)}$, $i=1, \dots ,r$, are independent 
$n\times n$\ random matrices, the entries in each of these matrices being
also independent, and consider the mean values of the form 
\begin{equation}
\mathrm{E}[\mathrm{tr}(X_{i_{1}}^{(n)} \dots X_{i_{p}}^{(n)})].
\label{rmmean}
\end{equation}
Under some mild regularity assumptions, for any given index set 
$i_{1}, \dots ,i_{p}$ and for $n\rightarrow \infty $, the quantity (\ref{rmmean}) 
will have a limiting value, and the collection of such mean values
corresponds to a random limiting object. Freeness expresses how the
independence of $X_{1}^{(n)}, \dots ,X_{r}^{(n)}$ is reflected in properties of
that object. It is now possible to develop a theory of free infinite
divisibility and free L\'{e}vy processes that to a large extent parallels
that of infinite divisibility and L\'{e}vy processes in classical
probability but also exhibit intriguing differences from the latter. There is,
in particular, a one-to-one correspondence between the class of infinitely
divisible laws in the classical sense and the class of the free infinitely
divisible laws, with the `free normal distribution' being the Wigner, or
semicircle, law which has probability density 
\begin{equation*}
\pi ^{-1}(1-x^{2}/2)^{1/2}.
\end{equation*}
This law was first derived by Wigner in the 1950's as the limiting law of
the distribution of eigenvalues of a random Hermitian matrix $X^{(n)}$ with
independent, complex Gaussian entries. Wigner's motivation for studying the
eigenvalue distribution was based on the supposition that the local
statistical behaviour of the energy levels of a sufficiently complex
physical system is approximately simulated by that of the eigenvalues of a
random matrix (Hamiltonian), see \citet{wigner58} and \citet{mehta67}.

More detailed summaries of the mathematical connections indicated above are
available in \citet{biane98a,biane98b} and 
\citet{barndorffnielsenthorbjoernsen01}. Furthermore, there are deep connections 
between the theory of random
matrices and that of longest increasing subsequences, see for instance
\citet{deift00}.
We also wish to draw attention to a recent paper by \citet{bianespeicher01} 
which introduces a concept of free Fisher information.

\paragraph{General framework and continuous-time measurements}

The generic mathematical description of the measurement process embodied in
formula (\ref{cpinstr}) applies, in particular, to situations where a quantum 
system is observed continuously over a time interval $[0,T]$. 
For each 
time point $t\in[0,T]$, a representation such as in (\ref{cpinstr}) is 
available for the data as available at that moment,
but it is a highly non-trivial task, carried out by 
\citet{loubenets99,loubenets00}, \citet{barndorffnielsenloubenets01}
to mesh these representations together in an 
interpretable and canonical fashion.
For simplicity, 
consider the case when the index $i$ in (\ref{cpinstr}) takes only one value
and hence can be omitted. Often the outcome of a
measurement of this type can be considered as the realisation of a cadlag
stochastic process $x_{0}^{T}=\{x_{t}:0\leq t\leq T\}$ on $\mathbb{R}$ and
the evolution of this and of the quantum system are determined by a
probability measure $\nu $ on $D[0,T]$ and a collection of mappings 
$W_{s}^{t}(x_{0}^{t})$, $0\leq s<t\leq T$ from $\mathcal{X}=D[0,T]$ to 
$\mathbb{B(\mathcal{H})}$, satisfying the normalisation relations 
\begin{equation*}
\int_{D[0,T]}W_{s}^{t}(x_{0}^{t})^{\ast }W_{s}^{t}(x_{0}^{t})\nu (\mathrm{d}
x_{s}^{t}|x_{0}^{s})=I
\end{equation*}
and the cocycle conditions 
\begin{equation*}
W_{s}^{t}(x_{0}^{t})=W_{\tau }^{t}(x_{0}^{t})W_{s}^{\tau }(x_{0}^{\tau })
\thinspace .
\end{equation*}
If the initial state of the quantum system in the Hilbert space $\mathcal{H}$
is a pure state $\psi _{0}$ then its evolutionary trajectory, conditional
on $x_{0}^{T}$, is given by
\begin{equation*}
\psi _{t}(x_{0}^{t})=W_{s}^{t}(x_{0}^{t})\psi _{0}  \thinspace .
\end{equation*}
Under suitable further conditions, the evolutions of $x_{t}$ and $\psi _{t}$
will be Markovian.

\subsection{Differential-geometric aspects}

In asymptotic parametric inference, differential geometry has proved to 
be an appropriate language for expressing various key concepts, 
see \citet[{\ }Chaps.\ 5--7]{barndorffnielsencox94}, 
\citet{kassvos97}. 
Likewise, several concepts in quantum mechanics have differential-geometric 
interpretations.
In particular, the quantum information $I(\theta)$ of a parametric quantum model is a 
Riemannian metric on the parameter space $\Theta$, as is the Fisher 
information $i(\theta ;M)$ obtained by a measurement $M$. 
There are many other Riemannian metrics of importance in quantum theory. 
A characterisation of a large class of them is given in 
\citet{petz96}. See also \citet{petzsudar99}.
Any (complex) Riemannian metric on the space $\mathbb{SA}(\cal{H})$ of 
self-adjoint operators on a finite-dimensional $\cal{H}$ (and 
satisfying some mild conditions)
yields an inequality analogous to Helstrom's quantum Cram\'er--Rao 
inequality (\ref{QCRB}). These inequalities and results on geometries 
obtained from suitable real-valued functions on $\Theta \times \Theta$ are 
given in \citet[{\ }Chap.\ 7]{amarinagaoka00}.
Some other differential-geometric aspects of quantum theory are considered in 
\citet{brodyhughston01}.

\subsection{Concluding Remarks}

This paper has sketched some main features of quantum statistical 
inference, and more generally, quantum stochastic modelling. The basic 
concepts for our paper coincide with the basic concepts of quantum 
computation, quantum cryptography, quantum information theory,
see \citet{gruska99,gruska01}, \citet{nielsenchuang00}.
We hope that many statisticians will venture into these areas too, 
as we are convinced 
that probabilistic modelling and statistical thinking will play 
major roles there, and should not be left purely to computer scientists 
or theoretical 
physicists.

\paragraph{Acknowledgements}

We gratefully acknowledge Mathematische Forschungsinstitut Oberwolfach 
for support through the Research in Pairs programme, and the European 
Science Foundation's programme on quantum information for supporting a 
working visit to the University of Pavia.

We have benefitted from conversations with many colleagues. 
We are particularly grateful to Elena Loubenets, Hans Maassen, 
Franz Merkl, Klaus M{\o}lmer and 
Philip Stamp.

\appendix

\section{Mathematics of Quantum Instruments}

Recall that an instrument $\mathcal N$ with outcomes $x$ in the measurable space
$(\mathcal X, \mathcal A)$, is defined through a collection of 
observables $\mathcal N(A)[Y]$, for each $A\in\mathcal A$ and each 
bounded self-adjoint $Y$. With $\pi(\mathrm d x;\rho,\mathcal N)$ 
denoting the
probability distribution of the outcome of the measurement, and 
$\sigma(x;\rho,\mathcal N)$ denoting the posterior state when the prior 
state is $\rho$ and the outcome of the measurement is $x$, we have
$$
\mathrm{tr}\{\rho \mathcal N(A)[Y]\}~=~\int_{A}\pi(\mathrm d x;\rho,\mathcal 
N)\mathrm{tr}\{(\sigma(x;\rho,\mathcal N) Y \}
$$
Thus if one `measures the instrument' on the state $\rho$, 
registers whether or not the outcome 
is in $A$, and subsequently measures the observable $Y$, the expected 
value of the outcome so obtained equals the expected value of the 
outcome of measuring directly the osbervable 
$\mathcal N(A)[Y]$.

\subsection{Complete Positivity}\label{aa:cp}

The observables $\mathcal N(A)[Y]$ are sigma-additive in $A$, 
linear in $Y$, nonnegative in $Y$ (map non-negative operators to non-negative 
operators), and normalised by $\mathcal N(\mathcal 
X)[\boldone]=\boldone$. Any collection satisfying these constraints 
is called a \emph{positive instrument}. Now given a positive 
instrument $\mathcal N$ defined on a 
Hilbert space $\mathcal H$, we can extend the instrument to the 
tensor product of this space with another Hilbert space $\mathcal K$ 
by defining $\mathcal N(A)[Y\otimes Z]=\mathcal N(A)[Y]\otimes Z$. 
This corresponds intuitively to measuring $\mathcal N$ on the first 
component of a quantum system in the product space, 
leaving the second component untouched. By linearity, once the 
extended instrument is defined on product observables like $Y\otimes 
Z$, it is defined on all observables of the product system.
An instrument $\mathcal N$ is called \emph{completely positive} if 
and only if every such extension (i.e., for any auxiliary system $\mathcal 
K$) remains positive. It turns out that one need only verify the 
positivity of the extensions for $\mathcal K$ of dimension
$2,3,\ldots,\mathrm{dim}(\mathcal H)+1$.

Here is a classic example of an instrument which is positive, but not 
completely positive, hence is not physically realisable.

\begin{example}[A positive, but not completely positive, 
instrument]\label{e:Qinstr}
Let the outcome space be trivial (consisting of a single element) so the 
instrument only transforms the incoming state, and does not generate 
any data. We therefore just specify an observable $\mathcal N[Y]$ for 
each observable $Y$: we define it by $\mathcal 
N[Y]=Y^{\top}$, the transpose of the observable $Y$. This corresponds 
to the outcome state $\sigma(\rho;\mathcal N)=\rho^{\top}$. Now 
take $\mathcal K=\mathcal H$, of finite dimension $d$, and define 
$\ket\psi=\frac 1 d  \sum_{i}\ket i \otimes \ket i$ where the 
vectors $\ket i$ form an orthonormal basis of $\mathcal H$, take 
$\rho=\ketbra\psi\psi$. Let $\sigma=\rho^{\top}$ denote the 
corresponding output state.
As a matrix operating on vectors,
$\sigma(\sum_{i}c_{i} \ket i \otimes 
\sum_{j} d_{j} \ket j)=(\sum_{i}d_{i}\ket i \otimes 
\sum_{j} c_{j}\ket j)$. Thus in particular, $\sigma$ 
maps $\ket i\otimes\ket j - \ket j 
\otimes\ket i$ to minus itself. Hence it has negative eigenvalues, and 
therefore cannot be a density matrix.
\hfill$\ \qedsymbol$\end{example}

Any dominated measurement $M$ can be embedded into an instrument. The 
simplest way is by taking the posterior states to be $m(x)^{\frac 
12}\rho m(x)^{\frac 12}/\mathrm{tr}(\rho m(x))$ for each outcome $x$ 
having a positive density $\mathrm{tr}(\rho m(x))$ with respect to 
the same measure $\nu$ which dominates $M$. This corresponds to there 
being only one index $i$ in (\ref{cpinstr}), and $W(x)=m(x)^{\frac 12}$.

The next example illustrates the need to allow unbounded operators 
$W_{i}(x)$ in (\ref{cpinstr}), even if the completely positive
instrument in question is 
bounded.

\begin{example}[Position measurement]\label{e:posinstr} 
As in Section \ref{ss:harmo}
take as Hilbert space 
$\mathcal H =L^2_{\mathbb C }(\mathbb R )$ and consider the PProM
corresponding to the position observable $Q$. Thus the operator $Q$ 
simply multiplies an $L^2$ function of $x$ by the identity function 
$x\mapsto x$. The PProM has elements $M(B)$, for each Borel subset 
$B$ of the real line, equal to the operator which multiplies an 
$L^2$ function by $1_{B}$, the indicator function of the set $B$. In 
other words, $M(B)$ projects onto the subspace of functions which are zero 
outside $B$. The intuitively natural way to consider this measurement 
as part of an instrument would be to take the posterior state, given 
that the outcome is $x\in\mathbb R$, to be a delta-function at the 
point $x$. This is not an element of $\mathcal 
H$. However, one can easily imagine the following instrument 
$\mathcal N$: measure 
$Q$, and replace the quantum system by a new particle in the fixed 
state $\rho_{0}$, independently of the outcome $x$. (We reconsider the 
original instrument, later).
By the physical interpretation of $\mathcal N(B)[Y]$, we must have, for any 
state $\rho$, that $\mathrm{tr}(\rho\mathcal N(B)[Y])
=\mathrm{tr}(\rho 1_{B})\mathrm{tr}(\rho_{0}Y)$. Suppose $\rho_{0}$ 
is the pure state with state vector $\ket{\psi_{0}}$. Then informally, 
in (\ref{cpinstr}), one should have a single index $i$, dominating 
measure $\nu$ equal to Lebesgue measure, and 
$W(x)=\ketbra x {\psi_{0}}$ where the $\ket x$ stands for the 
delta-function at $x$, thus is not a particular member of $\mathcal 
H$, but is defined through the formula $\braket x \psi = \psi(x)$.
Thus $W(x)$ is the operator defined on the subspace of continuous 
$L^{2}$ functions $\psi$ by $W(x)\psi = \psi(x) \psi_{0}$. It cannot 
be extended in a continuous way to all of $L^{2}$, and is therefore an 
unbounded operator. The 
instrument $\mathcal N$ can be written as $\mathcal{N}(\mathrm d x)[Y] = 
\ketbra {\psi_{0}}{\psi_{0}} \, \bra x Y \ket x \, \mathrm d x$, or
$\mathcal{N}(B)[Y] = 
\ketbra {\psi_{0}}{\psi_{0}} \bra {1_{B}} Y \ket {1_{B}}$, which is 
defined for all bounded operators $Y$ and arbitrary Borel sets $B$.

Reconsider the instrument $\mathcal N'$ defined formally by $W(x)=\ketbra x x$.
Formally, we should have $\mathcal{N}'(\mathrm d x)[Y] = 
\ketbra x x \bra x Y \ket x \mathrm d x$ and thus 
$\mathcal{N}'(B)[Y] = \int_{B} \ketbra x x \bra x Y \ket x \mathrm d x$.
This formula is supposed to represent an observable, i.e., a possibly 
unbounded operator on $\mathcal H$. To find out what 
it does, we manipulate with delta-functions to find
$\bra \phi \mathcal{N}'(B)[Y] \ket \psi = \int_{\mathrm R}1_{B}\overline 
\phi\psi\mathrm d\mu_{Y}$ where $\mu_{Y}$ is the finite measure 
on the real line defined by $\mu_{Y}(A)=\bra {1_{A}} Y \ket {1_{A}}$.
Note that $\mu_{Y}$ is absolutely continuous with repect to 
Lebesgue measure $\nu$.
Thus $\mathcal{N}'(B)[Y]$ is defined on the subspace of $L^{2}$ functions, 
square integrable on $B$ with respect to $\mu_{Y}$, and on that subspace it 
acts by multiplying by the function $1_{B}\cdot \mathrm d \mu_{Y}/\mathrm 
d \nu$. The instrument $\mathcal N'$ is unbounded. It has an 
informal representation (\ref{cpinstr}) involving objects $W$ which 
cannot even be considered as unbounded operators, and
there does not exist a posterior state for each outcome $x$ of the 
instrument. There is a well-defined posterior state given the 
outcome lies in a set $B$ of positive probability 
$\pi(B;\rho)=\mathrm{tr}(\rho 1_{B})$.
It is formally defined by
$\sigma(B;\rho)=\int_{B}\ketbra x x \pi(\mathrm d x | B ;\rho)$.
\hfill$\ \qedsymbol$\end{example}

\subsection{Projection and Dilation of Measurements}\label{a:naimark}

Let $\Pi:\mathcal{H}^{\prime}\rightarrow\mathcal{H}$
be the orthogonal projection of a Hilbert space 
$\mathcal{H}^{\prime}$ onto a subspace $\mathcal{H}$.
Then $\Pi$ induces a map
$$
\Pi^{\ast}: \mathrm{OProM}(\mathcal{X},\mathcal{H}^{\prime}) 
\rightarrow
\mathrm{OProM}(\mathcal{X},\mathcal{H})
$$
by
\begin{equation}
(\Pi^{\ast}(M))(A)=\Pi M(A)\Pi^{\ast}\quad\quad A\in\mathcal{A}
\thinspace . 
\label{Pi}
\end{equation}
In the physical literature, the OProM $M$ is said to be a 
\emph{dilation} or \emph{extension} of $\Pi^{\ast}(M)$. 

The following theorem shows that every OProM can be obtained from some PProM
by the above construction: every generalised measure\-ment can be 
dilated to a simple measurement.

\begin{thm}[\citealt{naimark40}] 
Given $M$ in $\mathrm{OProM}(\mathcal{X}, \mathcal{H})$, there is 
(i) a Hilbert space $\mathcal{H}^{\prime}$ containing
$\mathcal{H}$, (ii) a projection-valued probability meas\-ure 
$M^{\prime}$ in
$\mathrm{PProM}(\mathcal{X},\mathcal{H}^{\prime})$, such 
that
$$
\Pi^{\ast}\left(  M^{\prime}\right)  = M
$$
(in the sense of (\ref{Pi})), where $\Pi:\mathcal{H}^{\prime}\rightarrow
\mathcal{H}$ is the orthogonal projection.
\end{thm}

The theorem of Naimark shows how to extend a generalised measurement
to a simple measurement on a larger space. There is also an obvious way
to consider a state on the smaller space as a state on the larger 
space, concentrating on the subspace.
These two extensions together do not have the same
statistical behaviour as the original pair of state and measurement.
Adapting the proof of Naimark's theorem one can show how to
extend an arbitrary state on the smaller space to a state on a 
larger space, in a way which matches the extension of the measurement,
and together reproduces the statistics of the original set-up.
This is taken care of by Holevo's theorem, Theorem \ref{t:holevo}
at the end of subsection \ref{ss:meas}.

\section{The Braunstein--Caves Argument}\label{a:bc}

A measurement $M$ with density $m$ with respect to a sigma-finite 
measure $\nu$ is given. 
Its outcome has density 
$p(x;\theta)=\mathrm{tr}\{\rho(\theta)m(x)\}$ with respect to $\nu$.
In the argument below, $\theta$ is also 
fixed. Define $\mathcal X_{+}=\{x:p(x;\theta)>0\}$ and $\mathcal 
X_{0}=\{x:p(x;\theta)=0\}$. 
Define $A=A(x)=m(x)^{\frac12}\rho_{/\!\!/\theta}\rho^{\frac12}$,
$B=B(x)=m(x)^{\frac12}\rho^{\frac12}$, and $z=\mathrm{tr}\{ A^* B\}$.
Note that $p(x;\theta)=\mathrm{tr}\{ B^* B\}$.

The proof of (\ref{qinfoineq}) given below consists of three inequality steps.
The first will be an application of the trivial inequality
$\Re(z)^2\le |z|^2$ with equality if and only if $\Im(z)=0$. 
The second will be an application of the Cauchy--Schwarz inequality 
$| \mathrm{tr} \{A^* B\} |^2 \le \mathrm{tr} \{A^* A\} \mathrm{tr} \{B^* B\} $ 
with equality if and only if
$A$ and $B$ are linearly dependent over the complex numbers.
The last step consists of replacing an integral of a nonnegative function
over $\mathcal X_{+}$ by an integral over $\mathcal X$. Here they are:
\begin{align}
  i(\theta;M)  ~ &= ~  \int_{\mathcal X_{+}}
        p(x;\theta)^{-1} ( \Re\,\mathrm{tr} (\rho\rho_{/\!\!/\theta} m(x) )^2 \nu(\mathrm d  x) 
               \nonumber \\
  ~ &\le ~ \int_{\mathcal X_{+}} p(x;\theta)^{-1} |\mathrm{tr} (\rho\rho_{/\!\!/\theta} 
               m(x))|^2  \nu(\mathrm d  x) 
                                                                \nonumber \\
  ~ &= ~  \int_{\mathcal X_{+}} \left| 
         \mathrm{tr}  \left( \,  m(x)^\frac12  \rho^\frac12  )^* \,
                  ( m(x)^\frac12  \rho_{/\!\!/\theta} \rho^\frac12 \, \right)
   \right|^2 (\mathrm{tr} (\rho m(x)))^{-1}\nu(\mathrm d  x)   \nonumber  \\
  ~ &\le ~  \int_{\mathcal X_{+}} 
  \mathrm{tr} ( m(x) \rho_{/\!\!/\theta} \rho \rho_{/\!\!/\theta} ) 
                \nu(\mathrm d  x)  \nonumber \\
  ~ &\le ~ \int_{\mathcal X}\mathrm{tr} ( m(x) \rho_{/\!\!/\theta} 
             \rho \rho_{/\!\!/\theta} ) \nu(\mathrm d  x)  \nonumber \\
  ~ &=  ~  I(\theta) .
\label{QIineq}
\end{align}
The necessary and sufficient conditions for equality at each of the 
three steps are therefore:
\begin{align*}
\Im(\mathrm{tr}\{ A(x)^* B(x)\} )~ &= ~0, \\
\alpha(x) A(x)+\beta(x) B(x)~ &= ~\boldzero, \\
\int_{\mathcal X_0}\mathrm{tr} \{ A(x)^* A(x) \} \nu(\mathrm d  x)~ &= ~0,
\end{align*}
where 
$\alpha(x)$ and $\beta(x)$ are arbitrary complex numbers, 
not both equal to zero, and the first two equalities are supposed to hold
$\nu$-almost everywhere where $p(x;\theta)$ is positive, while in the 
third equality $\mathcal X_{0}$ is precisely the set where $p(x;\theta)$ is zero.

Now if $A(x)=r(x)B(x)$ for real $r(x)$, for $\nu$ almost all $x$,
then $A^{*}B=rB^{*}B$ and its trace is real. Hence the first and second 
conditions are satisfied. Moreover, we then also have
$\mathrm{tr}\{A(x)A^{*}(x)\}=r(x)^{2}p(x;\theta)$
so the third condition is also satisfied.

Conversely, suppose all three conditions are satisfied. Since
$p(x;\theta)=\mathrm{tr}\{B(x)^{*}B(x)\}$, on $\mathcal X_{+}$ we must have 
$B$ non-zero and hence $\alpha$ non-zero. So (still on $\mathcal X_{+}$)
$A\propto B$ and the first condition implies that the proportionality constant 
must be real. The third condition implies that 
$\mathrm{tr}\{A(x)A(x)^{*}\}$ and hence $A(x)$ is almost everywhere 
zero where $p(x;\theta)=\mathrm{tr}\{B(x)^{*}B(x)\}=0$, i.e., where 
$B(x)=\boldzero$. So certainly one may write $A(x)=r(x)B(x)$ for some real $r(x)$ 
there, too.
 
In Braunstein and Caves' somewhat sketchy proof,  it seems to be 
assumed that $p(x;\theta)$ is everywhere positive, hence only two 
inequality steps are involved. We note that the main ingredient of these 
proofs is the Cauchy--Schwarz inequality. This is also the main step 
in proving Helstrom's quantum Cram\'er--Rao bound, and of course in proving 
the classical Cram\'er--Rao bound.

\bibliographystyle{chicago}

\bibliography{qiread9}



\end{document}